\documentclass[fleqn,usenatbib]{mnras}

\usepackage{newtxtext,newtxmath}

\usepackage[T1]{fontenc}

\DeclareRobustCommand{\VAN}[3]{#2}
\let\VANthebibliography\thebibliography
\def\thebibliography{\DeclareRobustCommand{\VAN}[3]{##3}\VANthebibliography}

\usepackage{graphicx}	
\usepackage{amsmath}	
\usepackage{url}
\usepackage{float}
\usepackage{physics}
\usepackage{parskip}
\usepackage{anyfontsize}
\usepackage{subcaption}
\usepackage{booktabs}
\usepackage{pdflscape}
\usepackage{longtable}
\usepackage{adjustbox}
\usepackage{caption}
\usepackage{xspace}
\usepackage{hhline}
\newcommand{\teff}{\ensuremath{T_{\textup{eff}}}\xspace}
\newcommand{\feh}{\ensuremath{[\text{Fe/H}]}\xspace}
\newcommand{\meh}{\ensuremath{[\text{M/H}]}\xspace}

\newcommand{\logg}{\ensuremath{\log g}\xspace}
\usepackage{enumitem}

\title[Homogeneous stellar characterisation]{Homogeneous stellar parameters for 717,807 TESS FGK stars using Gaia DR3}

\author[F. M. Waines et al.]{
Francesca M. Waines,$^{1}$\thanks{E-mail: francesca.waines.19@ucl.ac.uk}
Angharad Weeks$^{2}$ and
Vincent Van Eylen$^{1}$
\\
$^{1}$Mullard Space Science Laboratory, University College London, Dorking, Surrey, RH5 6NT\\
$^{2}$Sydney Institute for Astronomy, School of Physics, University of Sydney NSW 2006, Australia
}

\date{Accepted XXX. Received YYY; in original form ZZZ}

\pubyear{\the\year{}}

\begin{document}
\label{firstpage}
\pagerange{\pageref{firstpage}--\pageref{lastpage}}
\maketitle

\begin{abstract}
Precise homogeneous stellar characterisation is crucial for our understanding of the physical properties of exoplanets, their demographics and the environment from which they are formed. We present a homogeneous catalogue of 717,807 TESS FGK dwarfs and early subgiants, making use of isochrones along with Gaia DR3 inputs of photometry, parallax and spectroscopic temperature and metallicity, thus providing one of the largest homogeneous catalogues of stellar ages for TESS stars to date. We determine values for distance, \logg, \meh, \teff, radius, mass and age. For our best fit values, we calculate median fractional errors of 4.29\%, 4.49\%, 1.74\% and 50.00\% on radius, mass, temperature and age respectively. We compare and validate our catalogue values to various literature sources which employ other isochrone grids and asteroseismology. In addition, we identify 278 TESS exoplanet hosts and 915 candidates and recalculate the planet radii for such systems. These homogeneous parameters provide a state-of-the art sample to probe the effect of physical stellar parameters on exoplanet characteristics and architectures. 
\\
\end{abstract}




\section{Introduction}


Characterising the physical and atmospheric properties of stars is vital for providing insight into their structure and evolution. Our knowledge and understanding of stars has been revolutionised by large scale ground-based surveys, providing either photometric (e.g., SDSS, \citealt[][]{Kollmeier_2026}; and WISE, \citealt[][]{Wright_2010}) or spectroscopic information (e.g., RAVE, \citealt[][]{RAVE_1,RAVE_2}; GALAH, \citealt[][]{GALAH_1,GALAH_2}; GALEX, \citealt[][]{GALEX}; APOGEE, \citealt[][]{APOGEE}; and LAMOST, \citealt[][]{LAMOST}). In more recent times, space-based surveys such as Kepler \citep[][]{Koch_2010}, TESS \citep[][]{Ricker_2014} and Gaia \citep[][]{Gaia_collab_2016} have contributed greatly to the field, observing the positions and brightness of millions of stars allowing for the precise mapping of the Milky Way. For example, the latest Gaia data release \citep[DR3,][]{Gaia_collab_2016,Gaia_collab_2021,Gaia_collab_2023} includes 1.46 billion sources with full astrometric solutions and 1.806 billion sources with measured G magnitudes. 


Large catalogues are important in the advancement of numerous astrophysical fields. For example, Galactic Archaeology has benefitted from recent homogeneous and accurate measurements of stellar metallicities, masses and ages \citep[e.g.,][]{Frebel2010,Deason_Belukorov_2024,Kawate_2026}. These measurements have provided crucial insights into the beginnings and subsequent evolution of our galaxy, \citep[e.g.][]{casagrande_2011,Miglio_2013,Nissen_2020,Ciuca_2021,Jofre_2021,Anders_2023}. Furthermore, comparisons of observed parameters to those predicted from theoretical stellar models can reveal differences and thus help to test and improve the physical modelling of stellar interiors \citep[e.g.][]{kjeldsen_1995,Huber_2012,Yu_2018}.

The availability of reliable, homogeneous stellar parameters is also critical for exoplanet studies. In many cases, planetary properties such as mass and radius are derived indirectly through observations of the host star. Furthermore, linking exoplanet properties with those of their host stars is an important aspect of studying their formation and evolution. Host stars are often characterised on a case-by-case basis, employing a range of different methods, input parameters, and stellar models, therefore key catalogues are heterogeneous \citep[e.g., the NASA exoplanet archive,][]{NASA_exo_archive}. There are a number of homogeneous sub-samples of exoplanet hosts \citep[e.g.][]{Santos_2013,silva-aguirre_2015,Magrini_2022,Bonomo_2023,MacDougall_2023,Berger_2026}, however, it is challenging to use these catalogues in conjunction due to the differences in model grids and microphysics which can lead to offsets \citep[][]{tayar_2022}. Such offsets propagate into derived planetary properties, which may introduce spurious scatter or artificial trends when comparing planets across a heterogeneously characterised sample. Homogeneous characterisation ensures that any systematic uncertainties are consistent across the stellar sample, allowing relative trends between stars to be interpreted with greater confidence \citep[e.g.,][]{berger_2020b,Chen_2026}.

At the demographic level, stellar parameters are vital for understanding the influence of host star parameters on their constituent exoplanetary systems.  For example, hot Jupiters are found around more metal-rich stars \citep[][]{Fischer_2005,Mortier_2013}, and rocky planet composition depends on host star chemistry \citep[][]{Adibekyan_2021,Brinkman_2024,Brinkman_2025}. In addition, more massive, short-period planets seem to preferentially orbit multiple star systems \citep[][]{Mazeh_2003,eggenberger_2004,Eeles-nolle_2025}. Key demographic features such as the radius valley \citep{Fulton_2017,Vaneylen_2018} have also been shown to vary clearly with host star parameters
\citep[][]{David_2021,Chen_2022,Petigura_2022,Ho_2024,Venturini_2024}. These studies consist of smaller planet host samples, some employing their own stellar characterisation or making use of current homogeneous catalogues.
Recently, occurrence rate studies have performed more extensive investigations, identifying how common different planet populations are around a range of different stars. Such studies typically rely on large stellar samples of the order $\sim$100,000
\citep[][]{Yang_2020,Beleznay_Kunimoto_2022,Yee_2023,Gaidos_2024,Bryant_2025}. These studies require stellar parameters for all stars in the sample, not just those which host exoplanets. Both homogeneous and inhomogeneous sources are used for stellar properties for samples of this scale.  


One crucial parameter is stellar age, as it allows us to consider the history and evolution of exoplanet systems. Several evolutionary processes may occur over Gyr timescales, such as atmospheric loss \citep[][]{Owen_2013,Owen_2017} or dynamical interactions and orbital decay \citep[][]{Penev_2018,Hamer_Schlaufman_2019}. Additionally, several trends between exoplanet properties and stellar age have been investigated \citep[e.g.][]{Winter_2020,Chen_2021,PAST_2021,weeks_2025}. Notably, a number of studies have investigated exoplanet occurrence as a function of stellar age \citep[e.g.,][]{Bashi_2022,Sayeed_2025,PAST_2023}, however they are limited due to the difficulty in obtaining reliable stellar ages for large samples.


For many population studies, parameters from the TESS Input Catalog \citep[TIC,][]{Stassun_2018,Stassun_2019} are commonly used due to its size, along with TESS data \citep[e.g.][]{Beleznay_Kunimoto_2022,Bryant_2025}. TESS has monitored the brightness of millions of stars and the TIC provides various stellar properties such as temperature, radius, mass and \logg. Version 8 of the TIC \citep[TICv8][]{Stassun_2019} includes 1.7 billion point sources. However, stellar parameters from this catalogue involve a degree of inhomogeneity. Parameters are first adopted from specially curated catalogues \citep[e.g., Cool Dwarfs,][]{Muirhead_2018} if available. If no spectroscopic temperatures are available, temperature is estimated from a spline relation with Gaia G$_{BP}$ - G$_{RP}$ colours. Stellar radius is calculated from Gaia parallax with the Stefan Boltzmann relation, along with mass from a spline relation with temperature. Both physical properties are thus highly dependent on temperature, which is inhomogeneously derived. The TIC also does not include values for stellar age. 


Stellar ages are typically challenging to quantify as they cannot be directly determined by observations so have to be inferred through empirical relations or model fitting \citep[][]{Soderblom_2010}. Gyrochronology utilises the relationship between the stellar rotation period and age \citep[][]{Skumanich_1972,Barnes_2003,Barnes_2007,Mamajek_2008}, however this is only suitable for young stars, and requires calibration against asteroseismic results for later ages.
Kinematic ages can be determined from the velocity dispersion of a stellar group \citep[][]{Spitzer_1951,Sellwood_Carlberg_1984} rather than for individual stars. In some cases, asteroseismology can provide relatively precise stellar ages \citep[e.g.,][]{Lebreton_2009,silva-aguirre_2015,Li_2020}, using oscillations. However, it is often challenging to detect these oscillations in main-sequence stars due to their lower amplitudes, making asteroseismology difficult to apply to large samples. In this work we elect to use a Bayesian grid-based modelling approach, allowing for homogeneous characterisation of our full sample of FGK dwarfs, including ages, from more readily available stellar observables, e.g., photometry and spectroscopy. We compare stellar observations i.e. luminosity, metallicity and temperature to a grid of stellar models to infer their physical parameters such as age \citep[][]{Pont_2004,Jorgensen_2005,Angus_2019}. This method is applicable to a broad range of temperatures, metallicities and evolutionary states, making it ideal for large sample sizes of field stars. Specifically we use the BAyesian STellar Algorithm \citep[BASTA,][]{Aguirre_B_rsen_Koch_2021} to fit pre-computed stellar models \citep[using BaSTI isochrones,][]{Hidalgo_2018}, as this tool has shown success in a range of other studies \citep[][]{Persson_2022,Hon_2023,weeks_2025}.
%

We homogeneously characterise $\sim$700,000 TESS FGK dwarfs and early subgiants, fitting isochrones to Gaia DR3 photometry (G, G$_{BP}$, G$_{RP}$), parallax and spectroscopic \teff and \feh. We obtain \teff, \logg, \meh, radius, mass and age, and present one of the largest catalogue of homogeneous stellar parameters, including ages, for TESS stars to date, spanning a mass range of 0.6 -- 1.6 M$_\odot$. This paper is structured as follows: We define our sample selection in Section \ref{sec:sample selection}. In Section \ref{sec: BASTA fitting} we discuss our methodology in characterising the sample, detailing the preparation of inputs and model fitting. We compare and validate our best fit parameters to existing literature sources in Section \ref{sec:validation}, and discuss our results in Section \ref{sec:results}. We summarise our main findings in Section \ref{sec:conclusion}.

\section{Sample selection}
\label{sec:sample selection}

We define our sample by identifying stars that have been observed by the Transiting Exoplanet Survey Satellite \citep[TESS,][]{Ricker_2014}. TESS is a wide-field survey that has been monitoring the sky since its launch in 2018 with the primary goal of searching for planets orbiting the nearest and brightest stars. Using four identical wide-field cameras, the telescope monitors a 24 x 96 degree strip of the sky for 27.4 days, dubbed a sector whereby two orbits around Earth are completed. Comprising 1.7 billion point sources \citep[TICv8,][]{Stassun_2019} TESS provides an ideal initial sample to consider for stellar characterisation via isochrones, not only for further stellar studies but also for exoplanet demographics. We consider stars which have Quick Look Pipeline light curves available \citep[QLP,][]{Huang_2020a,Huang_2020b}. The QLP produces light curves for stars brighter than TESS magnitude 13.5 from the Full Frame Images (FFIs), being made publicly available from Mikulski Archive for Space Telescopes (MAST) as a High Level Science Product (HLSP)\footnote{\url{https://archive.stsci.edu/hlsp/qlp}}. We choose to restrict to QLP light curves from the first and second extended missions only (sectors 27-83) \citep[][]{Kunimoto_2021,Kunimoto_2022} since the FFIs are observed at higher cadence of 10-minutes and 200 seconds respectively compared to the primary mission (sectors 1-26) observed at 30-minute cadence.

We start by querying the TESS Input catalogue (TICv8), selecting both dwarfs (\logg $\geq$ 4.1) and a fraction of early subgiants (3.8 $\leq$ \logg < 4.1). We focus mainly on main sequence stars since they are more amenable to the detection of transiting exoplanets (Waines et al. in prep.) whilst also providing a large enough sample for isochrone characterisation. Additionally, we choose stars with a TIC stellar effective temperature of 4000 K $\leq$ \teff $\leq$ 6500 K, excluding M dwarfs since their isochrone ages are typically harder to determine and therefore less reliable \citep[][]{Mann_2015,Mann_2019,Berger_2020}.  

We next limit the sample to stars with parameters available in the  Gaia GSP-Spec catalogue. GSP-Spec provides values for \teff, \logg, \meh, [$\alpha$/Fe] and the abundances of various chemical species for 5.6 million stars, derived from the combined radial velocity spectrometer \citep[RVS,][]{Cropper_2018} spectra of single stars \citep[][]{Recio_Blanco_2023}. This is done using the Matisse-Gauguin algorithm \citep[][]{Recio_Blanco_2016,Creevey_2023}, where the best fit value for each parameter is calculated as the median value from 50 Monte-Carlo realisations. The uncertainties are determined as the 16th and the 84th quartiles of the resulting parameter distributions. We restrict to GSP-Spec inputs only since this catalogue provides reliable, homogeneous values for \teff and \meh which we can utilise as inputs in our fitting methodology (detailed in Section \ref{sec: BASTA fitting}). Additionally, such values from GSP-Spec are most accurate in the temperature range 4000 K -- 7000 K, since the spectra of cooler stars tend to have more atomic and molecular lines present, making them challenging to parametrise \citep[][]{Recio_Blanco_2023}. Similarly, this temperature range is best for stellar characterisation via evolutionary models; below $\sim$4000 K there is little difference between the isochrone tracks for stars on the main sequence, leading to degeneracies in the colour-magnitude plane which makes it harder to determine precise ages.  

To further limit the stellar sample to the best stars for isochrone characterisation, we filter the sample using the following error ranges for the GSP-Spec \meh, \teff and \logg:
\begin{enumerate}[leftmargin=*]
    \item $\Delta \sigma_{\meh} <$ 0.5
    \item $\Delta \sigma_{\teff} <$ 750
    \item $\Delta \sigma_{\logg} <$ 1
\end{enumerate}

 After querying the Gaia DR3 catalogue, we obtain the GSP-Spec \meh, \logg and \teff values along with the GSP-Spec flags. Such flags indicate the quality of the data, and thus any biases in fitted parameters. This is given by a number from 0 to 9, with 0 being the best and 9 the worst. To ensure the removal of potentially problematic stars which may cause issues with stellar characterisation, we only keep stars which have values of 0 for the first 6 flags. These flags are indicative of potential bias in the derived temperature, metallicity and \logg values which arise from rotational line broadening or uncertainties in radial velocity therefore affecting the quality of the spectra, where highly flagged systems represent low quality spectra and poor spectroscopic solutions.

\begin{figure}
    \hspace*{-0.3cm}
	\includegraphics[width=\columnwidth]{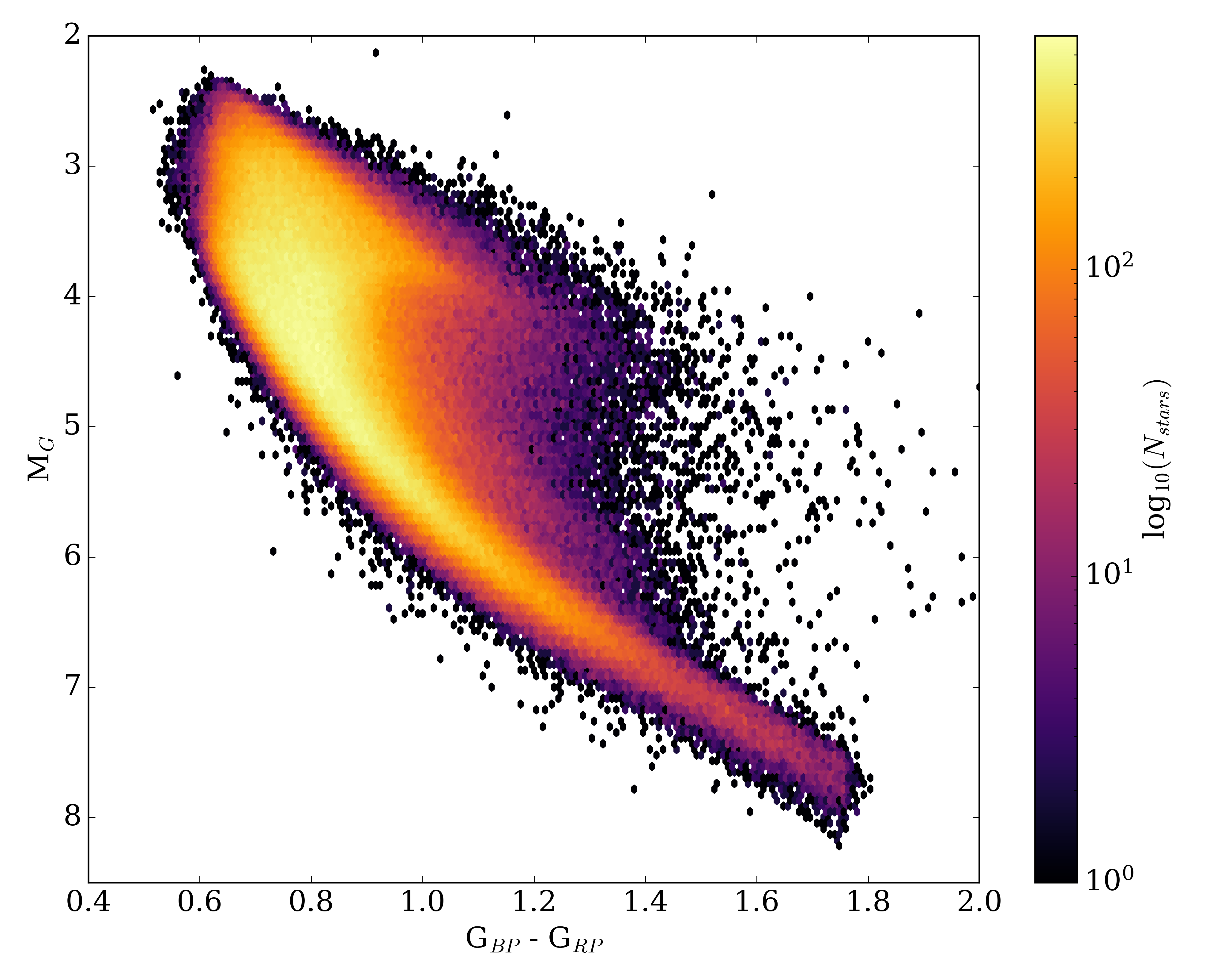}
    \caption{Hertzsprung-Russel diagram for the 755,847 stars used in this study, using Gaia BP-RP colour and absolute G magnitude. To ensure clarity we bin the sample into points and colour code by the log of number of stars in each bin.}
    \label{fig:BLS_sample}
\end{figure}

\begin{figure}
    \centering
    \includegraphics[width=\columnwidth]{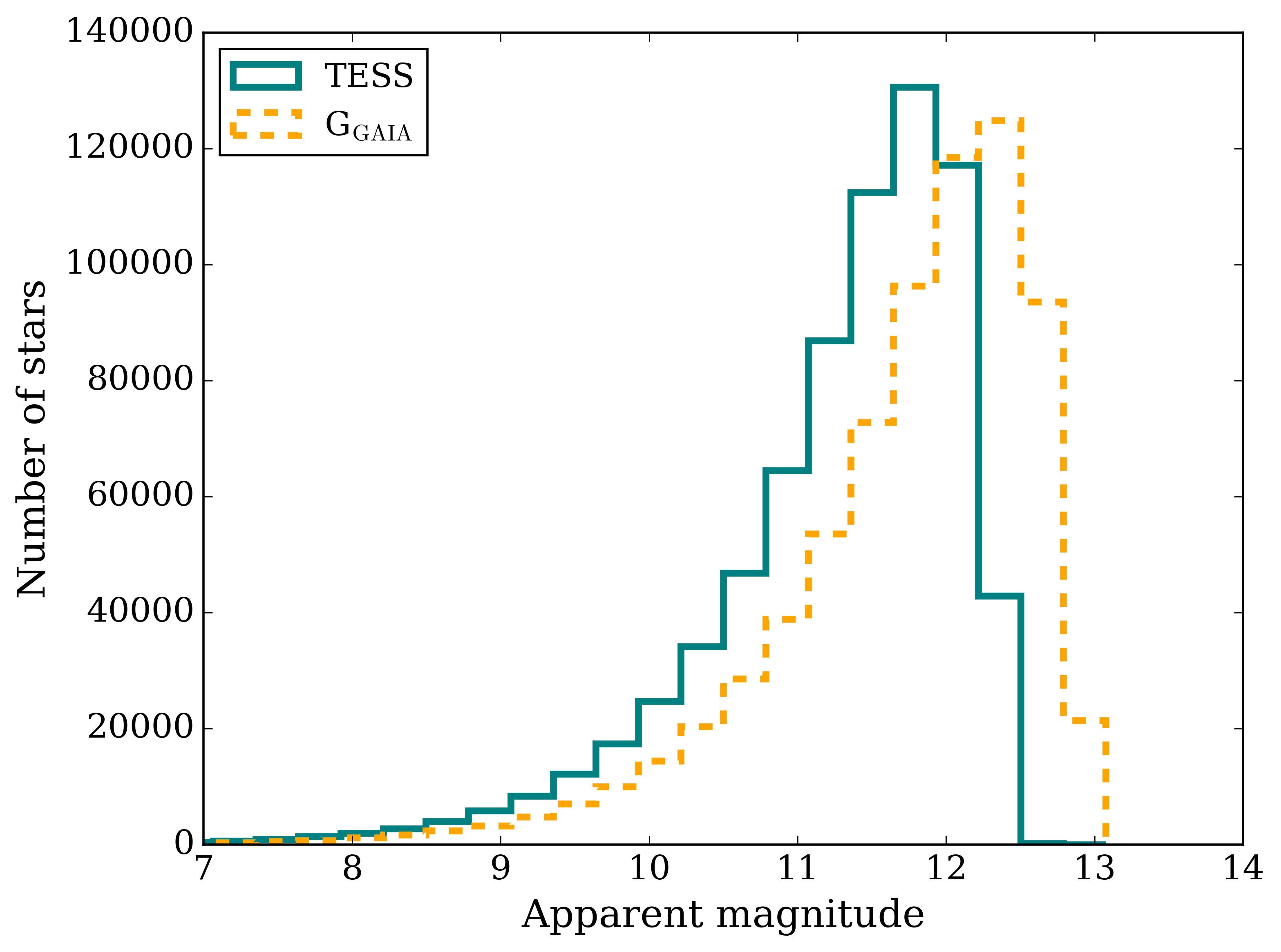}
    \caption{Histogram showing the distributions of the TESS (solid green) and Gaia G band (dashed orange) magnitudes for our stellar sample.}
    \label{fig:Tmag hist}
\end{figure}

We finally remove stars which do not have a Gaia parallax, as well as those with large temperature discrepancies between the TIC and Gaia DR3 temperatures ($\Delta T_{\textnormal{eff}}$ > 1000K), since such stars will likely have inaccurate parameters. Crossmatching this list with stars that have available QLP light curves yields a final stellar sample of 755,847 stars. The HR diagram of the final sample is shown in Figure~\ref{fig:BLS_sample}. Our sample makes up fairly bright stars, ranging from magnitude 6 to 13 with a median of 11.4 in TESS magnitude (see Figure~\ref{fig:Tmag hist}), and 11.9 in Gaia G magnitude. A sharp cut off at $\sim$13 is a result of our GSP-Spec selection since parameters from this module are only published for stars with spectra of a high enough signal to noise ratio (S/N > 20) \citep[][]{Kordopatis_2023}. 

\section{Stellar characterisation}
\label{sec: BASTA fitting}
\subsection{Input parameters}
Our final set of inputs comprise Gaia DR3 photometry in G, B$_\textnormal{P}$ and R$_\textnormal{P}$ bands, GSP-Spec metallicity (\meh) and effective temperature (\teff) , and parallax ($\varpi$). We correct for the known zero-point offset in parallax values using the \verb|gaiadr3-zeropoint| package \citep[][]{Lindegren_2021}. 

Gaia DR3 also provides parameters from the GSP-Phot module, which are fitted from BP/RP spectra \citep[][]{Bailer-Jones_2013,Carrasco_2021,DeAngeli_2023} for 471 million stars. This catalogue offers alternative values of \teff and \meh, however we employ the GSP-Spec values from the higher-resolution RVS spectra. Tests conducted on simulated data prior to the launch of Gaia showed that bright dwarfs ($G$ $\leq$ 12.5) are best parametrised in \teff, \logg and \meh from the RVS spectra, than from BP/RP spectra \citep[][]{Recio_Blanco_2016}. This is ideal for our sample, since only a small proportion of stars have $G_{\mathrm{Gaia}}$ > 13 ($\sim$ 0.3\%).

To ensure we use good quality photometric data, we check the BP and RP excess factor (\verb|phot_bp_rp_excess_factor|) defined as the ratio between the total BP and RP flux, and the G-band flux \citep[][]{Evans_2018}. This has been utilised as an indicator of problematic photometry, and it is recommended to filter out sources with large values. However, due to the strong colour dependence observed which can complicate interpretability, we calculate the corrected flux excess factor, $C^*$ following methodology from \cite{Riello_2021}. Good quality photometric data is reflected by a $C^*$ close to 0. For our sample, we find only 111 stars to have values outside of the range (-0.15,0.15), 82 of which also have RUWE > 1.2 that we assign a flag to in our final catalogue (see Section \ref{sec:validation} and \ref{sec:results}).

\begin{table*}
    \centering
    \caption{Input parameters of RA, Dec, temperature, metallicity, photometry in G, BP, RP bands and parallax that we use for our isochrone fitting detailed in Section \ref{sec: BASTA fitting}. Note that this is a subset of the full input parameter table which we show for clarity. The full table can be accessed online.}
    \begin{adjustbox}{width=0.9\textwidth}
        \begin{tabular}{ccccccccc}
            \hline
            TIC & RA & Dec & T$_{\textnormal{eff}}$ [K] & [M/H] [dex] & G [mag] & G$_{BP}$ [mag] & G$_{RP}$ [mag] & $\varpi$ [mas]\\
            \hline
            293164624 & 66.24 & -57.97 & 5944 $\pm$ 107 & -0.05 $\pm$ 0.20 & 11.5087 $\pm$ 0.0028 & 11.7843 $\pm$ 0.0029 & 11.0734 $\pm$ 0.0038 & 2.081 $\pm$ 0.014 \\
            396721891 & 65.79 & -57.76 & 6553 $\pm$ 174 & 0.23 $\pm$ 0.26 & 12.0346 $\pm$ 0.0028 & 12.3119 $\pm$ 0.0029 & 11.5978 $\pm$ 0.0038 & 2.456 $\pm$ 0.009 \\
            396721876 & 65.67 & -57.72 & 5169 $\pm$ 98 & -0.16 $\pm$ 0.13 & 10.7906 $\pm$ 0.0028 & 11.2232 $\pm$ 0.0028 & 10.1940 $\pm$ 0.0038 & 10.383 $\pm$ 0.010 \\
            293163171 & 62.69 & -56.02 & 5818 $\pm$ 263 & -0.14 $\pm$ 0.26 & 12.0521 $\pm$ 0.0028 & 12.3308 $\pm$ 0.0029 & 11.6074 $\pm$ 0.0038 & 2.262 $\pm$ 0.048 \\
            197932179 & 59.21 & -57.97 & 6095 $\pm$ 165 & 0.37 $\pm$ 0.17 & 12.1373 $\pm$ 0.0028 & 12.4732 $\pm$ 0.0028 & 11.6339 $\pm$ 0.0038 & 2.568 $\pm$ 0.010 \\
            197932161 & 59.06 & -57.84 & 6718 $\pm$ 200 & -0.11 $\pm$ 0.27 & 12.1920 $\pm$ 0.0028 & 12.4482 $\pm$ 0.0028 & 11.7754 $\pm$ 0.0038 & 1.723 $\pm$ 0.010 \\
            209109109 & 209.73 & -56.01 & 6050 $\pm$ 111 & 0.36 $\pm$ 0.13 & 9.6528 $\pm$ 0.0028 & 9.9797 $\pm$ 0.0028 & 9.1566 $\pm$ 0.0038 & 6.779 $\pm$ 0.053 \\
            95598703 & 110.42 & -15.20 & 5767 $\pm$ 108 & -0.08 $\pm$ 0.14 & 11.3002 $\pm$ 0.0028 & 11.6639 $\pm$ 0.0030 & 10.7720 $\pm$ 0.0039 & 5.809 $\pm$ 0.022\\
            102914932 & 155.82 & -46.09 & 6141 $\pm$ 158 & 0.09 $\pm$ 0.20 & 11.4362 $\pm$ 0.0028 & 11.7100 $\pm$ 0.0028 & 11.0058 $\pm$ 0.0038 & 2.375 $\pm$ 0.016 \\
            233789478 & 303.08 & 56.18 & 5805 $\pm$ 270 & -0.40 $\pm$ 0.26 & 11.9138 $\pm$ 0.0028 & 12.2530 $\pm$ 0.0029 & 11.4040 $\pm$ 0.0038 & 4.172 $\pm$ 0.017 \\
            \hline
            \label{tab:input params}
        \end{tabular}
    \end{adjustbox}
\end{table*}

We determine symmetric uncertainties on our input effective temperature and metallicity by calculating the mean error associated with each value. Uncertainties reported for each measurement are small and not representative of the overall error since they do not take into account systematic uncertainties. An average systematic offset of 2\% $\pm$ 0.5\% was reported in \cite{tayar_2022}, along with \cite{Recio_Blanco_2023} finding a scatter of 90\,K for their best quality sample. Therefore, we adjust the errors on temperature by adding 90\,K in quadrature to the given errors. Similarly, a median scatter of 0.13 dex was found for GSP-Spec \meh values \citep[][]{Recio_Blanco_2023}, so we add this in quadrature with the reported values. Additionally, GSP-Spec metallicities show slightly underestimated values for giants, whereas dwarfs have overestimated values. Therefore, such values are calibrated using the following equation from \citep{Recio_Blanco_2023}:
\begin{equation}
    \begin{aligned}
     \relax{[M/H]}_{\textnormal{calibrated}} = [M/H] + 0.274 - 0.1373\logg \\
    - 0.0050\textnormal\logg^2 + 0.0048\logg^3
    \end{aligned}
\end{equation}

\noindent where \logg is the uncalibrated surface gravity. We display a shortened version of our input parameters in Table \ref{tab:input params}.


\subsection{Model fitting with BASTA}
To characterise the full sample of 755,847 stars we use the BAyesian STellar Algorithm, known as \verb|BASTA| \citep[][]{Aguirre_B_rsen_Koch_2021} alongside BaSTI\footnote{\url{http://basti-iac.oa-abruzzo.inaf.it}} isochrones \citep[][]{Hidalgo_2018,Pietrinferni_2024}. \verb|BASTA| is a python-based tool that determines stellar properties given a pre-computed grid of stellar models along with various priors and inputs of astrophysical observables. \verb|BASTA| has been widely used to characterise exoplanet hosts and provide reliable stellar properties \citep[][]{Persson_2022,Hon_2023,Knudstrup_2023,Osborne_2024,weeks_2025,Bryant_2025} using a Bayesian framework. The BaSTI grid covers 2100 masses ranging between 0.1~M$_\odot$ and 15~M$_\odot$, 31 initial chemical composition [Fe/H] values between -3.20 and +0.45 and 229 ages from 0.2 Gyr to 20 Gyr, covering all major evolutionary stages (pre-main-sequence to white dwarf). BaSTI has been used with different algorithms to fit ages with typical uncertainties of 10\% for a sample of subgiant and main sequence turn-off stars \citep[][]{Casamiquela_2024}.

\verb|BASTA| uses a Bayesian framework, consisting of model stellar parameters $\boldsymbol{\Theta}$ such as mass, radius and age in combination with data \textbf{\textit{D}} such as effective temperature and metallicity. The probability of observing the model parameters given the data can be expressed as:
\begin{equation}
    P(\boldsymbol{\Theta}|\textbf{\textit{D}}) = \frac{P(\textbf{\textit{D}}|\boldsymbol{\Theta})P(\boldsymbol{\Theta})}{P(\textbf{\textit{D}})}
\end{equation}

\noindent From this, the full likelihood can be defined as the product of likelihoods of groups of observables, $D_i$:
\begin{equation}
    P(\textbf{\textit{D}}|\boldsymbol{\Theta}) = \prod_{i} P(\textbf{\textit{D}$_i$}|\boldsymbol{\Theta})
\end{equation}

\noindent whereby the group likelihoods are determined using:
\begin{equation}
    P(\textbf{\textit{D}$_i$}|\boldsymbol{\Theta}) = \frac{1}{\sqrt{2\pi | \vb{C}_i}} \textnormal{exp}(-\chi_i^2/2)
\end{equation}
\noindent with |$\vb{C}_i$| as the determinant of the covariance matrix. We can derive an expression for the marginalized posterior for a stellar model parameter $\theta$ from the computed posterior probability:
\begin{equation}
    P(\theta | \textbf{\textit{D}}) = \int P(\theta, \boldsymbol{\Theta'}|\textbf{\textit{D}}) w_{\boldsymbol{\Theta}}d\boldsymbol{\Theta'}
\end{equation}

\noindent where $\boldsymbol{\Theta'}$ represents all model parameters except $\theta$. 

We include parallax as a fitting parameter, which follows a slightly different approach since the assumption of parameters following a normal distribution weakens. In this case, BASTA creates prior distributions in distance and apparent magnitude using the observed values. The distance distribution is calculated from measured parallax and associated errors as described in \cite{Bailer-Jones_2015} and \cite{Astraatmadja_2016}. The distance and apparent magnitude distributions are sampled to calculate the reddening for each target, using the \texttt{Bayestar} dust map \citep[][]{Green_2019} for targets with a declination north of -30$^{\circ}$, or the \cite{Schlegel_1998} map if the target falls outside of this coverage. Extinction coefficients are computed as described in \cite{Casagrande_2014}, and are dependent on temperature and metallicity.


We use the Salpeter Initial Mass Function \citep[][]{Salpeter_1955} as an informative prior.
This IMF prescription reflects a mass distribution with a higher number of low-mass stars. Additionally, we use inputs of Gaia DR3 photometry in G, B$_\textnormal{P}$ and R$_\textnormal{P}$ bands, GSP-Spec metallicity [M/H] and effective temperature (T$_{\textnormal{eff}}$), and parallax ($\varpi$). We set flat priors on temperature and metallicity such that likelihood values are only calculated for models within 5$\sigma$ of the observed \teff, and within 0.25 dex of \meh.


There are several microphysical processes that can be considered when fitting for stellar age using \verb|BASTA|, such as convective overshooting, mass loss and atomic diffusion. These processes are important to consider when characterising stars, since the exclusion of such will systematically affect stellar isochrone ages \citep[e.g.][]{Claret_2019,Nsamba_2018}. For example, convective overshooting gives rise to extra mixing in layers above the core, providing more fuel hence stellar models with longer main-sequence lifetimes \citep[][]{Claret_2016}. Similarly, atomic diffusion can alter the surface composition of the star over its lifetime. If this is not taken into consideration and the current metallicity is assumed to be the initial, bulk metallicity, fitted stellar ages can be overestimated by up to 20\% \citep[][]{Dotter_2017}. 

Additionally, depending on the mass, different processes will be more important within the star. For stars with masses larger than $\sim 1.2~M_\odot$ the effects of atomic diffusion due to radiative forces or macroscopic turbulent transport counteract those of atomic diffusion due to concentration and thermal gradients \citep[][]{Mowlavi_2012}. Therefore, for higher mass stars the effects of atomic diffusion are not appropriate to include in the stellar modelling grid.

Furthermore, several other parameters in stellar models can affect the characterisation. For example, the specific definition of the solar mixture chosen affects the bulk metallicity and initial hydrogen mass fraction, and therefore the stellar age. The BaSTI models parametrise convection through mixing length theory \citep[][]{Bohm-Vitense_1958,cox_1968}, setting the mixing-length parameter, $\alpha_{\mathrm{ML}}$, to the solar model calibration of 2.006. We use the equation of state by A. Irwin\footnote{The EOS code is made publicly available at \url{http://freeeos.sourceforge.net/}
under the GNU General Public License.} \citep[][]{Cassisi_2003}, specifically adopting the `EOS1' option in the code which provides the best match to the OPAL EOS \citep[][]{Rogers_2002}. The models use opacities from the OPAL calculations for high temperatures \citep[][]{Iglesias_1996}, and those from \cite{Ferguson_2005} for the low temperature regime. Nuclear reaction rates from NACRE \citep[][]{Angulo_1999} are used except for $^3\mathrm{He}(^4\mathrm{He},\gamma)^7\mathrm{Be}$, $^{14}\mathrm{N(p},\gamma)^{15}\mathrm{O}$ and $^{12}\mathrm{C}(\alpha,\gamma)^{16}\mathrm{O}$ for which rates are taken from \cite{Cyburt_2008}, \cite{Formicola_1004} and \cite{Hammer_2005} respectively. The solar metal mixture by \cite{Caffau_2011} is adopted in the models with Z$_\odot$ = 0.0153. This definition slightly varies from the \cite{Grevesse_2007} prescription used in Gaia GSP-Spec when fitting parameters, with Z$_\odot$ = 0.012. Since we use these parameters as inputs we expect a slight systematic shift towards older ages, however this will likely not be a significant contributor to the overall systematic error associated with the cumulative effect of the various choices discussed above.

The resolution of the grid chosen can alter fitted stellar properties, and it is important that the parameter space is sufficiently sampled so posterior distributions can be resolved. To investigate this, we compare the BaSTI grid spacing in both the parameter space spanned by our sample and the prior-windows set to the average posterior distributions. BaSTI has non-uniform spacing in \meh of mean 0.079 dex (`overshoot' grid) and 0.147 dex (`diffusion' grid) within the sample range, and a mean age spacing of $\sim$97 Myr across the range 1-13.5 Gyr. Since we have inflated the input uncertainties on \teff and \meh, this allows for the expansion of the likelihood evaluated on the grid. The posteriors for age are well resolved in both grids, spanning an average of 11 and 27 grid steps for the overshoot and diffusion models respectively. Considering the prior window on \meh of $\pm$0.25 dex, we find that around 6 and 3 grid points are within this range. Although the sampling of \meh is more coarse in the diffusion grid, the median uncertainty on our input \meh values (0.174 dex) is comparable to the grid spacing (0.147 dex), allowing the likelihood to vary smoothly across the grid points evaluated. Furthermore, BASTA marginalises the likelihood across the evaluated grid points instead of choosing a best fit model, reducing sensitivity to grid discretisation. Therefore, we consider BaSTI's grid resolution to be sufficient for our sample.

We carry out two sets of grid-based fits, with two separate grids of BaSTI isochrones. One includes the effects of overshoot only (the `overshoot case') and the other includes overshoot, diffusion and mass loss (the `diffusion case'). One strategy would be to first split the sample into high mass ($M_\star > 1.2~M_\odot$) and low mass ($M_\star \leq 1.2~M_\odot$) stars using some prior value such as the TIC mass, however if a star is assigned the wrong group the isochrone fit may be invalid, causing edge cases which cannot be rectified without looking through the individual fits, which is not feasible for such a large sample. Therefore, we run both science cases for the entire sample. We decide the best fit parameters for each star using a similar approach to \cite{silva-aguirre_2015,silva-aguirre_2017} whereby we implement the following criteria on the fitted mass: 

\begin{enumerate}[leftmargin=*]
    \item if M$_* >$ 1.2M$_\odot$ for both sets of fits, we take the set of parameters from the overshoot case
    \item if M$_* \leq$ 1.2M$_\odot$ for both sets of fits, we take the set of parameters from the diffusion case
    \item if masses are in different mass groups, we take the set of parameters from the case with the highest log likelihood
\end{enumerate}

Additionally, we identify stars whereby the fits have failed to converge. Such stars are indicated by the following criteria, and are removed from our sample:
\begin{enumerate}[leftmargin=*]
    \item Age $\leq$ 1 Gyr
    \item Age = NaN
    \item $\sigma_{Age}$ $\leq$ 0.2 Gyr
    \item $\sigma_{Age}$ = NaN
\end{enumerate}
Following this, we remove 38,040 stars with invalid fits, leaving 717,807 stars which make up our final homogeneous sample. Of our final sample, 562,970 stars are assigned best fit values from the atomic diffusion case and 154,837 from the overshoot case. 

\section{Validation of fitted parameters}
\label{sec:validation}

\begin{table*}
    \centering
    \caption{Summary of the input physics and fitting choices used for the comparison samples discussed in Section \ref{sec:validation}. OPAL: \protect\cite{OPAL}, F05: \protect\cite{F05}, Aesopus: \protect\cite{aesopus}, FreeEOS: \protect\cite{Cassisi_2003}, OPAL02: \protect\cite{OPAL02}, SCVH: \protect\cite{SCVH}, M12: \protect\cite{M12}, Mihalas-Hummer-Dappen: \protect\cite{Mihalas-Hummer-Dappen}, NACRE: \protect\cite{NACRE}, JINA REACLIB: \protect\cite{JINA_REACLIB}, A11: \protect\cite{A11}, T94: \protect\cite{Thoul_1994}, P86: \protect\cite{Paquette_1986}, C11: \protect\cite{Caffau_2011}, GS98: \protect\cite{Grevesse_1998}, A09: \protect\cite{Asplund_2009}, GN93: \protect\cite{Grevesse_1993}}
    \begin{adjustbox}{width=\textwidth}
        \begin{tabular}{cccccccccccc}
            \hline
            Source & Modelling package & Grid & Opacity & EOS & Nucl. rates & $\alpha_{\mathrm{ML}}$ & Diffusion & Overshoot & Solar Mixture & Spectroscopic inputs & Spectroscopic inputs source \\
            \hline
            This Work & BASTA & BaSTI & OPAL/F05 & FreeEOS & NACRE & 2.006 & T94 (M $\leq$ 1.2M$_\odot$) & Yes & C11 & \teff and \meh & Gaia RVS \\
            LEGACY & BASTA & GARSTEC & OPAL/F05 & OPAL02 & NACRE & 1.791 & T94 (M $\leq$ 1.2M$_\odot$) & Yes & GS98 & \teff and \meh & TRES\\
            \cite{Berger_2026} & isoclassify & PARSEC & OPAL/Aesopus & FreeEOS & JINA REACLIB & 1.74 & T94 & Yes & C11 & \meh & Gaia BP/RP \\
            TESS-Keck & isoclassify & MIST & OPAL/F05 & OPAL02/SCVH/M12 & JINA REACLIB & 1.82 & T94/P86 & Yes & A09 & \teff, \feh and \logg & HIRES \\
            \cite{weeks_2025} & BASTA & GARSTEC & OPAL/F05 & OPAL02/Mihalas-Hummer-Dappen & NACRE & 1.6-2 & T94 (M $\leq$ 1.2M$_\odot$) & Yes & A09 & \teff and \meh & Gaia RVS\\
            StarHorse &  StarHorse & PARSEC & OPAL/Aesopus & FreeEOS & JINA REACLIB & 1.74 & T94 & Yes & C11 & \teff, \meh and \logg & Gaia RVS \\
            APOKASC & BeSPP & GARSTEC & OPAL/F05 & FreeEOS & A11 & 1.798 & T94 + extra mixing & Yes & GN93 & \meh & APOGEE ASPCAP \\
            \hline         
            \label{table:input physics}
        \end{tabular}
    \end{adjustbox}
\end{table*}

\begin{table*}
    \centering
    \caption{A comparison between this work and several stellar catalogues used in literature. We list the number of stars which are common to our sample and the comparison sample, N$_{\mathrm{stars}}$ and calculate the median residual scatter along with the median fractional error on our values for each overlapping sample. Note that for the LEGACY catalogue, we remove two outliers (detailed in Section \ref{sec:validation}). Additionally, 659,576 stars are used when calculating statistical values for mass and temperature for the StarHorse sample, whereas 350,166 are used for age.}
    \begin{adjustbox}{width=0.95\textwidth}
        \begin{tabular}{cccccccccc}
            \hline
            Literature source & N$_{\mathrm{stars}}$ & \multicolumn{2}{c}{Radius} 
            & \multicolumn{2}{c}{Mass} & \multicolumn{2}{c}{Age} & \multicolumn{2}{c}{Temperature}\\
             & &  S$_{\mathrm{med}}$ & Median error & S$_{\mathrm{med}}$ & Median error & S$_{\mathrm{med}}$ & Median error & S$_{\mathrm{med}}$ & Median error \\
             \hline
             TIC & 717,807 & 4.43\% & 4.29\% & 6.94\% & 4.49\% & & 50.00\% & 1.79\% & 1.74\% \\
             SWEET-Cat & 905 & 5.00\% & 3.60\% & 4.82\% & 4.01\% & & 48.21\% & 1.17\% & 1.37\% \\
             NASA exoplanet archive & 782 & 3.51\% & 3.56\% & 4.17\% & 3.98\% & & 50.00\% & 1.23\% & 1.36\%\\
             \cite{Berger_2026} & 989 & 2.76\% & 3.51\% & 3.77\% & 4.02\% & 37.90\% & 56.10\% & 1.30\% & 1.37\% \\
             StarHorse & 659,576 & & & 3.39\% & 4.57\% & 14.45\% & 27.59\% & 0.88\% & 1.78\% \\
             APOKASC & 170 & 3.31\% & 4.23\% & 3.99\% & 4.40\% & 17.72\% & 21.81\% & 1.37\% & 1.63\% \\
             Kepler LEGACY & 45 & 3.49\% & 3.75\% & 2.46\% & 4.01\% & 21.93\% & 25.60\% & 0.95\% & 1.38\% \\\
             TESS-Keck & 68 & 1.87\% & 3.15\% & 4.29\% & 4.11\% & 35.55\% & 56.98\% & 0.67\% & 1.22\% \\
             \cite{weeks_2025} & 26 & 2.47\% & 3.78\% & 4.21\% & 4.24\% & 16.20\% & 49.14\% & 0.87\% & 1.41\% \\
            \hline         
            \label{table:comp stats}
        \end{tabular}
    \end{adjustbox}
\end{table*}

\subsection{Validating stellar mass, radius and temperature}
\begin{figure*}
    \centering
    \includegraphics[width=\textwidth]{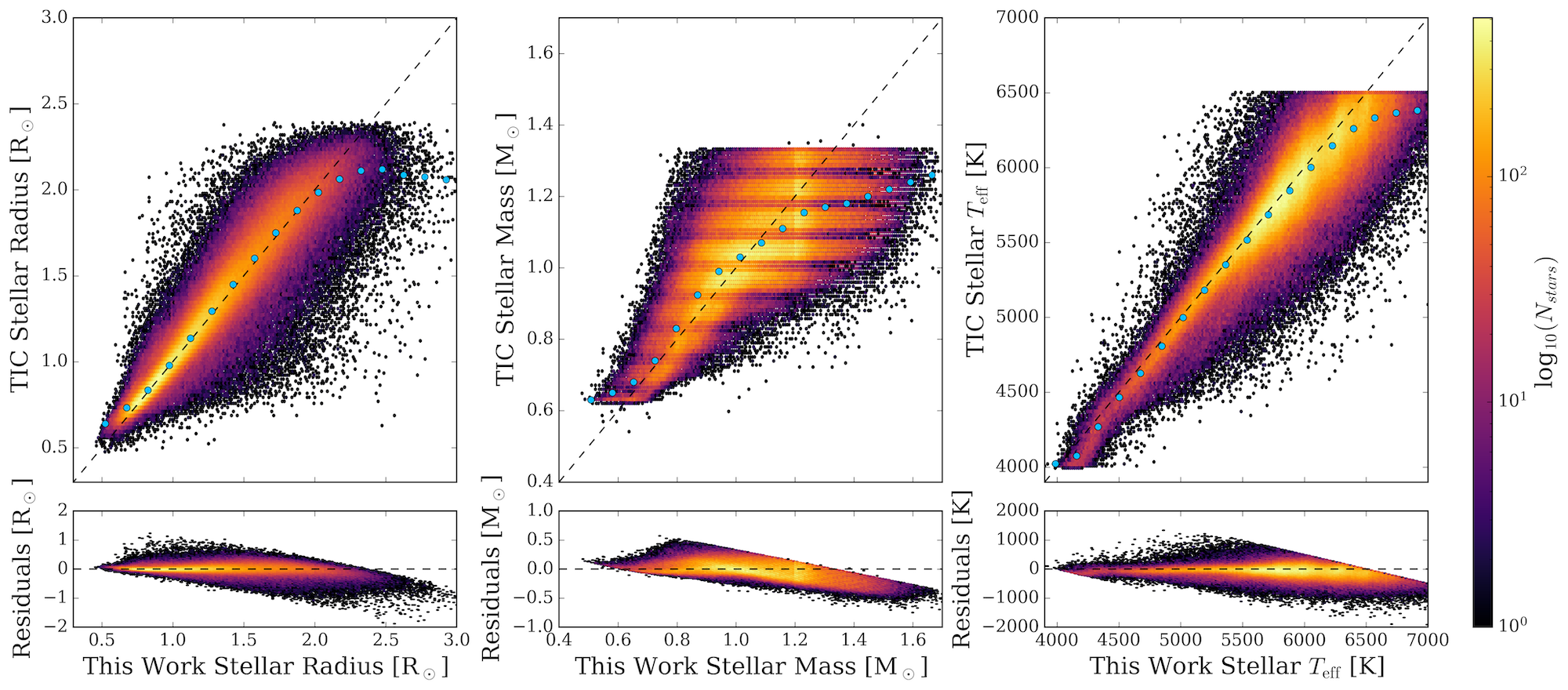}
        \caption{Stellar radius, mass and temperature from this work compared to the TIC radius, mass and temperature. Due to the size of the sample, points are colour coded by the log of the number of stars in each bin. The median trend is plotted in blue circles. We see good general agreement, with a median residual scatter of 4.43\%, 6.94\% and 1.79\% for radius, mass and temperature respectively. The dashed line represents the 1:1 line.}
    \label{fig:TIC_comp}
\end{figure*}

\begin{figure*}
    \centering
    \includegraphics[width=\textwidth]{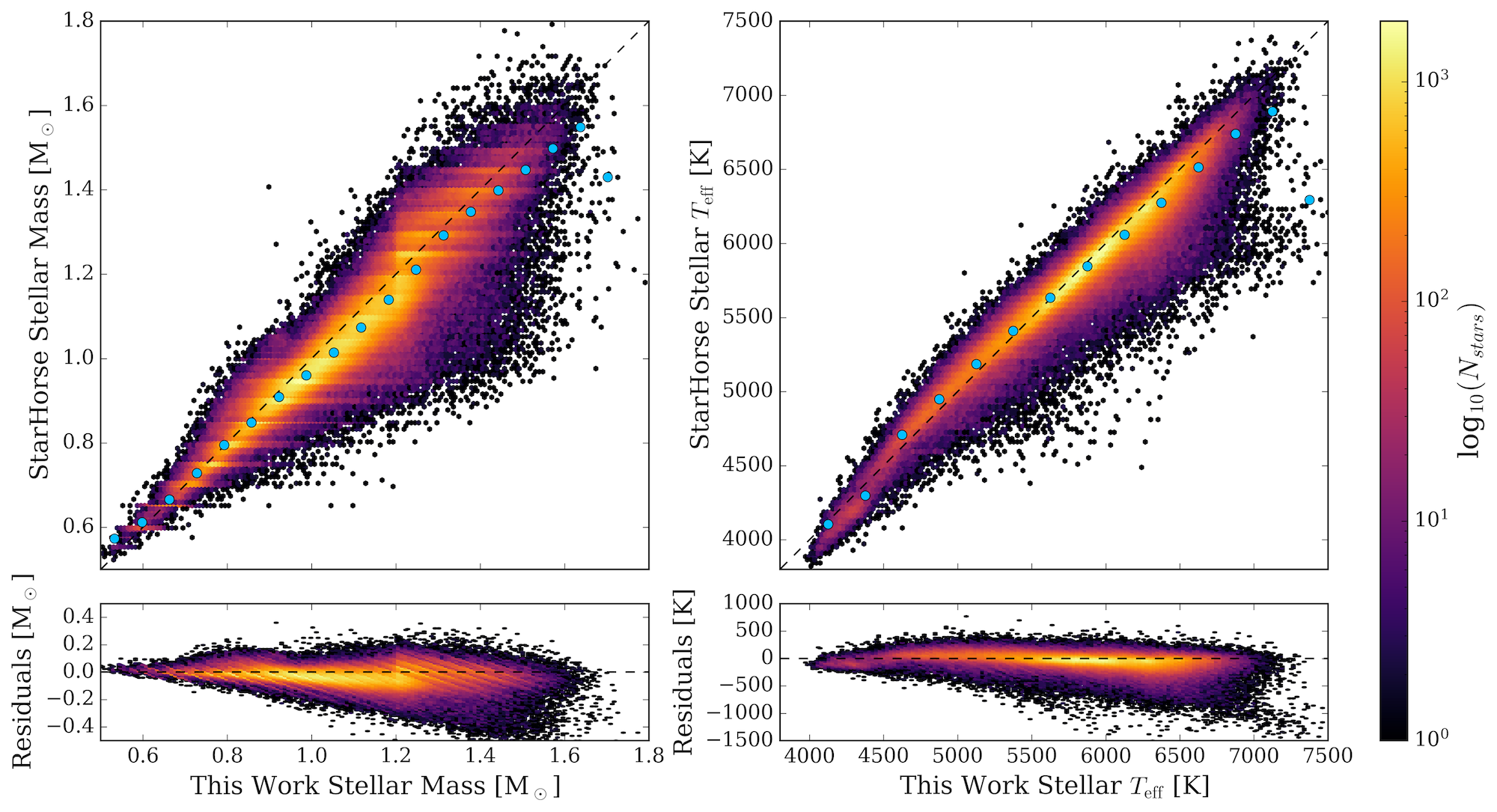}
        \caption{Stellar mass and temperature from this work compared to the StarHorse mass and temperature. Due to the size of the sample, points are colour coded by the log of the number of stars in each bin, and the median trend is plotted in the blue circles. We see very good general agreement, with a median residual scatter of 3.39\% and 0.88\% for mass and temperature respectively. The dashed line represents the 1:1 line.}
    \label{fig:starhorse_comp}
\end{figure*}

We validate our best fit stellar radius, mass and temperature by comparison to several surveys, ranging in size and homogeneity. We calculate several statistical values to be used as an indicator of how well our values agree with these sources, which we define here. The first is the absolute median residual offset, given by:
\begin{equation}
    \Delta_{\mathrm{med}} = \mathrm{Median}_i(|x_{\textnormal{lit},i} - x_{\textnormal{cat},i|})
\end{equation}

\noindent where x$_{\textnormal{lit},i}$ is the literature source being used as a comparison sample, and x$_{\textnormal{cat},i}$ are values from this study. 


Additionally, the median residual scatter (S$_{\mathrm{med}}$) is defined as:
\begin{equation}
    S_{\textnormal{med}} = \mathrm{Median}_i\left(\frac{|{x_{\textnormal{lit},i} - x_{\textnormal{cat},i}}|}{x_{\textnormal{cat},i}}\right)
\end{equation}

We first compare our stellar parameters to the TIC \citep[][]{Stassun_2018,Stassun_2019}, since this is the largest catalogue available, sharing all stars in our sample. The TIC is inhomogeneous in nature, since stellar parameters from specially curated catalogues, e.g., Cool Dwarfs \citep[][]{Muirhead_2018}, Hot Subdwarfs \citep[][]{Stassun_2019} take priority when adopting values in the TIC. Similarly, temperature is first taken from various spectroscopic catalogues if available, otherwise values are estimated from a spline relation with Gaia G$_{BP}$ - G$_{RP}$ colours. Stellar radius is calculated from Gaia parallax using the Stefan-Boltzmann relation, along with mass from a spline with temperature based on eclipsing binary measurements \citep[][]{Torres_2010}. Finally, \logg is determined from the TIC mass and radii, therefore all such parameters are dependent on the temperature chosen. 

From the density plot in Figure \ref{fig:TIC_comp}, we see overall there is good general agreement with the TIC values considering the number of stars in the sample. We calculate a median residual scatter of 4.43\%, 6.94\% and 1.79\% in radius, mass and temperature respectively. We calculate the median fractional errors on radius, mass and temperature for our best fit values for comparison to infer the degree of agreement between the two sources (See Table \ref{table:comp stats} for all sources). We calculate a median error of 4.29\%, 4.49\% and 1.74\% for radius, mass and temperature respectively, which are similar to the median residual scatter for radius and temperature. However, stellar mass is more discrepant, reflected by several features which can be seen in Figure \ref{fig:TIC_comp}. A band of higher density at 1.2 M$_\odot$ is apparent, resulting from the mass we use as a boundary to consider different grids when assigning best fit values for each star. There are also clear discretised bands in the TIC mass, due to the spline relation used. Additionally, there is a sharp boundary in TIC stellar radius and mass at $\sim$2.4 R$_\odot$ and $\sim$1.3 M$_\odot$, due to the \logg cut we made at 3.8 when defining our sample (see Section~\ref{sec:sample selection}), which in turn is calculated from the stellar radius and mass. 

We next compare mass and temperature to the data from the StarHorse catalogue \citep[][]{Queiroz_2023}. StarHorse is a Bayesian isochrone-fitting code which estimates age, mass and extinction for individual stars \citep[][]{StarHorse}. Inputs of photometry, parallax \logg, \teff, \meh along with the PARSEC grid \citep[][]{PARSEC} have been used to derive stellar ages for 2.5 million stars. StarHorse has estimated parameters for several surveys, however we only consider values determined for stars in the Gaia DR3 RVS survey, comprising 659,576 stars which overlap with our catalogue. This sample is fitted using inputs of Gaia DR3 photometry and parallax, and RVS \logg, \teff and \meh. From the comparison in Figure \ref{fig:starhorse_comp} we see excellent agreement in both mass and temperature, reflected by a median residual scatter of 3.39\% and 0.88\% respectively. This is likely due to the similarities between fitting techniques used by BASTA and StarHorse, as well as the input parameters. However, no significant systematic differences are seen despite different grids and input physics being used (see Table \ref{table:input physics}) which suggests consistency between the implementation of the PARSEC and BaSTI models for these studies.

We next compare our mass, radius and temperature to the Stars With ExoplanETs catalogue \citep[SWEET-Cat,][]{Santos_2013,sousa_2021} and the \cite{Berger_2026} catalogue. These are smaller, more homogeneous surveys which have fitted more precise stellar parameters for exoplanet hosts. SWEET-Cat utilises high-resolution spectra either observed by the team or taken from the ESO archive to extract stellar properties such as \teff, \logg and \feh. Stellar mass is also calculated using the calibration in \cite{Torres_2010} with further correction, providing parameters for 4369 stars. However, it should be noted that the catalogue is not entirely homogeneous - 29\% of stars have completely homogeneous parameters \citep[][]{Antoniadis-Karnavas_2024}. Similarly, the \cite{Berger_2026} Gaia-Kepler-TESS host stellar properties catalogue provides homogeneous parameters for 10022 exoplanet hosts using isochrone fitting techniques in combination with Gaia DR3 inputs of G$_{BP}$ and G$_{RP}$ photometry along with GSP-Phot metallicities which are corrected using the California-Kepler Survey \citep[CKS,][]{Petigura_2017}.

\begin{figure*}
    \centering
    \includegraphics[width=\textwidth]{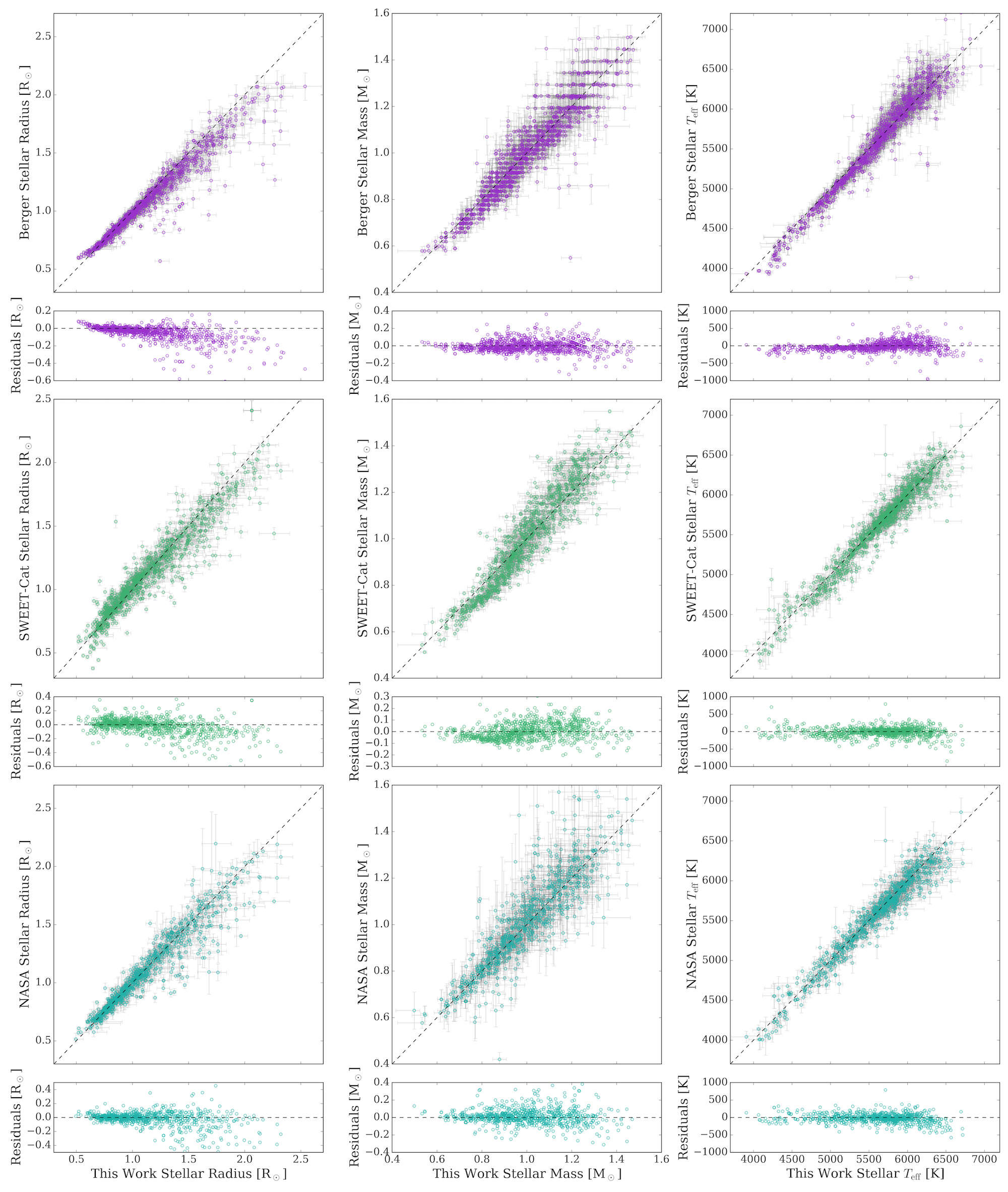}
        \caption{Stellar radius (left), mass (middle) and temperature (right) we fit with our methodology compared to the \protect \cite{Berger_2026} (top), SWEET-Cat (middle) and NASA exoplanet archive (bottom) catalogues. We only take systems from the NASA exoplanet archive with a radius and mass precision greater than 20\%. The 1:1 line is represented by the dashed line. We calculate a median residual scatter of 2.76\%, 3.77\% and 1.30\% in radius, mass and temperature respectively for our comparison to the \protect\cite{Berger_2026} catalogue, 5.00\%, 4.82\% and 1.17\% for SWEET-Cat and 3.51\%, 4.17\% and 1.23\% for NASA exoplanet archive systems.}
    \label{fig:Berger_sweetcat_comp_all}
\end{figure*}

We compare the parameters for the 905 stars that overlap in our catalogue and SWEET-Cat. We calculate a median residual scatter of 5.00\%, 4.82\% and 1.17\% for radius, mass and temperature respectively. The scatter is smaller than our median error of 1.37\% for temperature, however values are slightly larger than the median errors of 3.60\% and 4.01\% on our fitted radius and mass, reflected by the offset in the mass comparison plot in Figure \ref{fig:Berger_sweetcat_comp_all}. A similar trend is seen in \cite{Santos_2013} between masses derived from Padova isochrones \citep[][]{Padova_2000} and the \cite{Torres_2010} calibration, following which they apply a correction by fitting a quadratic. Therefore, the offset we see between our masses and SWEET-Cat is likely a result of this differing methodology for mass determination. 

We also plot the same parameters for the 989 overlapping stars in \cite{Berger_2026} relative to our sample, from which we find a median residual scatter of 2.76\%, 3.77\% and 1.30\% for radius, mass and temperature respectively, which are smaller than our median error values of 3.51\%, 4.02\% and 1.37\% indicating good agreement. An offset in radius starting from $\sim1~R_\odot$ is apparent, reflected by a median residual scatter of 1.65\% for stars with R $\leq$ 1 R$_\odot$ and 4.50\% for R > 1 R$_\odot$. Such an offset is not apparent in our comparisons to other literature sources. There are many differences in the choices used for our stellar characterisation and \cite{Berger_2026} which may be the cause of the trends seen in Figure \ref{fig:Berger_sweetcat_comp_all}. For example, \cite{Berger_2026}  uses a different modelling package \citep[isoclassify,][]{Huber_2017,Berger_2020} and grid \citep[PARSEC,][]{PARSEC} to this study, as well as differing inputs and their calibration (see Table \ref{table:input physics}). The effect of fitting technique was quantified in \cite{silva-aguirre_2015} whereby three different fitting pipelines were used and compared to the standard \texttt{BASTA} grid. Estimates of mass and age were found to differ between pipelines in some cases up to $\sim$ 15\% in mass and 40\% in age.

\begin{figure*}
    \centering
    \includegraphics[width=\textwidth]{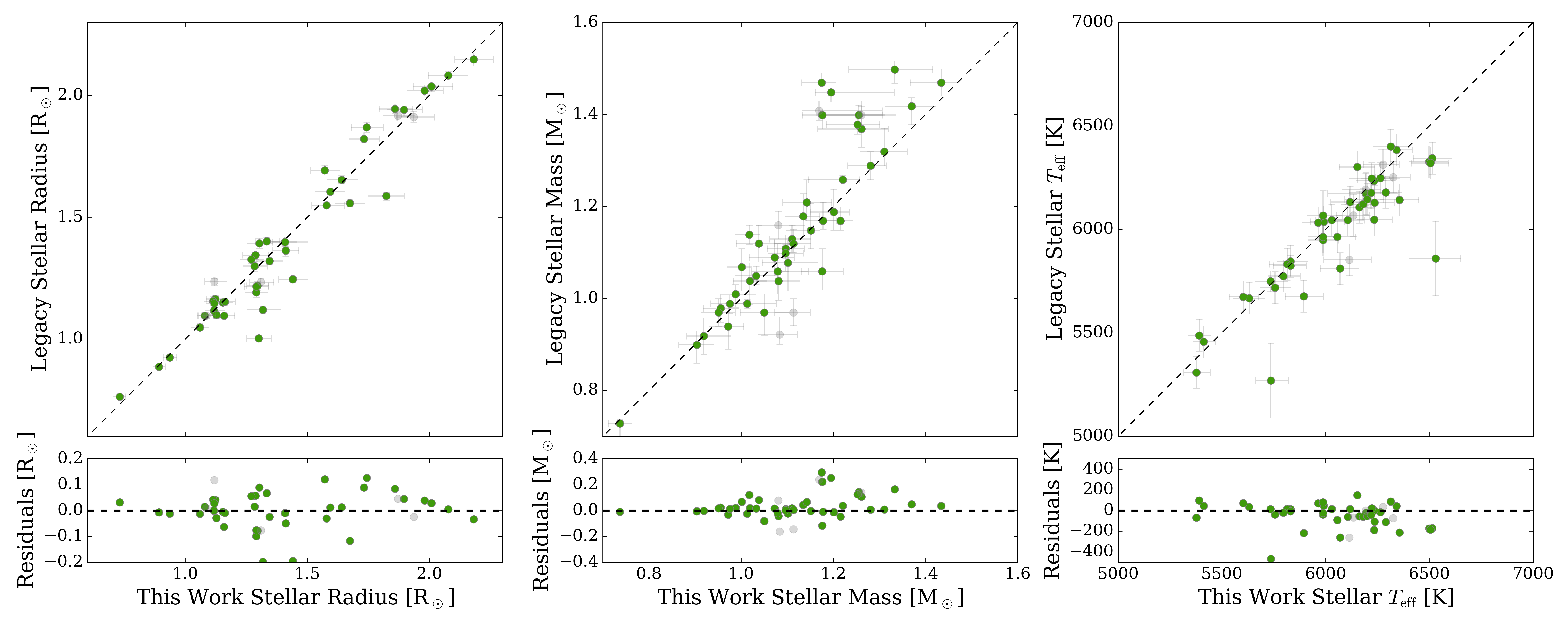}
        \caption{Values of stellar radius (left), mass(middle) and temperature (right) we derive with our methodology compared to overlapping stars in the Kepler LEGACY survey. We plot stars with discrepant metallicity between Gaia and LEGACY values in grey ($\Delta_{\textnormal{met}} > 0.2$). We calculate a median residual scatter of 3.49\%, 2.46\% and 0.95\% on radius, mass and temperature respectively. The dotted line represents the 1:1 line.}
    \label{fig:legacy_basti}
\end{figure*}

Continuing our validation, we investigate how our parameters compare to those from a heterogeneous sample of exoplanet hosts. These stars are typically studied on a case-by-case basis, employing different methodology and modelling, however a large number of planet hosts are expected to have reliable parameters since this is important for exoplanet characterisation. We choose to compare to literature values from the NASA exoplanet archive composite table\footnote{\url{https://exoplanetarchive.ipac.caltech.edu/cgi-bin/TblView/nph-tblView?app=ExoTbls&config=PSCompPars}}, considering values for systems which have a radius and mass precision greater than 20\%, 728 of which are in our sample. We compare the two sets of values in Figure \ref{fig:Berger_sweetcat_comp_all}, and find good agreement between our values and literature, with median residual scatter of 3.51\%, 4.17\% and 1.23\% in radius, mass and temperature respectively, which are similar to our median errors of 3.56\%, 3.98\% and 1.36\%. 

Due to the various trends seen in our comparisons with the \cite{Berger_2026}, SWEET-Cat and NASA exoplanet archive samples, we investigate how the absolute residual standard deviation changes as a function of radius, mass and temperature. We find that for radius, there is a general increase in the spread towards larger radii (R > 1.5 R$_\odot$) for all catalogues, with a maximum change of 0.22 for the \cite{Berger_2026} sample, reflected by the offset discussed previously. However, for the comparison to SWEET-Cat we see an initial increase in the range 0.5 R$_\odot$ to 2 R$_\odot$ followed by a decrease for R > 2 R$_\odot$, likely due to the small number of stars in this bin. Considering mass, there is a slight increase towards higher masses for the \cite{Berger_2026} (0.02 M$_\odot$ for M < 0.8 M$_\odot$ to 0.06 M$_\odot$ for M > 1.4 M$_\odot$) and NASA samples (0.05 M$_\odot$ to 0.07 M$_\odot$). We see a more sinusoidal trend for the SWEET-Cat residual spread, representative of the pattern seen in Figure \ref{fig:Berger_sweetcat_comp_all}. For temperature, an increase towards higher temperature is apparent for all samples. Additionally, we calculate how the median fractional error on our measurements change with these parameters, finding a general increase with all three parameters, likely driving the changes in residual spread. The largest increase is seen for radius, changing from 2.88\% for 0.5 R$_\odot$ $\leq$ R $\leq$ 1 R$_\odot$ to 5.29\% for 1.5 R$_\odot$ $\leq$ R $\leq$ 2 R$_\odot$.
                               
We finally compare our stellar parameters for the 47 stars which are included in the Kepler LEGACY survey. This study used precise asteroseismic frequencies with detailed stellar model grids using a number of different pipelines \citep[][]{silva-aguirre_2017}. For our comparisons, we use the values from LEGACY which are fitted with the BASTA pipeline. This pipeline makes use of the \verb|GARSTEC| modelling grid with inputs of spectroscopic temperature and \feh, along with the two frequency ratios $r_{010}$ and $r_{02}$ \citep[][]{GARSTEC}. Additionally, two grids are used depending on the stellar mass, one including the effects of diffusion for low mass stars (M $\leq$ 1.15 M$_\odot$) and the other with overshoot (M > 1.15 M$_\odot$). We found three stars to have largely discrepant values in metallicity between Gaia and LEGACY which we plot as grey points in Figure \ref{fig:legacy_basti}, however we include these in our statistical calculations. It is also to be noted that two stars in this sample, KIC 9414417 and KIC 3632418 were found to be outliers in the LEGACY survey due to a large difference in stellar properties between the standard and overshoot grids used \citep[][]{silva-aguirre_2015}, therefore we also plot these in grey and remove them from our analysis. We calculate a median residual scatter of 3.49\%, 2.46\% and 0.95\% for radius, mass and temperature respectively, reflecting excellent agreement with our values, which have median errors of 3.75\%, 4.01\% and 1.38\%. In particular, we find very similar values for stars with $\sim$M < 1.2M$_\odot$ shown in Figure \ref{fig:legacy_basti} with little scatter in residuals. However, we see an increased scatter in mass for stars $\sim$M > 1.2 M$_\odot$, due to the change in grids in both LEGACY and this work and therefore a difference in specific microphysics such as the implementation of convective overshoot. We calculate a median residual scatter of 2.45\% for stars fitted with the diffusion grid (M $\leq$ 1.2 M$_\odot$), and a much higher scatter of 8.51\% for stars we fit with the overshoot grid (M > 1.2 M$_\odot$). Furthermore, although LEGACY makes use of the BASTA pipeline, the GARSTEC tracks are adopted rather than BaSTI which involve a different equation of state, mixing length parameter and solar mixture definition.

Overall, we see good agreement between our best fit values of radius, mass and temperature with various other samples/catalogues, differing in size, homogeneity and methodology. We observe different trends/offsets, reflecting the effect of the chosen inputs, stellar modelling grids and microphysics, different combinations of which are used for each catalogue. Recent work has suggested that current uncertainties on measurements set an error floor of 4.2\% and 2.4\% in radius and temperature. Additionally, values of stellar mass fitted depend on the grid chosen, with typical uncertainties of order 5\% across commonly used model grids \citep[][]{tayar_2022}. Considering these errors, the values we derive are in good agreement with the various catalogues we have used as comparison.

\subsection{Validating stellar age}
\label{sec:age validation}
\begin{figure*}
    \centering
    \includegraphics[width=\textwidth]{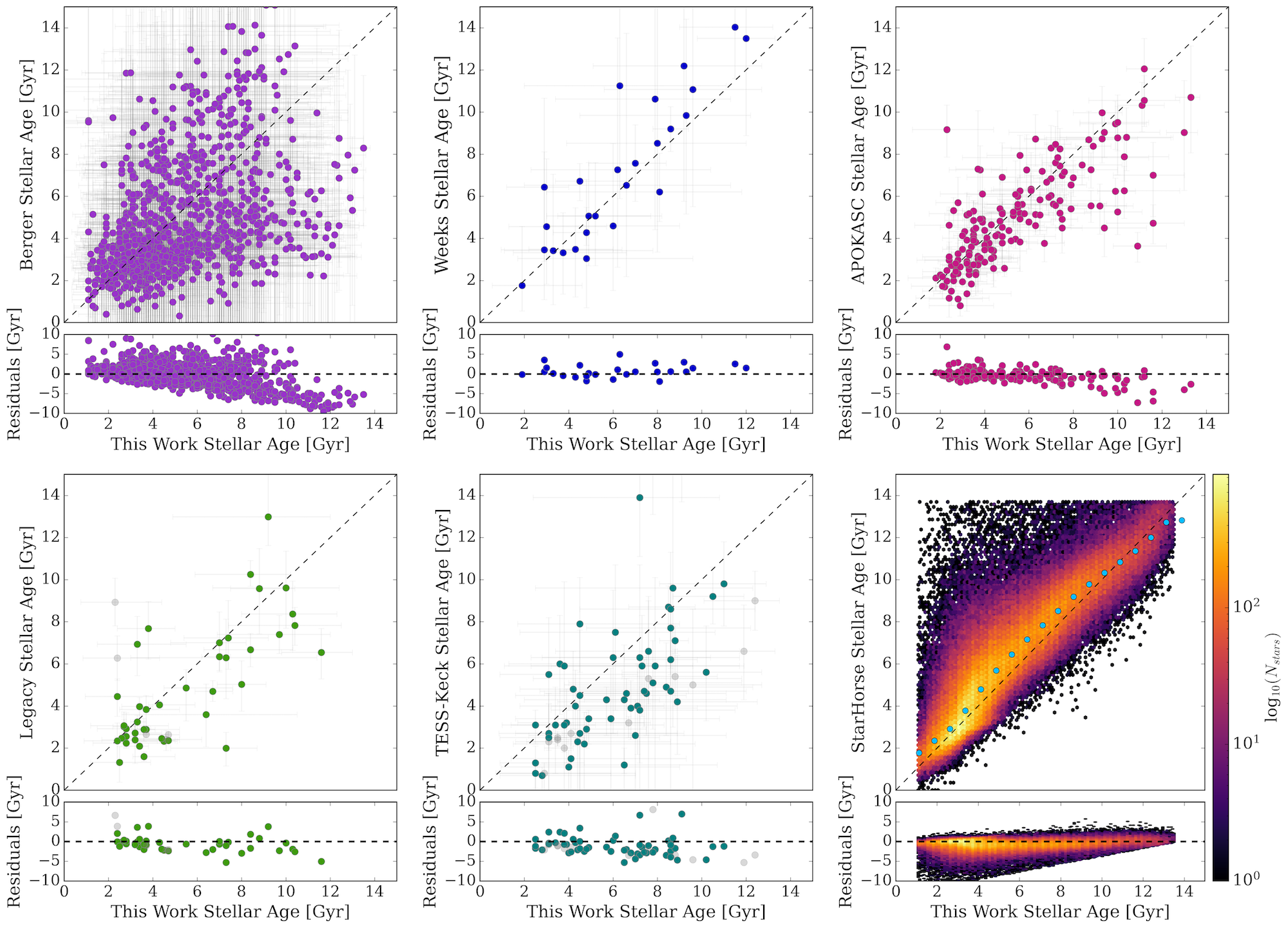}
        \caption{Values of stellar age we fit with our isochrone analysis compared to ages from \protect\cite{Berger_2026} (top left), \protect\cite{weeks_2025} (top middle), APOKASC (top right), Kepler LEGACY (bottom left), TESS-Keck (bottom middle) and StarHorse (bottom right). We calculate a median residual scatter of 37.90\%, 16.20\%, 21.93\% and 35.55\% for \protect\cite{Berger_2026}, \protect\cite{weeks_2025}, Kepler LEGACY and TESS-Keck respectively. We plot stars with discrepant metallicity values in grey for the Kepler LEGACY and TESS-Keck samples. The median trend for the comparison to StarHorse is plotted in blue circles. The 1:1 line is indicated by the black dotted line.}
    \label{fig:age_comp}
\end{figure*}

\begin{table*}
    \centering
    \caption{Best fit values for distance, \teff, \logg, \meh, radius, mass and age from our isochrone fitting detailed in Section \ref{sec: BASTA fitting}. We include the grid used for each star under the `Case' column, where we either consider overshooting or atomic diffusion, denoted by `overshoot' or `diffusion'. We also include the Gaia DR3 RUWE values as an indicator of reliability. Note that this is a subset of the full best fit parameter table which we show for clarity. The full table can be accessed online.}
    \begin{adjustbox}{width=0.9\textwidth}
        \begin{tabular}{ccccccccccc}
            \hline
            TIC & Distance [pc] & T$_{\textnormal{eff}}$ [K] & logg & M/H [dex] & Radius [R$_\odot$] & Mass [M$_\odot$] & Age [Gyr] & Case & RUWE & Quality flag \\
            \hline
            350333784 & 425.5$^{+5.4}_{-5.7}$ & 6200$^{+125}_{-140}$ & 4.22$^{+0.06}_{-0.05}$ & 0.25$^{+0.05}_{-0.10}$ & 1.44$^{+0.08}_{-0.07}$ & 1.26$^{+0.05}_{-0.05}$ & 3.40$^{+1.0}_{-1.3}$ & Overshoot & 0.95 & 000 \\
            350297081 & 679.8$^{+12.9}_{-11.9}$ & 6600$^{+146}_{-181}$ & 4.25$^{+0.06}_{-0.05}$ & 0.00$^{+0.10}_{-0.15}$ & 1.40$^{+0.08}_{-0.06}$ & 1.27$^{+0.07}_{-0.07}$ & 2.20$^{+1.20}_{-1.25}$ & Overshoot & 1.06 & 000 \\
            350824216 & 414.5$^{+5.6}_{-5.0}$ & 6029$^{+111}_{-124}$ & 3.82$^{+0.05}_{-0.04}$ & -0.52$^{+0.04}_{-0.20}$ & 2.09$^{+0.1}_{-0.09}$ & 1.07$^{+0.04}_{-0.05}$ & 5.80$^{+0.60}_{-0.50}$ & Diffusion & 0.86 & 000 \\
            350824257 & 204.5$^{+2.8}_{-2.8}$ & 6067$^{+104}_{-75}$ & 4.14$^{+0.05}_{-0.04}$ & 0.05$^{+0.02}_{-0.1}$ & 1.57$^{+0.06}_{-0.06}$ & 1.23$^{+0.03}_{-0.07}$ & 4.20$^{+1.70}_{-0.60}$ & Diffusion & 0.97 & 000 \\
            350767231 & 316.7$^{+4.7}_{-4.8}$ & 5873$^{+83}_{-90}$ & 4.03$^{+0.05}_{-0.04}$ & 0.30$^{+0.15}_{-0.05}$ & 1.82$^{+0.08}_{-0.07}$ & 1.31$^{+0.07}_{-0.09}$ & 4.50$^{+1.50}_{-0.60}$ & Overshoot & 0.95 & 000 \\
            350821546 & 632.8$^{+8.5}_{-8.7}$ & 5820$^{+163}_{-199}$ & 3.98$^{+0.07}_{-0.07}$ & -0.28$^{+0.12}_{-0.11}$ & 1.72$^{+0.14}_{-0.11}$ & 1.03$^{+0.04}_{-0.04}$ & 8.30$^{+0.80}_{-0.90}$ & Diffusion & 1.09 & 000\\
            350841387 & 91.8$^{+1.2}_{-1.2}$ & 6032$^{+76}_{-117}$ & 4.41$^{+0.07}_{-0.04}$ & -0.16$^{+0.14}_{-0.15}$ & 1.05$^{+0.05}_{-0.03}$ & 1.03$^{+0.05}_{-0.07}$ & 3.60$^{+3.70}_{-2.00}$ & Diffusion & 0.93 & 010\\
            350841389 & 95.8$^{+1.2}_{-1.2}$ & 4885$^{+60}_{-55}$ & 4.61$^{+0.03}_{-0.02}$ & -0.17$^{+0.15}_{-0.11}$ & 0.71$^{+0.02}_{-0.02}$ & 0.75$^{+0.03}_{-0.02}$ & 6.50$^{+4.70}_{-4.90}$ & Diffusion & 0.89 & 100 \\
            350767277 & 669.1$^{+11.3}_{-9.6}$ & 5900$^{+135}_{-128}$ & 4.08$^{+0.06}_{-0.06}$ & 0.30$^{+0.15}_{-0.05}$ & 1.71$^{+0.09}_{-0.09}$ & 1.29$^{+0.05}_{-0.10}$ & 4.40$^{+2.00}_{-0.70}$ & Overshoot & 1.06 & 000\\
            \hline
            \label{table:output params}
        \end{tabular}
    \end{adjustbox}
\end{table*}

Validating stellar ages for a large number of stars is challenging due to the lack of available catalogues of a similar size which provide reliable ages. More precise literature sources involve much smaller numbers, focusing on stars that are best for characterisation. We investigate how our ages compare to several catalogues as means of validation.

We first compare our stellar ages to the 989 stars in \cite{Berger_2026}. We see weak agreement, with a median residual scatter of 37.90\% compared to our median error on age of 56.1\%. We also calculate a median absolute offset of 1.86 Gyr, which can be seen as an apparent trend in the top left panel of Figure \ref{fig:age_comp} where we plot our best fit ages against those in \cite{Berger_2026}. As mentioned previously, there is little similarity between the spectroscopic inputs, modelling grid and fitting package used between this work and \cite{Berger_2026} which likely drives the differences in resulting stellar age. For example, BaSTI uses reaction rates from the NACRE database, whereas PARSEC uses JINA REACLIB. Additionally, compared to other studies which use the PARSEC grid e.g. StarHorse, \cite{Berger_2026} interpolates between the models to provide a finer grid with differing resolution. Many of these differences contribute to the overall systematic trend seen. 

Considering the StarHorse age values, we find 350,166 stars in common with our sample. Similar to the mass and temperature comparison, we see excellent agreement between the two samples, reflected by a median residual scatter of only 14.45\%, compared to an error on our measurements of 27.59\%. This is the smallest scatter calculated for all of our validation sources, likely as a result of StarHorse and our catalogue both using Gaia RVS inputs that have been calibrated using the same methodology.

Next, we compare our computed stellar ages to the 47 stars also in the Kepler LEGACY sample (where we compare to the LEGACY values calculated with the BASTA pipeline). Similarly to before, we plot metallicity discrepant stars in grey along with the two outliers from the LEGACY catalogue, excluding outliers only in our analysis. Our values have good agreement, with a median residual scatter of 21.93\% which is within our median error on age for the subset of 25.60\%. We calculate an absolute median residual offset of 1.18 Gyr which is apparent in our comparison plot in Figure~\ref{fig:age_comp}. A similar offset is seen in \cite{weeks_2025}, where a median residual of 1.28 Gyr was found. Such an offset is potentially attributed to the difference in the definition of the solar mixture. BaSTI uses the \cite{Caffau_2011} implementation with Z$_\odot$ = 0.0153, whereas GARSTEC uses \cite{Grevesse_1998} with Z$_\odot$ = 0.0171. Therefore, a star with \feh = 0 will appear more metal poor in the \cite{Grevesse_1998} scale compared to \cite{Caffau_2011}, resulting in systematically younger ages being fitted. Additionally, larger CNO values are defined in the \cite{Grevesse_1998} scale, which may result in shorter main sequence lifetimes therefore slightly younger ages. Furthermore, the effect of changing the mixing length parameter and solar mixture was investigated in \cite{silva-aguirre_2015}, ultimately quoting a reference systematic uncertainty from the difference in input physics of 9.6\% in age.

Similar to the LEGACY survey, the APOKASC catalogue provides spectroscopic (\teff,\meh) and asteroseismic ($\Delta \nu$, $\nu_{max}$) data for dwarfs and subgiants obtained from APOGEE spectra and Kepler respectively \citep[][]{Serenelli_2017}. Stellar properties are estimated from these parameters using grid-based modelling techniques with the Bellaterra Stellar Parameters Pipeline \citep[BeSPP,][]{Serenelli_2013} and GARSTEC grid. The current catalogue comprises 415 stars, 170 of which overlap with our sample. We take the central values determined using the SDSS \teff scale, since this was found to provide a better match to Gaia benchmark stars. Comparing these stars we see excellent agreement, reflected by a median residual scatter of 17.72\% well within our error of 21.81\%. We see a slight offset towards older ages, which may be an effect of the difference in the adoption of diffusion. The models used in APOKASC follow the same basic prescription from \cite{Thoul_1994}, however additional mixing below the convective envelope is included. This is more likely to affect lower mass stars, and therefore older ages.

We then use the TESS-Keck sample as a source for comparison. TESS-Keck uses a combination of photometry, high resolution spectroscopy and Gaia parallaxes to homogeneneously fit stellar properties \citep[][]{MacDougall_2023}. \teff, \feh and \logg are directly extracted from HIRES spectra at the W.M. Keck Observatory \citep[][]{vogt_1994}, along with isochrone modeling to fit radius, mass and age using \verb|isoclassify| and the MESA Isochrones and Stellar Tracks models (MIST). We find 68 stars which overlap, of which 13 have discrepant metallicity values between Gaia and TESS-Keck. We plot these in grey in the lower right panel of Figure \ref{fig:age_comp} which displays our comparison, however, all 68 stars are included in our statistical analysis. We find good agreement between our values and TESS-Keck, with a median residual scatter of 35.55\% which is well within our median error of 56.98\%. Similarly to LEGACY, we calculate a median residual offset of 2 Gyr. TESS-Keck is the only comparison sample which utilises the MIST grid \citep[][]{MIST}, sharing very little in common with our stellar characterisation choices. 

Lastly, to further validate our methodology and thus our stellar ages, we take the 26 stars used in \cite{weeks_2025} whereby the stellar properties of rocky exoplanet hosts were characterised, using asteroseismology when available. We prepare the inputs in the same way as explained in Section~\ref{sec: BASTA fitting}, however we first run the standard grid, not accounting for any additional microphysics. Additionally, we run the overshoot and diffusion models on the full sample. We then cut on the mass fitted from the standard grid run to determine our final set of values. For stars with M $\leq$ 1.2 M$_\odot$ we take values fitted using the diffusion grid, and for M > 1.2 M$_\odot$ from the overshoot run. Plotting our comparison in Figure \ref{fig:age_comp}, we find excellent agreement between the sets of values, with a median residual scatter of 16.20\% compared to the median age error on our values of 49.14\%, likely due to the same fitting package and spectroscopic inputs being used.

Overall, we find good agreement between the stellar ages we derive with our methodology and those from various samples. Considering the difficulties in validating stellar ages, our values compare well to smaller, precise catalogues which have selected targets best for stellar age determination. Although a precision of 50\% seems large, this is comparable to the fractional median errors on age for the literature sources, and in some cases, we calculate higher precision on our values which we discuss in Section \ref{sec:results}. Offsets may be explained by a combination of the difference in grids and pipelines used, along with differing input parameters as displayed in Table \ref{table:input physics}. Recent studies have investigated the effect of model grids, finding uncertainties of order $\sim$20\% on age for main sequence and subgiants \citep[][]{tayar_2022} when running isochrone fits using four widely utilised model grids. We note that we have not included any systematic errors that may arise from using different modelling grids in our best fit values and associated uncertainties in the catalogue. Therefore, we argue that we find reliable stellar parameters for a large proportion of our sample, thus providing a homogeneous catalogue of stellar ages for a large subset of TESS stars.

\subsection{The effect of RUWE and temperature}
We have not identified or treated any potential binaries with different methodology, therefore risk contamination from these systems. To investigate how this may effect our results, we use the Gaia Re-normalised Unit Weight Error (RUWE), a metric which measures the quality of the astrometric solution for a single star. Typically, RUWE values $\leq$ 1.4 indicate good astrometric solution and values larger than this could be caused by unresolved binarity. However RUWE can also be amplified by other factors such as stellar crowding or the presence of a circumstellar disk \citep[][]{Belokurov_2020,Fitton_2022}. We plot the age distribution of stars with good astrometric solution (RUWE $\leq$ 1.4, 591,546 stars) in blue, and those with poor solution (RUWE > 1.4, 126,261 stars) in black, seen in the histogram in Figure \ref{fig:BLS_sample_hists}. We see a slight shift towards older ages for the high RUWE stars, with a median age of 5.1 Gyr compared to 4.4 Gyr for low RUWE stars. The two distributions are similar until $\sim$ 7 Gyr, after which there are a consistently higher proportion of high RUWE stars. This is expected if such stars are binaries being treated as single stars, and have a mass ratio close to 1. For these cases the absolute magnitude M$_V$ is 0.75 brighter than for a single star, which may place it in the position of a turn-off star on the HR diagram, resulting in an overestimation of age \citep[][]{Jorgensen_2005,Casamiquela_2024}. We further investigate if removing stars with high RUWE improves comparison to the previously mentioned literature sources by calculating the median residual scatter. We find that for the majority of the sources there is no significant change in the median residual scatter for radius, mass or age since there are generally a small fraction of high RUWE stars in each overlapping sample, ranging from 0 to 13\% of the sample.
We opt not to remove any stars based on their RUWE from our sample, but we do provide the Gaia DR3 RUWE values in our final results table as an indicator of reliability, displayed in Table \ref{table:output params}. 

Similarly, despite the temperature cut at 4000~K when defining our sample (see Section \ref{sec:sample selection}), there is likely still a high degree of degeneracy in the isochrone grids at these low temperatures, and such stars have main sequence lifetimes that exceed the age of the universe which makes it difficult to precisely model them. We further check if a more conservative lower limit of 5000~K significantly changes our results, and recalculate the values of median residual scatter for stars with \teff $\geq$ 5000~K. Overall, we find that values for the median residual scatter do not change considerably, thus do not affect the outcomes of our validation. Therefore, we also do not remove any of these stars, but note that caution should be taken if using the parameters from this catalogue for stars of \teff < 5000~K.

\section{Results}
\label{sec:results}
\subsection{Full stellar sample}

\begin{table}
    \centering
    \caption{Median fractional errors of each literature catalogue we use as comparison in this work, detailed in Section \ref{sec:results}. We include the number of stars in each sample as N$_{\mathrm{stars}}$, whereby we have removed any NaN values on relevant measurements and errors, as well as removing any 0 values. Additionally, we remove uninformative ages from the \protect\cite{Berger_2026} sample, indicated by an asterisk. However, we note that the number of stars used in each precision calculation may differ slightly. For NASA exoplanet archive, 5439 stars are used to find precision on radius and mass, and 4472 for age. Additionally, for the \protect\cite{Berger_2026} sample 8259 stars are used for radius and mass precision, and 6576 for age. Finally, 4,210,235 stars are used for mass and 798,812 for age when calculating StarHorse precisions. The same sample of stars are used to calculate all precision values for all other literature sources. }
    \begin{adjustbox}{width=\columnwidth}
        \begin{tabular}{ccccc}
            \hline
            Literature Source & N$_{\mathrm{stars}}$ & \multicolumn{3}{c}{Median error} \\
            & & Radius & Mass & Age \\
            \hline
             This work & 717,807 & 4.29\% & 4.49\% & 50.00\% \\
             TIC & 1,727,987,580 & 6.95\% & 11.12\% & \\
             SWEET-Cat & 4205 & 3.28\% & 2.12\% \\
             NASA exoplanet archive & 5439 & 5.33\% & 5.12\% & 49.38\% \\
             \cite{Berger_2026} & 8259 & 4.06\% & 5.49\% & 84.37\% \\
             StarHorse & 4,210,235 & & 7.04\% & 19.85\% \\
             APOKASC & 408 & 2.29\% & 4.30\% & 15.61\% \\
             Kepler LEGACY & 66 & 1.09\% & 2.62\% & 13.47\% \\
             TESS-Keck & 85 & 1.76\% & 3.03\% & 41.18\% \\
             \cite{weeks_2025} & 26 & 3.76\% & 7.32\% & 52.34\% \\
             \hline
             \label{table: precision table}
        \end{tabular}
    \end{adjustbox}
\end{table}

\begin{figure}
    \centering
        \includegraphics[width=\linewidth]{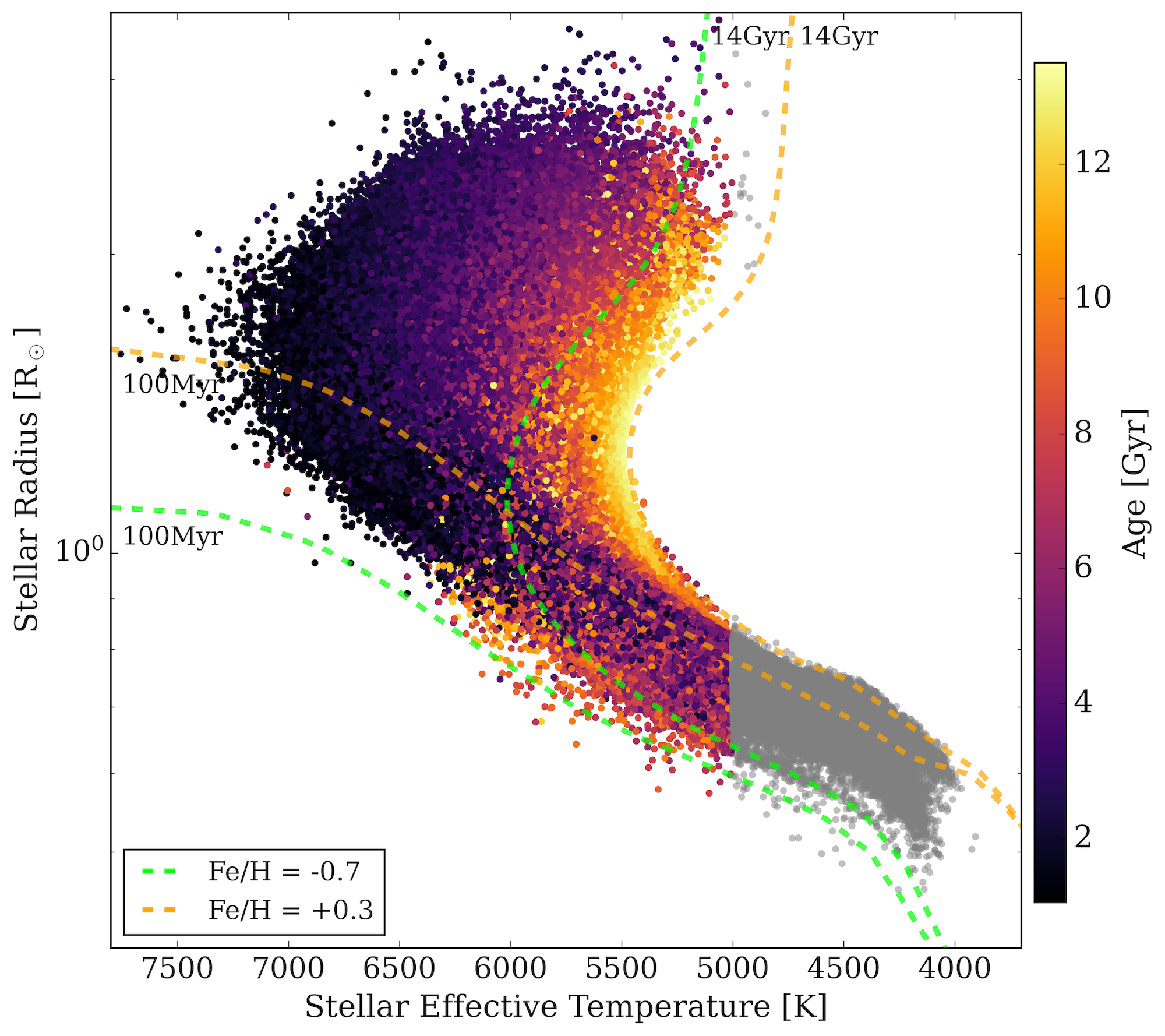}
    \caption{Hertzsprung-Russel diagram of our full stellar sample in stellar radius-temperature space, colour-coded by the best fit age. We plot stars with \teff < 5000 K in grey. Overplotted are the BaSTI tracks for Fe/H = -0.7 (green) and +0.3 (orange) for 100 Myr and 14 Gyr, showing the ranges in which our sample covers.}
    \label{fig:HR_diagram}
\end{figure}

\begin{figure}                                                                
    \centering
    \includegraphics[width=\linewidth]{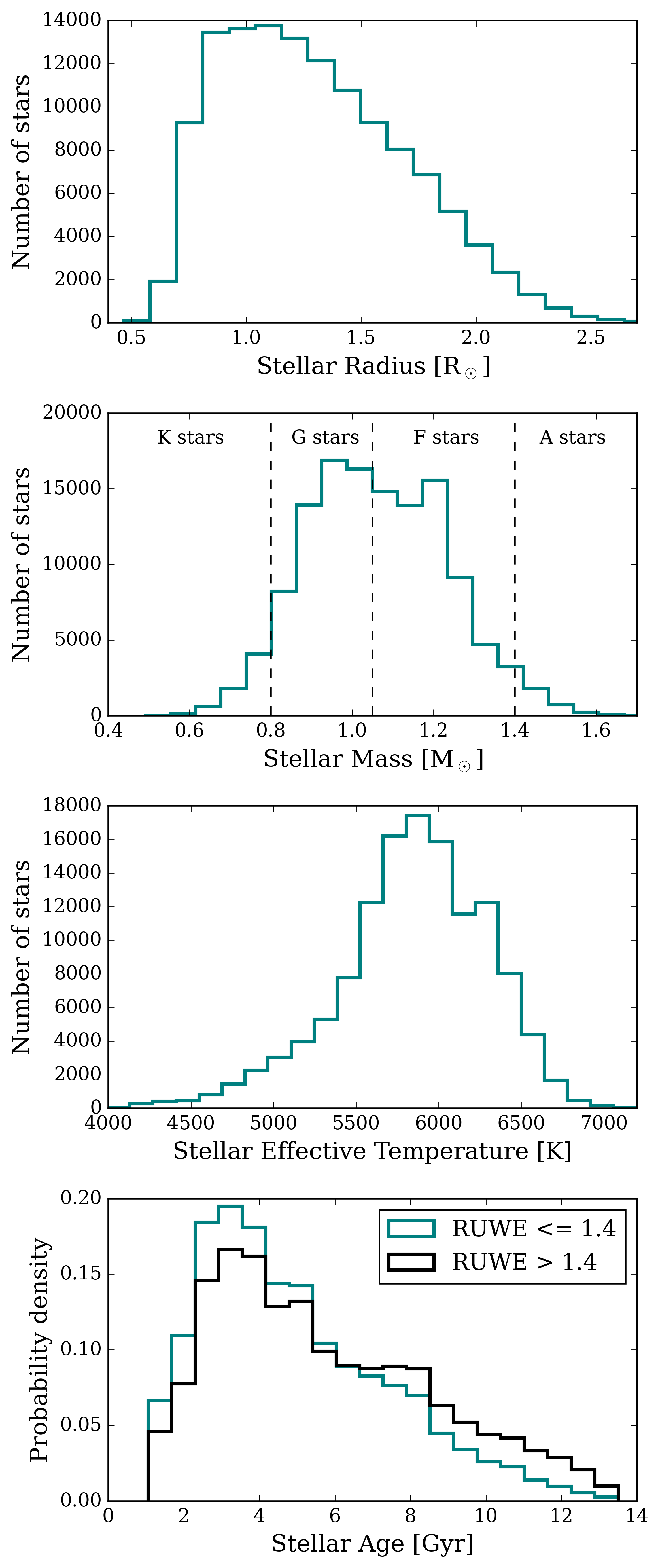}
        \caption{Distributions of the stellar radius, mass, age and temperature we fit for the sample of 717,899 stars used in this study. Stars with RUWE $\leq$ 1.4 are plotted in blue, and stars with RUWE > 1.4 in black for the age histogram.}
    \label{fig:BLS_sample_hists}
\end{figure}

\begin{figure}
    \centering
        \includegraphics[width=\linewidth]{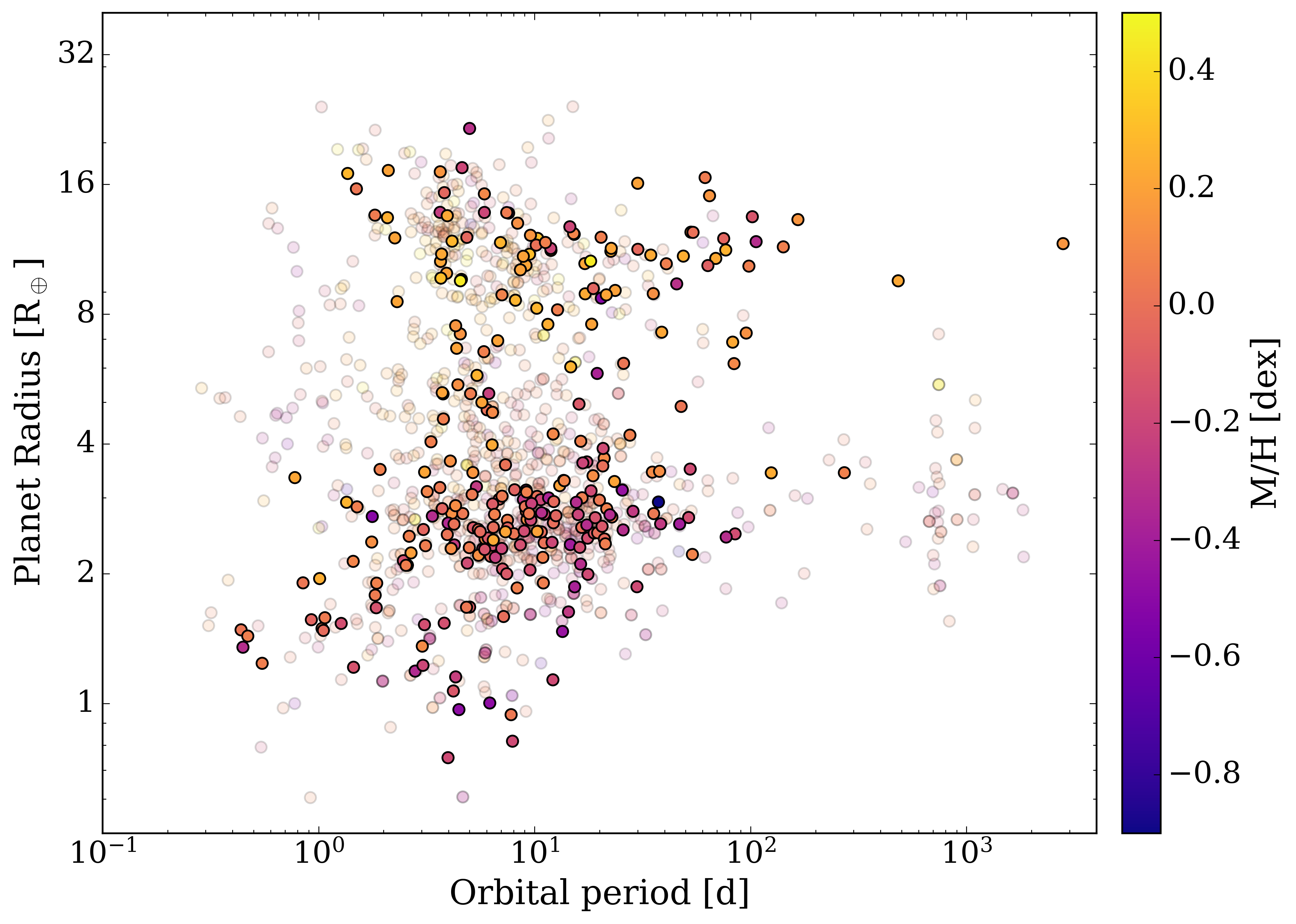}
    \caption{Recalculated planet radius against orbital period for the TESS confirmed planets and candidates in our sample. We plot candidates with a lower transparency and colour code by the best fit \meh.}
    \label{fig:per-rad}
\end{figure}

We have determined precise masses, radii, and ages for a sample of 717,807 stars. We list these parameters, together with other relevant stellar parameters, in Table~\ref{table:output params}. We plot the stellar radius as a function of temperature for our full sample in Figure \ref{fig:HR_diagram}, colour-coded based on their stellar age. As expected, a gradient in age is apparent, which increases towards cooler temperatures. 
We plot all stars with low temperatures (\teff < 5000 K) as grey points, since stellar ages are less reliable in this part of parameter space (see Section \ref{sec:validation}). Near the bottom of the main sequence, we see a track of $\sim$ 700 old, metal poor stars, which we expect are most likely cool subdwarfs. These stars are main sequence stars that are less massive than dwarfs at the same temperature, and are less luminous so they lie between the main sequence and white dwarfs on the colour-magnitude diagram \citep[][]{Marshall_2007}. They are believed to be Population II stars, so are metal-poor ([Fe/H] < -1) \citep[][]{Ziegler_2015}. We check the RUWE values for these stars, and find that only a small number ($\sim$ 200) have a RUWE $\geq$ 1.2. Therefore, these stars are most likely not binary stars that are misclassified.

We plot the distribution of our best fit radius, mass, age, and temperature in Figure \ref{fig:BLS_sample_hists}. Radius and mass peak around slightly super-solar values, with a median of 1.22 R$_\odot$ and 1.07 M$_\odot$, and a 1$\sigma$ range of 0.83 -- 1.61 R$_\odot$ and 0.90 -- 1.24 M$_\odot$ respectively. 
For our full sample we find median fractional uncertainties of 4.29\%, 4.49\%, 50.0\% and 1.74\% on stellar radius, mass, age and temperature respectively, reflecting a median absolute precision of 0.056 R$_\odot$, 0.048 M$_\odot$, 2.1 Gyr and 104~K. 
This consists of 40,691 K-type (0.45 $\leq$ M$_*$ < 0.8 M$_\odot$), 292,113 G-type (0.8 $\leq$ M$_*$ < 1.05 M$_\odot$), 363,281 F-type (1.05 $\leq$ < M$_*$ < 1.4 M$_\odot$) and 21,722 A-type (1.4 $\leq$ M$_*$ < 2.3 M$_\odot$) stars. We find that K stars have the largest errors on age (a median error of 4.55 Gyr), with error values ranging between 1 -- 6 Gyr. We note that the majority of these stars are also cooler than 5000 K, which as previously mentioned are the most challenging stars for age characterisation. G-type stars have a median age error of 3.55 Gyr, followed by 1.36 Gyr for F stars and 0.5 Gyr for A stars, with errors on age ranging from $\sim$0.3 -- 1 Gyr. Additionally, subgiants have smaller errors on age compared to dwarfs in our sample, with median errors of 0.9 Gyr and 2.7 Gyr respectively.  

As discussed in the previous section, age estimates using grid-based modelling are more suitable in certain parameter ranges. We have included a quality flag in our catalogue consisting of three numbers which are either 0 or 1. The first number indicates if the star has a \teff < 5000 K or RUWE > 1.2 (1), or not (0). The second number indicates if the star has an age precision > 75\% (1) or not (0). The third indicates if the input values for \teff and \meh disagree with the best fit values within the uncertainties (1) or if they agree (0). We encourage that these flags should be considered when using values from this catalogue.

For comparison, we calculate the precision given by the median fractional error on stellar radius, mass and age (when available) for several samples commonly used, and list the results in Table~\ref{table: precision table}. Considering the entire TIC, we find values of 6.95\% and 11.12\% on radius and mass, compared to a precision of 4.29\% and 4.49\% on our measurements. We see a large improvement on precision in both parameters, but particularly for stellar mass for which we have refined the precision by a factor of $\sim$2.5. Such an improvement may be beneficial for a number of research areas, and in particular for exoplanet population studies where homogeneous parameters such as stellar masses are needed for large samples of stars, irrespective of whether or not they are known to host transiting planets. Compared to the NASA exoplanet archive, we find a very similar level of precision for radius, mass, and age, noting that this sample is more than two orders of magnitude smaller than ours.
Furthermore, comparing to measurements from \cite{Berger_2026}, we see that we achieve a similar precision for radius and mass, and a significantly improved age precision (84.37\% for \cite{Berger_2026}, and 50.00\% for our sample), again noting our sample is $\sim$ two orders of magnitude larger. Overall, we typically match the level of precision on radius, mass and age or see a large improvement on our values when comparing to samples commonly used in the literature, additionally obtaining such parameters for a much larger number of stars.

One exception for isochrone-based catalogues used in this study is the StarHorse sample. Although we see an improvement on our mass precision compared to this catalogue (4.49\% for our sample and 7.04\% for StarHorse) a better age precision is apparent on the StarHorse age of 19.85\% compared to our median precision of 50\%. However, StarHorse quotes age values for a subset of the full sample, restricting to main sequence turn off and subgiant branch stars only. Such stars will provide more reliable isochrone age determination, since the evolutionary tracks vary more significantly with age in this region. This is apparent from a much improved precision of 27.59\% on our age measurements for the stars which overlap with StarHorse only. We provide ages for our full sample, spanning a larger range of evolutionary states.

When we compare to smaller, precise asteroseimology catalogues such as Kepler LEGACY and APOKASC, we see that our values have larger median fractional errors overall. For example, a precision on radius, mass and age of 1.09\%, 2.62\% and 13.47\% is seen for Kepler LEGACY, compared to our precision of 4.29\%, 4.49\% and 50.00\%. However, these sources consist of stars which are the most suitable targets for characterisation. This is reflected by a higher precision on our measurements for the stars which overlap with Kepler LEGACY (see Section~\ref{sec:validation}). For example, such stars are among the brightest in our sample and the most amenable for age determination, reflected by a much higher precision of 25.60\% on our best fit age for these overlapping stars. Our catalogue allows for the availability of stellar ages for a much larger number of stars, increasing the sample size by a factor of $\sim$10,000 relative to these samples.

\subsection{Exoplanet hosts}
To put our sample into the exoplanet context, we recalculate the planet radii for the TESS confirmed planets and TOIs which overlap. We use the values of $R_p/R_\star$ and orbital period from the NASA Exoplanet Archive, from which we calculate the revised planet radius using our best fit stellar radii. For TESS planet candidates we use the TESS project candidate table from NASA Exoplanet Archive, however this does not provide a radius ratio. Therefore, we calculate $R_P/R_*$ using the values from the TOI table first, before recalculating planet radius using our stellar radii values. Our sample is found to host 278 known TESS planets and 915 candidates, of which we plot the recalculated radii and orbital periods in Figure \ref{fig:per-rad}. Two populations of planets can clearly be seen, with a separation between hot Jupiters and smaller planets comprising super-Earths and sub-Neptunes. We do not see a clear radius valley \citep[][]{Fulton_2017,Vaneylen_2018} for the smaller planets, which is most likely due to the precision on the radius ratio and the lack of super-Earths in the sample. A more blurred valley has been commonly observed for TESS planets \citep[e.g.,][]{Mulders_2025,Berger_2026}, so this is unsurprising. We colour code by the fitted \meh for each star, which clearly shows the differences in metallicity between the hot Jupiters and smaller, rocky planets, confirming the well known concept that hot Jupiters orbit more metal-rich stars \citep[][]{Gonzalez_1997,Fischer_2005,Mortier_2013}.

\section{Conclusion}
\label{sec:conclusion}
Here, we have presented a catalogue of 717,807 TESS FGK dwarfs and early subgiants which we have characterised using a homogeneous methodology. We made use of the \verb|BASTA| package along with the BaSTI isochrone grid, using Gaia DR3 inputs of photometry in the G, BP and RP bands, parallax, temperature and metallicity. Additionally, we have considered different microphysical processes depending on the stellar mass, invoking the effects of either convective overshoot or atomic diffusion to allow for careful isochrone analysis. 

As a result, we have derived values of distance, \teff, \meh, \logg, radius, mass and age, making up one of the biggest homogeneous stellar catalogues to date. Our final catalogue provides stellar properties for a large fraction of TIC stars, spanning several stellar types and two evolutionary stages. Most importantly, we provide stellar ages for these stars, which are not available in the TIC. We fit radius, mass and ages with precisions of 4.29\%, 4.49\% and 50.00\% respectively, which are either similar to or an improvement on the equivalent precision for several current, widely-used catalogues. The availability of this new catalogue will aid many future studies for a range of astrophysical avenues such as exoplanet occurrence rates, characterisation and stellar structure, whereby stellar parameters for a large number of stars is warranted.

Additionally, we validated our best fit parameters to several sources of literature for catalogues which provide precise properties, ranging in sample size, homogeneity and characterisation technique. We find good general agreement for all sources on a statistical level for radius, mass, temperature and age, however offsets are apparent for different catalogues. Such offsets are a reflection of the model grid chosen or methodology used, highlighting the typical issues experienced in stellar characterisation. We also identify TESS exoplanet hosts within our sample, finding 278 known planet hosts and 915 candidates. We recalculate the planet radii for these systems, and observe the well known giant planet-metallicity correlation using our fitted \meh values. We provide our input parameters and best fit parameter catalogues as machine-readable tables.

\section*{Acknowledgements}
F.W would like to thank the Science and Technology Facilities Council (STFC) for funding support through a PhD studentship (grant number ST/Y509784/1). V.V.E. has been supported by UK’s Science \& Technology Facilities Council through the STFC grants ST/W001136/1 and ST/S000216/1, and UKRI1180. A.W. was supported by UK’s Science \& Technology Facilities Council through the STFC grants ST/W001136/1, and by the Australian Research Council 
(DP 250102562). We also thank the referee for their constructive comments which helped to improve the quality of the manuscript. 

\section*{Data availability}
This research has made use of the NASA Exoplanet Archive, which is operated by the California Institute of Technology, under contract with the National Aeronautics and Space Administration under the Exoplanet Exploration Program. This work has made use of data from the European Space Agency (ESA) mission
{\it Gaia} (\url{https://www.cosmos.esa.int/gaia}), processed by the {\it Gaia}
Data Processing and Analysis Consortium (DPAC,
\url{https://www.cosmos.esa.int/web/gaia/dpac/consortium}). Funding for the DPAC
has been provided by national institutions, in particular the institutions
participating in the {\it Gaia} Multilateral Agreement.\\
This study makes use of numpy (\hyperlink{numpy}{https://numpy.org}), pandas \citep[][]{pandas}, matplotlib \citep[][]{matplotlib}, astropy \citep[][]{astropy}, astroquery \citep[][]{astroquery} and BASTA (\hyperlink{basta}{https://basta.readthedocs.io}). The full input parameter and best fit parameter catalogues are available as machine-readable tables at CDS via anonymous ftp to cdsarc.u-strasbg.fr (130.79.128.5) or via \url{https://cdsarc.cds.unistra.fr/viz-bin/cat/J/MNRAS}.



\bibliographystyle{mnras}
\bibliography{example} 

@INPROCEEDINGS{Ricker_2014,
       author = {{Ricker}, George R. and {Winn}, Joshua N. and {Vanderspek}, Roland and {Latham}, David W. and {Bakos}, G{\'a}sp{\'a}r. {\'A}. and {Bean}, Jacob L. and {Berta-Thompson}, Zachory K. and {Brown}, Timothy M. and {Buchhave}, Lars and {Butler}, Nathaniel R. and {Butler}, R. Paul and {Chaplin}, William J. and {Charbonneau}, David and {Christensen-Dalsgaard}, J{\o}rgen and {Clampin}, Mark and {Deming}, Drake and {Doty}, John and {De Lee}, Nathan and {Dressing}, Courtney and {Dunham}, E.~W. and {Endl}, Michael and {Fressin}, Francois and {Ge}, Jian and {Henning}, Thomas and {Holman}, Matthew J. and {Howard}, Andrew W. and {Ida}, Shigeru and {Jenkins}, Jon and {Jernigan}, Garrett and {Johnson}, John A. and {Kaltenegger}, Lisa and {Kawai}, Nobuyuki and {Kjeldsen}, Hans and {Laughlin}, Gregory and {Levine}, Alan M. and {Lin}, Douglas and {Lissauer}, Jack J. and {MacQueen}, Phillip and {Marcy}, Geoffrey and {McCullough}, P.~R. and {Morton}, Timothy D. and {Narita}, Norio and {Paegert}, Martin and {Palle}, Enric and {Pepe}, Francesco and {Pepper}, Joshua and {Quirrenbach}, Andreas and {Rinehart}, S.~A. and {Sasselov}, Dimitar and {Sato}, Bun'ei and {Seager}, Sara and {Sozzetti}, Alessandro and {Stassun}, Keivan G. and {Sullivan}, Peter and {Szentgyorgyi}, Andrew and {Torres}, Guillermo and {Udry}, Stephane and {Villasenor}, Joel},
        title = "{Transiting Exoplanet Survey Satellite (TESS)}",
    booktitle = {Space Telescopes and Instrumentation 2014: Optical, Infrared, and Millimeter Wave},
         year = 2014,
       editor = {{Oschmann}, Jr., Jacobus M. and {Clampin}, Mark and {Fazio}, Giovanni G. and {MacEwen}, Howard A.},
       series = {Society of Photo-Optical Instrumentation Engineers (SPIE) Conference Series},
       volume = {9143},
        month = aug,
          eid = {914320},
        pages = {914320},
          doi = {10.1117/12.2063489},
archivePrefix = {arXiv},
       eprint = {1406.0151},
 primaryClass = {astro-ph.EP},
       adsurl = {https://ui.adsabs.harvard.edu/abs/2014SPIE.9143E..20R}
}

@ARTICLE{Aguirre_B_rsen_Koch_2021,
       author = {{Aguirre B{\o}rsen-Koch}, V. and {R{\o}rsted}, J.~L. and {Justesen}, A.~B. and {Stokholm}, A. and {Verma}, K. and {Winther}, M.~L. and {Knudstrup}, E. and {Nielsen}, K.~B. and {Sahlholdt}, C. and {Larsen}, J.~R. and {Cassisi}, S. and {Serenelli}, A.~M. and {Casagrande}, L. and {Christensen-Dalsgaard}, J. and {Davies}, G.~R. and {Ferguson}, J.~W. and {Lund}, M.~N. and {Weiss}, A. and {White}, T.~R.},
        title = "{The BAyesian STellar algorithm (BASTA): a fitting tool for stellar studies, asteroseismology, exoplanets, and Galactic archaeology}",
      journal = {\mnras},
         year = 2022,
        month = jan,
       volume = {509},
       number = {3},
        pages = {4344-4364},
          doi = {10.1093/mnras/stab2911},
archivePrefix = {arXiv},
       eprint = {2109.14622},
 primaryClass = {astro-ph.SR},
       adsurl = {https://ui.adsabs.harvard.edu/abs/2022MNRAS.509.4344A}
}

@ARTICLE{Salpeter_1955,
       author = {{Salpeter}, Edwin E.},
        title = "{The Luminosity Function and Stellar Evolution.}",
      journal = {\apj},
         year = 1955,
        month = jan,
       volume = {121},
        pages = {161},
          doi = {10.1086/145971},
       adsurl = {https://ui.adsabs.harvard.edu/abs/1955ApJ...121..161S}
}

@ARTICLE{Lindegren_2021,
       author = {{Lindegren}, L. and {Bastian}, U. and {Biermann}, M. and {Bombrun}, A. and {de Torres}, A. and {Gerlach}, E. and {Geyer}, R. and {Hern{\'a}ndez}, J. and {Hilger}, T. and {Hobbs}, D. and {Klioner}, S.~A. and {Lammers}, U. and {McMillan}, P.~J. and {Ramos-Lerate}, M. and {Steidelm{\"u}ller}, H. and {Stephenson}, C.~A. and {van Leeuwen}, F.},
        title = "{Gaia Early Data Release 3. Parallax bias versus magnitude, colour, and position}",
      journal = {\aap},
         year = 2021,
        month = may,
       volume = {649},
          eid = {A4},
        pages = {A4},
          doi = {10.1051/0004-6361/202039653},
archivePrefix = {arXiv},
       eprint = {2012.01742},
 primaryClass = {astro-ph.IM},
       adsurl = {https://ui.adsabs.harvard.edu/abs/2021A&A...649A...4L}
}

@ARTICLE{Recio_blanco_2023,
       author = {{Recio-Blanco}, A. and {de Laverny}, P. and {Palicio}, P.~A. and {Kordopatis}, G. and {{\'A}lvarez}, M.~A. and {Schultheis}, M. and {Contursi}, G. and {Zhao}, H. and {Torralba Elipe}, G. and {Ordenovic}, C. and {Manteiga}, M. and {Dafonte}, C. and {Oreshina-Slezak}, I. and {Bijaoui}, A. and {Fr{\'e}mat}, Y. and {Seabroke}, G. and {Pailler}, F. and {Spitoni}, E. and {Poggio}, E. and {Creevey}, O.~L. and {Abreu Aramburu}, A. and {Accart}, S. and {Andrae}, R. and {Bailer-Jones}, C.~A.~L. and {Bellas-Velidis}, I. and {Brouillet}, N. and {Brugaletta}, E. and {Burlacu}, A. and {Carballo}, R. and {Casamiquela}, L. and {Chiavassa}, A. and {Cooper}, W.~J. and {Dapergolas}, A. and {Delchambre}, L. and {Dharmawardena}, T.~E. and {Drimmel}, R. and {Edvardsson}, B. and {Fouesneau}, M. and {Garabato}, D. and {Garc{\'\i}a-Lario}, P. and {Garc{\'\i}a-Torres}, M. and {Gavel}, A. and {Gomez}, A. and {Gonz{\'a}lez-Santamar{\'\i}a}, I. and {Hatzidimitriou}, D. and {Heiter}, U. and {Jean-Antoine Piccolo}, A. and {Kontizas}, M. and {Korn}, A.~J. and {Lanzafame}, A.~C. and {Lebreton}, Y. and {Le Fustec}, Y. and {Licata}, E.~L. and {Lindstr{\o}m}, H.~E.~P. and {Livanou}, E. and {Lobel}, A. and {Lorca}, A. and {Magdaleno Romeo}, A. and {Marocco}, F. and {Marshall}, D.~J. and {Mary}, N. and {Nicolas}, C. and {Pallas-Quintela}, L. and {Panem}, C. and {Pichon}, B. and {Riclet}, F. and {Robin}, C. and {Rybizki}, J. and {Santove{\~n}a}, R. and {Silvelo}, A. and {Smart}, R.~L. and {Sarro}, L.~M. and {Sordo}, R. and {Soubiran}, C. and {S{\"u}veges}, M. and {Ulla}, A. and {Vallenari}, A. and {Zorec}, J. and {Utrilla}, E. and {Bakker}, J.},
        title = "{Gaia Data Release 3. Analysis of RVS spectra using the General Stellar Parametriser from spectroscopy}",
      journal = {\aap},
         year = 2023,
        month = jun,
       volume = {674},
          eid = {A29},
        pages = {A29},
          doi = {10.1051/0004-6361/202243750},
archivePrefix = {arXiv},
       eprint = {2206.05541},
 primaryClass = {astro-ph.GA},
       adsurl = {https://ui.adsabs.harvard.edu/abs/2023A&A...674A..29R}
}

@ARTICLE{Green_2019,
       author = {{Green}, Gregory M. and {Schlafly}, Edward and {Zucker}, Catherine and {Speagle}, Joshua S. and {Finkbeiner}, Douglas},
        title = "{A 3D Dust Map Based on Gaia, Pan-STARRS 1, and 2MASS}",
      journal = {\apj},
         year = 2019,
        month = dec,
       volume = {887},
       number = {1},
          eid = {93},
        pages = {93},
          doi = {10.3847/1538-4357/ab5362},
archivePrefix = {arXiv},
       eprint = {1905.02734},
 primaryClass = {astro-ph.GA},
       adsurl = {https://ui.adsabs.harvard.edu/abs/2019ApJ...887...93G}
}

@ARTICLE{Schlegel_1998,
       author = {{Schlegel}, David J. and {Finkbeiner}, Douglas P. and {Davis}, Marc},
        title = "{Maps of Dust Infrared Emission for Use in Estimation of Reddening and Cosmic Microwave Background Radiation Foregrounds}",
      journal = {\apj},
         year = 1998,
        month = jun,
       volume = {500},
       number = {2},
        pages = {525-553},
          doi = {10.1086/305772},
archivePrefix = {arXiv},
       eprint = {astro-ph/9710327},
 primaryClass = {astro-ph},
       adsurl = {https://ui.adsabs.harvard.edu/abs/1998ApJ...500..525S}
}

@ARTICLE{Claret_2019,
       author = {{Claret}, Antonio and {Torres}, Guillermo},
        title = "{The Dependence of Convective Core Overshooting on Stellar Mass: Reality Check and Additional Evidence}",
      journal = {\apj},
         year = 2019,
        month = may,
       volume = {876},
       number = {2},
          eid = {134},
        pages = {134},
          doi = {10.3847/1538-4357/ab1589},
archivePrefix = {arXiv},
       eprint = {1904.02714},
 primaryClass = {astro-ph.SR},
       adsurl = {https://ui.adsabs.harvard.edu/abs/2019ApJ...876..134C}
}

@INPROCEEDINGS{Nsamba_2018,
       author = {{Nsamba}, Benard and {Campante}, Tiago L. and {Monteiro}, M{\'a}rio J.~P.~F.~G. and {Cunha}, Margarida S. and {Rendle}, Ben M. and {Reese}, Daniel R. and {Verma}, Kuldeep},
        title = "{Model physics in low-mass solar-type stars: atomic diffusion and metallicity mixture}",
    booktitle = {PHysics of Oscillating STars},
         year = 2018,
        month = sep,
          eid = {11},
        pages = {11},
          doi = {10.5281/zenodo.1468510},
archivePrefix = {arXiv},
       eprint = {1812.00431},
 primaryClass = {astro-ph.SR},
       adsurl = {https://ui.adsabs.harvard.edu/abs/2018phos.confE..11N}
}

@ARTICLE{Claret_2016,
       author = {{Claret}, A.},
        title = "{Theoretical gravity darkening as a function of optical depth. A first approach to fast rotating stars}",
      journal = {\aap},
         year = 2016,
        month = apr,
       volume = {588},
          eid = {A15},
        pages = {A15},
          doi = {10.1051/0004-6361/201527336},
archivePrefix = {arXiv},
       eprint = {1606.00834},
 primaryClass = {astro-ph.SR},
       adsurl = {https://ui.adsabs.harvard.edu/abs/2016A&A...588A..15C}
}

@ARTICLE{Mowlavi_2012,
       author = {{Mowlavi}, N. and {Eggenberger}, P. and {Meynet}, G. and {Ekstr{\"o}m}, S. and {Georgy}, C. and {Maeder}, A. and {Charbonnel}, C. and {Eyer}, L.},
        title = "{Stellar mass and age determinations . I. Grids of stellar models from Z = 0.006 to 0.04 and M = 0.5 to 3.5 M$_{☉}$}",
      journal = {\aap},
         year = 2012,
        month = may,
       volume = {541},
          eid = {A41},
        pages = {A41},
          doi = {10.1051/0004-6361/201117749},
archivePrefix = {arXiv},
       eprint = {1201.3628},
 primaryClass = {astro-ph.SR},
       adsurl = {https://ui.adsabs.harvard.edu/abs/2012A&A...541A..41M}
}

@ARTICLE{Hidalgo_2018,
       author = {{Hidalgo}, Sebastian L. and {Pietrinferni}, Adriano and {Cassisi}, Santi and {Salaris}, Maurizio and {Mucciarelli}, Alessio and {Savino}, Alessandro and {Aparicio}, Antonio and {Silva Aguirre}, Victor and {Verma}, Kuldeep},
        title = "{The Updated BaSTI Stellar Evolution Models and Isochrones. I. Solar-scaled Calculations}",
      journal = {\apj},
         year = 2018,
        month = apr,
       volume = {856},
       number = {2},
          eid = {125},
        pages = {125},
          doi = {10.3847/1538-4357/aab158},
archivePrefix = {arXiv},
       eprint = {1802.07319},
 primaryClass = {astro-ph.GA},
       adsurl = {https://ui.adsabs.harvard.edu/abs/2018ApJ...856..125H}
}

@ARTICLE{Tayar_2022,
       author = {{Tayar}, Jamie and {Claytor}, Zachary R. and {Huber}, Daniel and {van Saders}, Jennifer},
        title = "{A Guide to Realistic Uncertainties on the Fundamental Properties of Solar-type Exoplanet Host Stars}",
      journal = {\apj},
         year = 2022,
        month = mar,
       volume = {927},
       number = {1},
          eid = {31},
        pages = {31},
          doi = {10.3847/1538-4357/ac4bbc},
archivePrefix = {arXiv},
       eprint = {2012.07957},
 primaryClass = {astro-ph.EP},
       adsurl = {https://ui.adsabs.harvard.edu/abs/2022ApJ...927...31T}
}

@ARTICLE{Stassun_2019,
       author = {{Stassun}, Keivan G. and {Oelkers}, Ryan J. and {Paegert}, Martin and {Torres}, Guillermo and {Pepper}, Joshua and {De Lee}, Nathan and {Collins}, Kevin and {Latham}, David W. and {Muirhead}, Philip S. and {Chittidi}, Jay and {Rojas-Ayala}, B{\'a}rbara and {Fleming}, Scott W. and {Rose}, Mark E. and {Tenenbaum}, Peter and {Ting}, Eric B. and {Kane}, Stephen R. and {Barclay}, Thomas and {Bean}, Jacob L. and {Brassuer}, C.~E. and {Charbonneau}, David and {Ge}, Jian and {Lissauer}, Jack J. and {Mann}, Andrew W. and {McLean}, Brian and {Mullally}, Susan and {Narita}, Norio and {Plavchan}, Peter and {Ricker}, George R. and {Sasselov}, Dimitar and {Seager}, S. and {Sharma}, Sanjib and {Shiao}, Bernie and {Sozzetti}, Alessandro and {Stello}, Dennis and {Vanderspek}, Roland and {Wallace}, Geoff and {Winn}, Joshua N.},
        title = "{The Revised TESS Input Catalog and Candidate Target List}",
      journal = {\aj},
         year = 2019,
        month = oct,
       volume = {158},
       number = {4},
          eid = {138},
        pages = {138},
          doi = {10.3847/1538-3881/ab3467},
archivePrefix = {arXiv},
       eprint = {1905.10694},
 primaryClass = {astro-ph.SR},
       adsurl = {https://ui.adsabs.harvard.edu/abs/2019AJ....158..138S}
}

@ARTICLE{Stassun_2018,
       author = {{Stassun}, Keivan G. and {Oelkers}, Ryan J. and {Pepper}, Joshua and {Paegert}, Martin and {De Lee}, Nathan and {Torres}, Guillermo and {Latham}, David W. and {Charpinet}, St{\'e}phane and {Dressing}, Courtney D. and {Huber}, Daniel and {Kane}, Stephen R. and {L{\'e}pine}, S{\'e}bastien and {Mann}, Andrew and {Muirhead}, Philip S. and {Rojas-Ayala}, B{\'a}rbara and {Silvotti}, Roberto and {Fleming}, Scott W. and {Levine}, Al and {Plavchan}, Peter},
        title = "{The TESS Input Catalog and Candidate Target List}",
      journal = {\aj},
         year = 2018,
        month = sep,
       volume = {156},
       number = {3},
          eid = {102},
        pages = {102},
          doi = {10.3847/1538-3881/aad050},
archivePrefix = {arXiv},
       eprint = {1706.00495},
 primaryClass = {astro-ph.EP},
       adsurl = {https://ui.adsabs.harvard.edu/abs/2018AJ....156..102S}
}

@ARTICLE{Santos_2013,
       author = {{Santos}, N.~C. and {Sousa}, S.~G. and {Mortier}, A. and {Neves}, V. and {Adibekyan}, V. and {Tsantaki}, M. and {Delgado Mena}, E. and {Bonfils}, X. and {Israelian}, G. and {Mayor}, M. and {Udry}, S.},
        title = "{SWEET-Cat: A catalogue of parameters for Stars With ExoplanETs. I. New atmospheric parameters and masses for 48 stars with planets}",
      journal = {\aap},
         year = 2013,
        month = aug,
       volume = {556},
          eid = {A150},
        pages = {A150},
          doi = {10.1051/0004-6361/201321286},
archivePrefix = {arXiv},
       eprint = {1307.0354},
 primaryClass = {astro-ph.SR},
       adsurl = {https://ui.adsabs.harvard.edu/abs/2013A&A...556A.150S}
}

@ARTICLE{sousa_2021,
       author = {{Sousa}, S.~G. and {Adibekyan}, V. and {Delgado-Mena}, E. and {Santos}, N.~C. and {Rojas-Ayala}, B. and {Soares}, B.~M.~T.~B. and {Legoinha}, H. and {Ulmer-Moll}, S. and {Camacho}, J.~D. and {Barros}, S.~C.~C. and {Demangeon}, O.~D.~S. and {Hoyer}, S. and {Israelian}, G. and {Mortier}, A. and {Tsantaki}, M. and {Monteiro}, M.~A.},
        title = "{SWEET-Cat 2.0: The Cat just got SWEETer. Higher quality spectra and precise parallaxes from Gaia eDR3}",
      journal = {\aap},
         year = 2021,
        month = dec,
       volume = {656},
          eid = {A53},
        pages = {A53},
          doi = {10.1051/0004-6361/202141584},
archivePrefix = {arXiv},
       eprint = {2109.04781},
 primaryClass = {astro-ph.EP},
       adsurl = {https://ui.adsabs.harvard.edu/abs/2021A&A...656A..53S}
}

@ARTICLE{Antoniadis-Karnavas_2024,
       author = {{Antoniadis-Karnavas}, A. and {Sousa}, S.~G. and {Delgado-Mena}, E. and {Santos}, N.~C. and {Andreasen}, D.~T.},
        title = "{ODUSSEAS: Upgraded version with new reference scale and parameter determinations for 82 planet-host M dwarf stars in SWEET-Cat}",
      journal = {\aap},
         year = 2024,
        month = oct,
       volume = {690},
          eid = {A58},
        pages = {A58},
          doi = {10.1051/0004-6361/202450722},
archivePrefix = {arXiv},
       eprint = {2408.15808},
 primaryClass = {astro-ph.SR},
       adsurl = {https://ui.adsabs.harvard.edu/abs/2024A&A...690A..58A}
}

@ARTICLE{weeks_2025,
       author = {{Weeks}, Angharad and {Van Eylen}, Vincent and {Huber}, Daniel and {Kawata}, Daisuke and {Stokholm}, Amalie and {Aguirre B{\o}rsen-Koch}, Victor and {Pinilla}, Paola and {R{\o}rsted}, Jakob Lysgaard and {Winther}, Mark Lykke and {Berger}, Travis},
        title = "{A link between rocky planet composition and stellar age}",
      journal = {\mnras},
         year = 2025,
        month = may,
       volume = {539},
       number = {1},
        pages = {405-421},
          doi = {10.1093/mnras/staf474},
archivePrefix = {arXiv},
       eprint = {2411.17358},
 primaryClass = {astro-ph.EP},
       adsurl = {https://ui.adsabs.harvard.edu/abs/2025MNRAS.539..405W}
}

@article{MacDougall_2023,
   title={The TESS-Keck Survey. XV. Precise Properties of 108 TESS Planets and Their Host Stars},
   volume={166},
   ISSN={1538-3881},
   url={http://dx.doi.org/10.3847/1538-3881/acd557},
   DOI={10.3847/1538-3881/acd557},
   number={1},
   journal={The Astronomical Journal},
   publisher={American Astronomical Society},
   author={MacDougall, Mason G. and Petigura, Erik A. and Gilbert, Gregory J. and Angelo, Isabel and Batalha, Natalie M. and Beard, Corey and Behmard, Aida and Blunt, Sarah and Brinkman, Casey and Chontos, Ashley and Crossfield, Ian J. M. and Dai, Fei and Dalba, Paul A. and Dressing, Courtney and Fetherolf, Tara and Fulton, Benjamin and Giacalone, Steven and Hill, Michelle L. and Holcomb, Rae and Howard, Andrew W. and Huber, Daniel and Isaacson, Howard and Kane, Stephen R. and Kosiarek, Molly and Lubin, Jack and Mayo, Andrew and Močnik, Teo and Akana Murphy, Joseph M. and Pidhorodetska, Daria and Polanski, Alex S. and Rice, Malena and Robertson, Paul and Rosenthal, Lee J. and Roy, Arpita and Rubenzahl, Ryan A. and Scarsdale, Nicholas and Turtelboom, Emma V. and Tyler, Dakotah and Van Zandt, Judah and Weiss, Lauren M. and Yee, Samuel W.},
   year={2023},
   month=jun, pages={33} }

@ARTICLE{silva-aguirre_2017,
       author = {{Silva Aguirre}, V{\'\i}ctor and {Lund}, Mikkel N. and {Antia}, H.~M. and {Ball}, Warrick H. and {Basu}, Sarbani and {Christensen-Dalsgaard}, J{\o}rgen and {Lebreton}, Yveline and {Reese}, Daniel R. and {Verma}, Kuldeep and {Casagrande}, Luca and {Justesen}, Anders B. and {Mosumgaard}, Jakob R. and {Chaplin}, William J. and {Bedding}, Timothy R. and {Davies}, Guy R. and {Handberg}, Rasmus and {Houdek}, G{\"u}nter and {Huber}, Daniel and {Kjeldsen}, Hans and {Latham}, David W. and {White}, Timothy R. and {Coelho}, Hugo R. and {Miglio}, Andrea and {Rendle}, Ben},
        title = "{Standing on the Shoulders of Dwarfs: the Kepler Asteroseismic LEGACY Sample. II.Radii, Masses, and Ages}",
      journal = {\apj},
         year = 2017,
        month = feb,
       volume = {835},
       number = {2},
          eid = {173},
        pages = {173},
          doi = {10.3847/1538-4357/835/2/173},
archivePrefix = {arXiv},
       eprint = {1611.08776},
 primaryClass = {astro-ph.SR},
       adsurl = {https://ui.adsabs.harvard.edu/abs/2017ApJ...835..173S}
}

@ARTICLE{silva-aguirre_2015,
       author = {{Silva Aguirre}, V. and {Davies}, G.~R. and {Basu}, S. and {Christensen-Dalsgaard}, J. and {Creevey}, O. and {Metcalfe}, T.~S. and {Bedding}, T.~R. and {Casagrande}, L. and {Handberg}, R. and {Lund}, M.~N. and {Nissen}, P.~E. and {Chaplin}, W.~J. and {Huber}, D. and {Serenelli}, A.~M. and {Stello}, D. and {Van Eylen}, V. and {Campante}, T.~L. and {Elsworth}, Y. and {Gilliland}, R.~L. and {Hekker}, S. and {Karoff}, C. and {Kawaler}, S.~D. and {Kjeldsen}, H. and {Lundkvist}, M.~S.},
        title = "{Ages and fundamental properties of Kepler exoplanet host stars from asteroseismology}",
      journal = {\mnras},
         year = 2015,
        month = sep,
       volume = {452},
       number = {2},
        pages = {2127-2148},
          doi = {10.1093/mnras/stv1388},
archivePrefix = {arXiv},
       eprint = {1504.07992},
 primaryClass = {astro-ph.SR},
       adsurl = {https://ui.adsabs.harvard.edu/abs/2015MNRAS.452.2127S}
}

@INPROCEEDINGS{vogt_1994,
       author = {{Vogt}, S.~S. and {Allen}, S.~L. and {Bigelow}, B.~C. and {Bresee}, L. and {Brown}, B. and {Cantrall}, T. and {Conrad}, A. and {Couture}, M. and {Delaney}, C. and {Epps}, H.~W. and {Hilyard}, D. and {Hilyard}, D.~F. and {Horn}, E. and {Jern}, N. and {Kanto}, D. and {Keane}, M.~J. and {Kibrick}, R.~I. and {Lewis}, J.~W. and {Osborne}, J. and {Pardeilhan}, G.~H. and {Pfister}, T. and {Ricketts}, T. and {Robinson}, L.~B. and {Stover}, R.~J. and {Tucker}, D. and {Ward}, J. and {Wei}, M.~Z.},
        title = "{HIRES: the high-resolution echelle spectrometer on the Keck 10-m Telescope}",
    booktitle = {Instrumentation in Astronomy VIII},
         year = 1994,
       editor = {{Crawford}, David L. and {Craine}, Eric R.},
       series = {Society of Photo-Optical Instrumentation Engineers (SPIE) Conference Series},
       volume = {2198},
        month = jun,
        pages = {362},
          doi = {10.1117/12.176725},
       adsurl = {https://ui.adsabs.harvard.edu/abs/1994SPIE.2198..362V}
}

@ARTICLE{Kordopatis_2023,
       author = {{Kordopatis}, G. and {Schultheis}, M. and {McMillan}, P.~J. and {Palicio}, P.~A. and {de Laverny}, P. and {Recio-Blanco}, A. and {Creevey}, O. and {{\'A}lvarez}, M.~A. and {Andrae}, R. and {Poggio}, E. and {Spitoni}, E. and {Contursi}, G. and {Zhao}, H. and {Oreshina-Slezak}, I. and {Ordenovic}, C. and {Bijaoui}, A.},
        title = "{Stellar ages, masses, extinctions, and orbital parameters based on spectroscopic parameters of Gaia DR3}",
      journal = {\aap},
         year = 2023,
        month = jan,
       volume = {669},
          eid = {A104},
        pages = {A104},
          doi = {10.1051/0004-6361/202244283},
archivePrefix = {arXiv},
       eprint = {2206.07937},
 primaryClass = {astro-ph.GA},
       adsurl = {https://ui.adsabs.harvard.edu/abs/2023A&A...669A.104K}
}

@ARTICLE{Gaia_collab_2023,
       author = {{Gaia Collaboration} and {Vallenari}, A. and {Brown}, A.~G.~A. and {Prusti}, T. and {de Bruijne}, J.~H.~J. and {Arenou}, F. and {Babusiaux}, C. and {Biermann}, M. and {Creevey}, O.~L. and {Ducourant}, C. and {Evans}, D.~W. and {Eyer}, L. and {Guerra}, R. and {Hutton}, A. and {Jordi}, C. and {Klioner}, S.~A. and {Lammers}, U.~L. and {Lindegren}, L. and {Luri}, X. and {Mignard}, F. and {Panem}, C. and {Pourbaix}, D. and {Randich}, S. and {Sartoretti}, P. and {Soubiran}, C. and {Tanga}, P. and {Walton}, N.~A. and {Bailer-Jones}, C.~A.~L. and {Bastian}, U. and {Drimmel}, R. and {Jansen}, F. and {Katz}, D. and {Lattanzi}, M.~G. and {van Leeuwen}, F. and {Bakker}, J. and {Cacciari}, C. and {Casta{\~n}eda}, J. and {De Angeli}, F. and {Fabricius}, C. and {Fouesneau}, M. and {Fr{\'e}mat}, Y. and {Galluccio}, L. and {Guerrier}, A. and {Heiter}, U. and {Masana}, E. and {Messineo}, R. and {Mowlavi}, N. and {Nicolas}, C. and {Nienartowicz}, K. and {Pailler}, F. and {Panuzzo}, P. and {Riclet}, F. and {Roux}, W. and {Seabroke}, G.~M. and {Sordo}, R. and {Th{\'e}venin}, F. and {Gracia-Abril}, G. and {Portell}, J. and {Teyssier}, D. and {Altmann}, M. and {Andrae}, R. and {Audard}, M. and {Bellas-Velidis}, I. and {Benson}, K. and {Berthier}, J. and {Blomme}, R. and {Burgess}, P.~W. and {Busonero}, D. and {Busso}, G. and {C{\'a}novas}, H. and {Carry}, B. and {Cellino}, A. and {Cheek}, N. and {Clementini}, G. and {Damerdji}, Y. and {Davidson}, M. and {de Teodoro}, P. and {Nu{\~n}ez Campos}, M. and {Delchambre}, L. and {Dell'Oro}, A. and {Esquej}, P. and {Fern{\'a}ndez-Hern{\'a}ndez}, J. and {Fraile}, E. and {Garabato}, D. and {Garc{\'\i}a-Lario}, P. and {Gosset}, E. and {Haigron}, R. and {Halbwachs}, J.-L. and {Hambly}, N.~C. and {Harrison}, D.~L. and {Hern{\'a}ndez}, J. and {Hestroffer}, D. and {Hodgkin}, S.~T. and {Holl}, B. and {Jan{\ss}en}, K. and {Jevardat de Fombelle}, G. and {Jordan}, S. and {Krone-Martins}, A. and {Lanzafame}, A.~C. and {L{\"o}ffler}, W. and {Marchal}, O. and {Marrese}, P.~M. and {Moitinho}, A. and {Muinonen}, K. and {Osborne}, P. and {Pancino}, E. and {Pauwels}, T. and {Recio-Blanco}, A. and {Reyl{\'e}}, C. and {Riello}, M. and {Rimoldini}, L. and {Roegiers}, T. and {Rybizki}, J. and {Sarro}, L.~M. and {Siopis}, C. and {Smith}, M. and {Sozzetti}, A. and {Utrilla}, E. and {van Leeuwen}, M. and {Abbas}, U. and {{\'A}brah{\'a}m}, P. and {Abreu Aramburu}, A. and {Aerts}, C. and {Aguado}, J.~J. and {Ajaj}, M. and {Aldea-Montero}, F. and {Altavilla}, G. and {{\'A}lvarez}, M.~A. and {Alves}, J. and {Anders}, F. and {Anderson}, R.~I. and {Anglada Varela}, E. and {Antoja}, T. and {Baines}, D. and {Baker}, S.~G. and {Balaguer-N{\'u}{\~n}ez}, L. and {Balbinot}, E. and {Balog}, Z. and {Barache}, C. and {Barbato}, D. and {Barros}, M. and {Barstow}, M.~A. and {Bartolom{\'e}}, S. and {Bassilana}, J.-L. and {Bauchet}, N. and {Becciani}, U. and {Bellazzini}, M. and {Berihuete}, A. and {Bernet}, M. and {Bertone}, S. and {Bianchi}, L. and {Binnenfeld}, A. and {Blanco-Cuaresma}, S. and {Blazere}, A. and {Boch}, T. and {Bombrun}, A. and {Bossini}, D. and {Bouquillon}, S. and {Bragaglia}, A. and {Bramante}, L. and {Breedt}, E. and {Bressan}, A. and {Brouillet}, N. and {Brugaletta}, E. and {Bucciarelli}, B. and {Burlacu}, A. and {Butkevich}, A.~G. and {Buzzi}, R. and {Caffau}, E. and {Cancelliere}, R. and {Cantat-Gaudin}, T. and {Carballo}, R. and {Carlucci}, T. and {Carnerero}, M.~I. and {Carrasco}, J.~M. and {Casamiquela}, L. and {Castellani}, M. and {Castro-Ginard}, A. and {Chaoul}, L. and {Charlot}, P. and {Chemin}, L. and {Chiaramida}, V. and {Chiavassa}, A. and {Chornay}, N. and {Comoretto}, G. and {Contursi}, G. and {Cooper}, W.~J. and {Cornez}, T. and {Cowell}, S. and {Crifo}, F. and {Cropper}, M. and {Crosta}, M. and {Crowley}, C. and {Dafonte}, C. and {Dapergolas}, A. and {David}, M. and {David}, P. and {de Laverny}, P. and {De Luise}, F. and {De March}, R.},
        title = "{Gaia Data Release 3. Summary of the content and survey properties}",
      journal = {\aap},
         year = 2023,
        month = jun,
       volume = {674},
          eid = {A1},
        pages = {A1},
          doi = {10.1051/0004-6361/202243940},
archivePrefix = {arXiv},
       eprint = {2208.00211},
 primaryClass = {astro-ph.GA},
       adsurl = {https://ui.adsabs.harvard.edu/abs/2023A&A...674A...1G}
}

@ARTICLE{Gaia_collab_2021,
       author = {{Gaia Collaboration} and {Brown}, A.~G.~A. and {Vallenari}, A. and {Prusti}, T. and {de Bruijne}, J.~H.~J. and {Babusiaux}, C. and {Biermann}, M. and {Creevey}, O.~L. and {Evans}, D.~W. and {Eyer}, L. and {Hutton}, A. and {Jansen}, F. and {Jordi}, C. and {Klioner}, S.~A. and {Lammers}, U. and {Lindegren}, L. and {Luri}, X. and {Mignard}, F. and {Panem}, C. and {Pourbaix}, D. and {Randich}, S. and {Sartoretti}, P. and {Soubiran}, C. and {Walton}, N.~A. and {Arenou}, F. and {Bailer-Jones}, C.~A.~L. and {Bastian}, U. and {Cropper}, M. and {Drimmel}, R. and {Katz}, D. and {Lattanzi}, M.~G. and {van Leeuwen}, F. and {Bakker}, J. and {Cacciari}, C. and {Casta{\~n}eda}, J. and {De Angeli}, F. and {Ducourant}, C. and {Fabricius}, C. and {Fouesneau}, M. and {Fr{\'e}mat}, Y. and {Guerra}, R. and {Guerrier}, A. and {Guiraud}, J. and {Jean-Antoine Piccolo}, A. and {Masana}, E. and {Messineo}, R. and {Mowlavi}, N. and {Nicolas}, C. and {Nienartowicz}, K. and {Pailler}, F. and {Panuzzo}, P. and {Riclet}, F. and {Roux}, W. and {Seabroke}, G.~M. and {Sordo}, R. and {Tanga}, P. and {Th{\'e}venin}, F. and {Gracia-Abril}, G. and {Portell}, J. and {Teyssier}, D. and {Altmann}, M. and {Andrae}, R. and {Bellas-Velidis}, I. and {Benson}, K. and {Berthier}, J. and {Blomme}, R. and {Brugaletta}, E. and {Burgess}, P.~W. and {Busso}, G. and {Carry}, B. and {Cellino}, A. and {Cheek}, N. and {Clementini}, G. and {Damerdji}, Y. and {Davidson}, M. and {Delchambre}, L. and {Dell'Oro}, A. and {Fern{\'a}ndez-Hern{\'a}ndez}, J. and {Galluccio}, L. and {Garc{\'\i}a-Lario}, P. and {Garcia-Reinaldos}, M. and {Gonz{\'a}lez-N{\'u}{\~n}ez}, J. and {Gosset}, E. and {Haigron}, R. and {Halbwachs}, J.-L. and {Hambly}, N.~C. and {Harrison}, D.~L. and {Hatzidimitriou}, D. and {Heiter}, U. and {Hern{\'a}ndez}, J. and {Hestroffer}, D. and {Hodgkin}, S.~T. and {Holl}, B. and {Jan{\ss}en}, K. and {Jevardat de Fombelle}, G. and {Jordan}, S. and {Krone-Martins}, A. and {Lanzafame}, A.~C. and {L{\"o}ffler}, W. and {Lorca}, A. and {Manteiga}, M. and {Marchal}, O. and {Marrese}, P.~M. and {Moitinho}, A. and {Mora}, A. and {Muinonen}, K. and {Osborne}, P. and {Pancino}, E. and {Pauwels}, T. and {Petit}, J.-M. and {Recio-Blanco}, A. and {Richards}, P.~J. and {Riello}, M. and {Rimoldini}, L. and {Robin}, A.~C. and {Roegiers}, T. and {Rybizki}, J. and {Sarro}, L.~M. and {Siopis}, C. and {Smith}, M. and {Sozzetti}, A. and {Ulla}, A. and {Utrilla}, E. and {van Leeuwen}, M. and {van Reeven}, W. and {Abbas}, U. and {Abreu Aramburu}, A. and {Accart}, S. and {Aerts}, C. and {Aguado}, J.~J. and {Ajaj}, M. and {Altavilla}, G. and {{\'A}lvarez}, M.~A. and {{\'A}lvarez Cid-Fuentes}, J. and {Alves}, J. and {Anderson}, R.~I. and {Anglada Varela}, E. and {Antoja}, T. and {Audard}, M. and {Baines}, D. and {Baker}, S.~G. and {Balaguer-N{\'u}{\~n}ez}, L. and {Balbinot}, E. and {Balog}, Z. and {Barache}, C. and {Barbato}, D. and {Barros}, M. and {Barstow}, M.~A. and {Bartolom{\'e}}, S. and {Bassilana}, J.-L. and {Bauchet}, N. and {Baudesson-Stella}, A. and {Becciani}, U. and {Bellazzini}, M. and {Bernet}, M. and {Bertone}, S. and {Bianchi}, L. and {Blanco-Cuaresma}, S. and {Boch}, T. and {Bombrun}, A. and {Bossini}, D. and {Bouquillon}, S. and {Bragaglia}, A. and {Bramante}, L. and {Breedt}, E. and {Bressan}, A. and {Brouillet}, N. and {Bucciarelli}, B. and {Burlacu}, A. and {Busonero}, D. and {Butkevich}, A.~G. and {Buzzi}, R. and {Caffau}, E. and {Cancelliere}, R. and {C{\'a}novas}, H. and {Cantat-Gaudin}, T. and {Carballo}, R. and {Carlucci}, T. and {Carnerero}, M.~I. and {Carrasco}, J.~M. and {Casamiquela}, L. and {Castellani}, M. and {Castro-Ginard}, A. and {Castro Sampol}, P. and {Chaoul}, L. and {Charlot}, P. and {Chemin}, L. and {Chiavassa}, A. and {Cioni}, M.-R.~L. and {Comoretto}, G. and {Cooper}, W.~J. and {Cornez}, T. and {Cowell}, S. and {Crifo}, F. and {Crosta}, M. and {Crowley}, C. and {Dafonte}, C. and {Dapergolas}, A. and {David}, M. and {David}, P.},
        title = "{Gaia Early Data Release 3. Summary of the contents and survey properties}",
      journal = {\aap},
         year = 2021,
        month = may,
       volume = {649},
          eid = {A1},
        pages = {A1},
          doi = {10.1051/0004-6361/202039657},
archivePrefix = {arXiv},
       eprint = {2012.01533},
 primaryClass = {astro-ph.GA},
       adsurl = {https://ui.adsabs.harvard.edu/abs/2021A&A...649A...1G}
}

@ARTICLE{Gaia_collab_2016,
       author = {{Gaia Collaboration} and {Prusti}, T. and {de Bruijne}, J.~H.~J. and {Brown}, A.~G.~A. and {Vallenari}, A. and {Babusiaux}, C. and {Bailer-Jones}, C.~A.~L. and {Bastian}, U. and {Biermann}, M. and {Evans}, D.~W. and {Eyer}, L. and {Jansen}, F. and {Jordi}, C. and {Klioner}, S.~A. and {Lammers}, U. and {Lindegren}, L. and {Luri}, X. and {Mignard}, F. and {Milligan}, D.~J. and {Panem}, C. and {Poinsignon}, V. and {Pourbaix}, D. and {Randich}, S. and {Sarri}, G. and {Sartoretti}, P. and {Siddiqui}, H.~I. and {Soubiran}, C. and {Valette}, V. and {van Leeuwen}, F. and {Walton}, N.~A. and {Aerts}, C. and {Arenou}, F. and {Cropper}, M. and {Drimmel}, R. and {H{\o}g}, E. and {Katz}, D. and {Lattanzi}, M.~G. and {O'Mullane}, W. and {Grebel}, E.~K. and {Holland}, A.~D. and {Huc}, C. and {Passot}, X. and {Bramante}, L. and {Cacciari}, C. and {Casta{\~n}eda}, J. and {Chaoul}, L. and {Cheek}, N. and {De Angeli}, F. and {Fabricius}, C. and {Guerra}, R. and {Hern{\'a}ndez}, J. and {Jean-Antoine-Piccolo}, A. and {Masana}, E. and {Messineo}, R. and {Mowlavi}, N. and {Nienartowicz}, K. and {Ord{\'o}{\~n}ez-Blanco}, D. and {Panuzzo}, P. and {Portell}, J. and {Richards}, P.~J. and {Riello}, M. and {Seabroke}, G.~M. and {Tanga}, P. and {Th{\'e}venin}, F. and {Torra}, J. and {Els}, S.~G. and {Gracia-Abril}, G. and {Comoretto}, G. and {Garcia-Reinaldos}, M. and {Lock}, T. and {Mercier}, E. and {Altmann}, M. and {Andrae}, R. and {Astraatmadja}, T.~L. and {Bellas-Velidis}, I. and {Benson}, K. and {Berthier}, J. and {Blomme}, R. and {Busso}, G. and {Carry}, B. and {Cellino}, A. and {Clementini}, G. and {Cowell}, S. and {Creevey}, O. and {Cuypers}, J. and {Davidson}, M. and {De Ridder}, J. and {de Torres}, A. and {Delchambre}, L. and {Dell'Oro}, A. and {Ducourant}, C. and {Fr{\'e}mat}, Y. and {Garc{\'\i}a-Torres}, M. and {Gosset}, E. and {Halbwachs}, J.-L. and {Hambly}, N.~C. and {Harrison}, D.~L. and {Hauser}, M. and {Hestroffer}, D. and {Hodgkin}, S.~T. and {Huckle}, H.~E. and {Hutton}, A. and {Jasniewicz}, G. and {Jordan}, S. and {Kontizas}, M. and {Korn}, A.~J. and {Lanzafame}, A.~C. and {Manteiga}, M. and {Moitinho}, A. and {Muinonen}, K. and {Osinde}, J. and {Pancino}, E. and {Pauwels}, T. and {Petit}, J.-M. and {Recio-Blanco}, A. and {Robin}, A.~C. and {Sarro}, L.~M. and {Siopis}, C. and {Smith}, M. and {Smith}, K.~W. and {Sozzetti}, A. and {Thuillot}, W. and {van Reeven}, W. and {Viala}, Y. and {Abbas}, U. and {Abreu Aramburu}, A. and {Accart}, S. and {Aguado}, J.~J. and {Allan}, P.~M. and {Allasia}, W. and {Altavilla}, G. and {{\'A}lvarez}, M.~A. and {Alves}, J. and {Anderson}, R.~I. and {Andrei}, A.~H. and {Anglada Varela}, E. and {Antiche}, E. and {Antoja}, T. and {Ant{\'o}n}, S. and {Arcay}, B. and {Atzei}, A. and {Ayache}, L. and {Bach}, N. and {Baker}, S.~G. and {Balaguer-N{\'u}{\~n}ez}, L. and {Barache}, C. and {Barata}, C. and {Barbier}, A. and {Barblan}, F. and {Baroni}, M. and {Barrado y Navascu{\'e}s}, D. and {Barros}, M. and {Barstow}, M.~A. and {Becciani}, U. and {Bellazzini}, M. and {Bellei}, G. and {Bello Garc{\'\i}a}, A. and {Belokurov}, V. and {Bendjoya}, P. and {Berihuete}, A. and {Bianchi}, L. and {Bienaym{\'e}}, O. and {Billebaud}, F. and {Blagorodnova}, N. and {Blanco-Cuaresma}, S. and {Boch}, T. and {Bombrun}, A. and {Borrachero}, R. and {Bouquillon}, S. and {Bourda}, G. and {Bouy}, H. and {Bragaglia}, A. and {Breddels}, M.~A. and {Brouillet}, N. and {Br{\"u}semeister}, T. and {Bucciarelli}, B. and {Budnik}, F. and {Burgess}, P. and {Burgon}, R. and {Burlacu}, A. and {Busonero}, D. and {Buzzi}, R. and {Caffau}, E. and {Cambras}, J. and {Campbell}, H. and {Cancelliere}, R. and {Cantat-Gaudin}, T. and {Carlucci}, T. and {Carrasco}, J.~M. and {Castellani}, M. and {Charlot}, P. and {Charnas}, J. and {Charvet}, P. and {Chassat}, F. and {Chiavassa}, A. and {Clotet}, M. and {Cocozza}, G. and {Collins}, R.~S. and {Collins}, P. and {Costigan}, G.},
        title = "{The Gaia mission}",
      journal = {\aap},
         year = 2016,
        month = nov,
       volume = {595},
          eid = {A1},
        pages = {A1},
          doi = {10.1051/0004-6361/201629272},
archivePrefix = {arXiv},
       eprint = {1609.04153},
 primaryClass = {astro-ph.IM},
       adsurl = {https://ui.adsabs.harvard.edu/abs/2016A&A...595A...1G}
}

@ARTICLE{Bailer-Jones_2013,
       author = {{Bailer-Jones}, C.~A.~L. and {Andrae}, R. and {Arcay}, B. and {Astraatmadja}, T. and {Bellas-Velidis}, I. and {Berihuete}, A. and {Bijaoui}, A. and {Carri{\'o}n}, C. and {Dafonte}, C. and {Damerdji}, Y. and {Dapergolas}, A. and {de Laverny}, P. and {Delchambre}, L. and {Drazinos}, P. and {Drimmel}, R. and {Fr{\'e}mat}, Y. and {Fustes}, D. and {Garc{\'\i}a-Torres}, M. and {Gu{\'e}d{\'e}}, C. and {Heiter}, U. and {Janotto}, A.-M. and {Karampelas}, A. and {Kim}, D.-W. and {Knude}, J. and {Kolka}, I. and {Kontizas}, E. and {Kontizas}, M. and {Korn}, A.~J. and {Lanzafame}, A.~C. and {Lebreton}, Y. and {Lindstr{\o}m}, H. and {Liu}, C. and {Livanou}, E. and {Lobel}, A. and {Manteiga}, M. and {Martayan}, C. and {Ordenovic}, Ch. and {Pichon}, B. and {Recio-Blanco}, A. and {Rocca-Volmerange}, B. and {Sarro}, L.~M. and {Smith}, K. and {Sordo}, R. and {Soubiran}, C. and {Surdej}, J. and {Th{\'e}venin}, F. and {Tsalmantza}, P. and {Vallenari}, A. and {Zorec}, J.},
        title = "{The Gaia astrophysical parameters inference system (Apsis). Pre-launch description}",
      journal = {\aap},
         year = 2013,
        month = nov,
       volume = {559},
          eid = {A74},
        pages = {A74},
          doi = {10.1051/0004-6361/201322344},
archivePrefix = {arXiv},
       eprint = {1309.2157},
 primaryClass = {astro-ph.IM},
       adsurl = {https://ui.adsabs.harvard.edu/abs/2013A&A...559A..74B}
}

@ARTICLE{Carrasco_2021,
       author = {{Carrasco}, J.~M. and {Weiler}, M. and {Jordi}, C. and {Fabricius}, C. and {De Angeli}, F. and {Evans}, D.~W. and {van Leeuwen}, F. and {Riello}, M. and {Montegriffo}, P.},
        title = "{Internal calibration of Gaia BP/RP low-resolution spectra}",
      journal = {\aap},
         year = 2021,
        month = aug,
       volume = {652},
          eid = {A86},
        pages = {A86},
          doi = {10.1051/0004-6361/202141249},
archivePrefix = {arXiv},
       eprint = {2106.01752},
 primaryClass = {astro-ph.IM},
       adsurl = {https://ui.adsabs.harvard.edu/abs/2021A&A...652A..86C}
}

@ARTICLE{DeAngeli_2023,
       author = {{De Angeli}, F. and {Weiler}, M. and {Montegriffo}, P. and {Evans}, D.~W. and {Riello}, M. and {Andrae}, R. and {Carrasco}, J.~M. and {Busso}, G. and {Burgess}, P.~W. and {Cacciari}, C. and {Davidson}, M. and {Harrison}, D.~L. and {Hodgkin}, S.~T. and {Jordi}, C. and {Osborne}, P.~J. and {Pancino}, E. and {Altavilla}, G. and {Barstow}, M.~A. and {Bailer-Jones}, C.~A.~L. and {Bellazzini}, M. and {Brown}, A.~G.~A. and {Castellani}, M. and {Cowell}, S. and {Delchambre}, L. and {De Luise}, F. and {Diener}, C. and {Fabricius}, C. and {Fouesneau}, M. and {Fr{\'e}mat}, Y. and {Gilmore}, G. and {Giuffrida}, G. and {Hambly}, N.~C. and {Hidalgo}, S. and {Holland}, G. and {Kostrzewa-Rutkowska}, Z. and {van Leeuwen}, F. and {Lobel}, A. and {Marinoni}, S. and {Miller}, N. and {Pagani}, C. and {Palaversa}, L. and {Piersimoni}, A.~M. and {Pulone}, L. and {Ragaini}, S. and {Rainer}, M. and {Richards}, P.~J. and {Rixon}, G.~T. and {Ruz-Mieres}, D. and {Sanna}, N. and {Sarro}, L.~M. and {Rowell}, N. and {Sordo}, R. and {Walton}, N.~A. and {Yoldas}, A.},
        title = "{Gaia Data Release 3. Processing and validation of BP/RP low-resolution spectral data}",
      journal = {\aap},
         year = 2023,
        month = jun,
       volume = {674},
          eid = {A2},
        pages = {A2},
          doi = {10.1051/0004-6361/202243680},
archivePrefix = {arXiv},
       eprint = {2206.06143},
 primaryClass = {astro-ph.IM},
       adsurl = {https://ui.adsabs.harvard.edu/abs/2023A&A...674A...2D}
}

@ARTICLE{Riello_2021,
       author = {{Riello}, M. and {De Angeli}, F. and {Evans}, D.~W. and {Montegriffo}, P. and {Carrasco}, J.~M. and {Busso}, G. and {Palaversa}, L. and {Burgess}, P.~W. and {Diener}, C. and {Davidson}, M. and {Rowell}, N. and {Fabricius}, C. and {Jordi}, C. and {Bellazzini}, M. and {Pancino}, E. and {Harrison}, D.~L. and {Cacciari}, C. and {van Leeuwen}, F. and {Hambly}, N.~C. and {Hodgkin}, S.~T. and {Osborne}, P.~J. and {Altavilla}, G. and {Barstow}, M.~A. and {Brown}, A.~G.~A. and {Castellani}, M. and {Cowell}, S. and {De Luise}, F. and {Gilmore}, G. and {Giuffrida}, G. and {Hidalgo}, S. and {Holland}, G. and {Marinoni}, S. and {Pagani}, C. and {Piersimoni}, A.~M. and {Pulone}, L. and {Ragaini}, S. and {Rainer}, M. and {Richards}, P.~J. and {Sanna}, N. and {Walton}, N.~A. and {Weiler}, M. and {Yoldas}, A.},
        title = "{Gaia Early Data Release 3. Photometric content and validation}",
      journal = {\aap},
         year = 2021,
        month = may,
       volume = {649},
          eid = {A3},
        pages = {A3},
          doi = {10.1051/0004-6361/202039587},
archivePrefix = {arXiv},
       eprint = {2012.01916},
 primaryClass = {astro-ph.IM},
       adsurl = {https://ui.adsabs.harvard.edu/abs/2021A&A...649A...3R}
}

@ARTICLE{Creevey_2023,
       author = {{Creevey}, O.~L. and {Sordo}, R. and {Pailler}, F. and {Fr{\'e}mat}, Y. and {Heiter}, U. and {Th{\'e}venin}, F. and {Andrae}, R. and {Fouesneau}, M. and {Lobel}, A. and {Bailer-Jones}, C.~A.~L. and {Garabato}, D. and {Bellas-Velidis}, I. and {Brugaletta}, E. and {Lorca}, A. and {Ordenovic}, C. and {Palicio}, P.~A. and {Sarro}, L.~M. and {Delchambre}, L. and {Drimmel}, R. and {Rybizki}, J. and {Torralba Elipe}, G. and {Korn}, A.~J. and {Recio-Blanco}, A. and {Schultheis}, M.~S. and {De Angeli}, F. and {Montegriffo}, P. and {Abreu Aramburu}, A. and {Accart}, S. and {{\'A}lvarez}, M.~A. and {Bakker}, J. and {Brouillet}, N. and {Burlacu}, A. and {Carballo}, R. and {Casamiquela}, L. and {Chiavassa}, A. and {Contursi}, G. and {Cooper}, W.~J. and {Dafonte}, C. and {Dapergolas}, A. and {de Laverny}, P. and {Dharmawardena}, T.~E. and {Edvardsson}, B. and {Le Fustec}, Y. and {Garc{\'\i}a-Lario}, P. and {Garc{\'\i}a-Torres}, M. and {Gomez}, A. and {Gonz{\'a}lez-Santamar{\'\i}a}, I. and {Hatzidimitriou}, D. and {Jean-Antoine Piccolo}, A. and {Kontiza}, M. and {Kordopatis}, G. and {Lanzafame}, A.~C. and {Lebreton}, Y. and {Licata}, E.~L. and {Lindstr{\o}m}, H.~E.~P. and {Livanou}, E. and {Magdaleno Romeo}, A. and {Manteiga}, M. and {Marocco}, F. and {Marshall}, D.~J. and {Mary}, N. and {Nicolas}, C. and {Pallas-Quintela}, L. and {Panem}, C. and {Pichon}, B. and {Poggio}, E. and {Riclet}, F. and {Robin}, C. and {Santove{\~n}a}, R. and {Silvelo}, A. and {Slezak}, I. and {Smart}, R.~L. and {Soubiran}, C. and {S{\"u}veges}, M. and {Ulla}, A. and {Utrilla}, E. and {Vallenari}, A. and {Zhao}, H. and {Zorec}, J. and {Barrado}, D. and {Bijaoui}, A. and {Bouret}, J.-C. and {Blomme}, R. and {Brott}, I. and {Cassisi}, S. and {Kochukhov}, O. and {Martayan}, C. and {Shulyak}, D. and {Silvester}, J.},
        title = "{Gaia Data Release 3. Astrophysical parameters inference system (Apsis). I. Methods and content overview}",
      journal = {\aap},
         year = 2023,
        month = jun,
       volume = {674},
          eid = {A26},
        pages = {A26},
          doi = {10.1051/0004-6361/202243688},
archivePrefix = {arXiv},
       eprint = {2206.05864},
 primaryClass = {astro-ph.GA},
       adsurl = {https://ui.adsabs.harvard.edu/abs/2023A&A...674A..26C}
}

@ARTICLE{Kunimoto_2021,
       author = {{Kunimoto}, Michelle and {Huang}, Chelsea and {Tey}, Evan and {Fong}, Willie and {Hesse}, Katharine and {Shporer}, Avi and {Guerrero}, Natalia and {Fausnaugh}, Michael and {Vanderspek}, Roland and {Ricker}, George},
        title = "{Quick-look Pipeline Lightcurves for 9.1 Million Stars Observed over the First Year of the TESS Extended Mission}",
      journal = {Research Notes of the American Astronomical Society},
         year = 2021,
        month = oct,
       volume = {5},
       number = {10},
          eid = {234},
        pages = {234},
          doi = {10.3847/2515-5172/ac2ef0},
archivePrefix = {arXiv},
       eprint = {2110.05542},
 primaryClass = {astro-ph.EP},
       adsurl = {https://ui.adsabs.harvard.edu/abs/2021RNAAS...5..234K}
}

@ARTICLE{Kunimoto_2022,
       author = {{Kunimoto}, Michelle and {Tey}, Evan and {Fong}, Willie and {Hesse}, Katharine and {Shporer}, Avi and {Fausnaugh}, Michael and {Vanderspek}, Roland and {Ricker}, George},
        title = "{Quick-look Pipeline Light Curves for 5.7 Million Stars Observed Over the Second Year of TESS' First Extended Mission}",
      journal = {Research Notes of the American Astronomical Society},
         year = 2022,
        month = nov,
       volume = {6},
       number = {11},
          eid = {236},
        pages = {236},
          doi = {10.3847/2515-5172/aca158},
archivePrefix = {arXiv},
       eprint = {2211.04386},
 primaryClass = {astro-ph.EP},
       adsurl = {https://ui.adsabs.harvard.edu/abs/2022RNAAS...6..236K}
}

@ARTICLE{Huang_2020a,
       author = {{Huang}, Chelsea X. and {Vanderburg}, Andrew and {P{\'a}l}, Andras and {Sha}, Lizhou and {Yu}, Liang and {Fong}, Willie and {Fausnaugh}, Michael and {Shporer}, Avi and {Guerrero}, Natalia and {Vanderspek}, Roland and {Ricker}, George},
        title = "{Photometry of 10 Million Stars from the First Two Years of TESS Full Frame Images: Part I}",
      journal = {Research Notes of the American Astronomical Society},
         year = 2020,
        month = nov,
       volume = {4},
       number = {11},
          eid = {204},
        pages = {204},
          doi = {10.3847/2515-5172/abca2e},
archivePrefix = {arXiv},
       eprint = {2011.06459},
 primaryClass = {astro-ph.EP},
       adsurl = {https://ui.adsabs.harvard.edu/abs/2020RNAAS...4..204H}
}

@ARTICLE{Huang_2020b,
       author = {{Huang}, Chelsea X. and {Vanderburg}, Andrew and {P{\'a}l}, Andras and {Sha}, Lizhou and {Yu}, Liang and {Fong}, Willie and {Fausnaugh}, Michael and {Shporer}, Avi and {Guerrero}, Natalia and {Vanderspek}, Roland and {Ricker}, George},
        title = "{Photometry of 10 Million Stars from the First Two Years of TESS Full Frame Images: Part II}",
      journal = {Research Notes of the American Astronomical Society},
         year = 2020,
        month = nov,
       volume = {4},
       number = {11},
          eid = {206},
        pages = {206},
          doi = {10.3847/2515-5172/abca2d},
       adsurl = {https://ui.adsabs.harvard.edu/abs/2020RNAAS...4..206H}
}

@ARTICLE{Berger_2020,
       author = {{Berger}, Travis A. and {Huber}, Daniel and {van Saders}, Jennifer L. and {Gaidos}, Eric and {Tayar}, Jamie and {Kraus}, Adam L.},
        title = "{The Gaia-Kepler Stellar Properties Catalog. I. Homogeneous Fundamental Properties for 186,301 Kepler Stars}",
      journal = {\aj},
         year = 2020,
        month = jun,
       volume = {159},
       number = {6},
          eid = {280},
        pages = {280},
          doi = {10.3847/1538-3881/159/6/280},
archivePrefix = {arXiv},
       eprint = {2001.07737},
 primaryClass = {astro-ph.SR},
       adsurl = {https://ui.adsabs.harvard.edu/abs/2020AJ....159..280B}
}

@ARTICLE{Recio_Blanco_2016,
       author = {{Recio-Blanco}, A. and {de Laverny}, P. and {Allende Prieto}, C. and {Fustes}, D. and {Manteiga}, M. and {Arcay}, B. and {Bijaoui}, A. and {Dafonte}, C. and {Ordenovic}, C. and {Ordo{\~n}ez Blanco}, D.},
        title = "{Stellar parametrization from Gaia RVS spectra}",
      journal = {\aap},
         year = 2016,
        month = jan,
       volume = {585},
          eid = {A93},
        pages = {A93},
          doi = {10.1051/0004-6361/201425030},
archivePrefix = {arXiv},
       eprint = {1510.00111},
 primaryClass = {astro-ph.GA},
       adsurl = {https://ui.adsabs.harvard.edu/abs/2016A&A...585A..93R}
}

@ARTICLE{Dotter_2017,
       author = {{Dotter}, Aaron and {Conroy}, Charlie and {Cargile}, Phillip and {Asplund}, Martin},
        title = "{The Influence of Atomic Diffusion on Stellar Ages and Chemical Tagging}",
      journal = {\apj},
         year = 2017,
        month = may,
       volume = {840},
       number = {2},
          eid = {99},
        pages = {99},
          doi = {10.3847/1538-4357/aa6d10},
archivePrefix = {arXiv},
       eprint = {1704.03465},
 primaryClass = {astro-ph.SR},
       adsurl = {https://ui.adsabs.harvard.edu/abs/2017ApJ...840...99D}
}

@ARTICLE{Torres_2010,
       author = {{Torres}, G. and {Andersen}, J. and {Gim{\'e}nez}, A.},
        title = "{Accurate masses and radii of normal stars: modern results and applications}",
      journal = {\aapr},
         year = 2010,
        month = feb,
       volume = {18},
       number = {1-2},
        pages = {67-126},
          doi = {10.1007/s00159-009-0025-1},
archivePrefix = {arXiv},
       eprint = {0908.2624},
 primaryClass = {astro-ph.SR},
       adsurl = {https://ui.adsabs.harvard.edu/abs/2010A&ARv..18...67T}
}

@ARTICLE{Petigura_2017,
       author = {{Petigura}, Erik A. and {Howard}, Andrew W. and {Marcy}, Geoffrey W. and {Johnson}, John Asher and {Isaacson}, Howard and {Cargile}, Phillip A. and {Hebb}, Leslie and {Fulton}, Benjamin J. and {Weiss}, Lauren M. and {Morton}, Timothy D. and {Winn}, Joshua N. and {Rogers}, Leslie A. and {Sinukoff}, Evan and {Hirsch}, Lea A. and {Crossfield}, Ian J.~M.},
        title = "{The California-Kepler Survey. I. High-resolution Spectroscopy of 1305 Stars Hosting Kepler Transiting Planets}",
      journal = {\aj},
         year = 2017,
        month = sep,
       volume = {154},
       number = {3},
          eid = {107},
        pages = {107},
          doi = {10.3847/1538-3881/aa80de},
archivePrefix = {arXiv},
       eprint = {1703.10400},
 primaryClass = {astro-ph.EP},
       adsurl = {https://ui.adsabs.harvard.edu/abs/2017AJ....154..107P}
}

@ARTICLE{Fitton_2022,
       author = {{Fitton}, Shannon and {Tofflemire}, Benjamin M. and {Kraus}, Adam L.},
        title = "{Disk Material Inflates Gaia RUWE Values in Single Stars}",
      journal = {Research Notes of the American Astronomical Society},
         year = 2022,
        month = jan,
       volume = {6},
       number = {1},
          eid = {18},
        pages = {18},
          doi = {10.3847/2515-5172/ac4bb7},
archivePrefix = {arXiv},
       eprint = {2206.02695},
 primaryClass = {astro-ph.SR},
       adsurl = {https://ui.adsabs.harvard.edu/abs/2022RNAAS...6...18F}
}

@ARTICLE{Belokurov_2020,
       author = {{Belokurov}, Vasily and {Penoyre}, Zephyr and {Oh}, Semyeong and {Iorio}, Giuliano and {Hodgkin}, Simon and {Evans}, N. Wyn and {Everall}, Andrew and {Koposov}, Sergey E. and {Tout}, Christopher A. and {Izzard}, Robert and {Clarke}, Cathie J. and {Brown}, Anthony G.~A.},
        title = "{Unresolved stellar companions with Gaia DR2 astrometry}",
      journal = {\mnras},
         year = 2020,
        month = aug,
       volume = {496},
       number = {2},
        pages = {1922-1940},
          doi = {10.1093/mnras/staa1522},
archivePrefix = {arXiv},
       eprint = {2003.05467},
 primaryClass = {astro-ph.SR},
       adsurl = {https://ui.adsabs.harvard.edu/abs/2020MNRAS.496.1922B}
}

@ARTICLE{Jorgensen_2005,
       author = {{J{\o}rgensen}, B.~R. and {Lindegren}, L.},
        title = "{Determination of stellar ages from isochrones: Bayesian estimation versus isochrone fitting}",
      journal = {\aap},
         year = 2005,
        month = jun,
       volume = {436},
       number = {1},
        pages = {127-143},
          doi = {10.1051/0004-6361:20042185},
       adsurl = {https://ui.adsabs.harvard.edu/abs/2005A&A...436..127J}
}

@ARTICLE{Casamiquela_2024,
       author = {{Casamiquela}, L. and {Reese}, D.~R. and {Lebreton}, Y. and {Haywood}, M. and {Di Matteo}, P. and {Anders}, F. and {Jash}, R. and {Katz}, D. and {Cerqui}, V. and {Boin}, T. and {Kordopatis}, G.},
        title = "{New stellar age estimates using SPInS based on Gaia DR3 photometry and LAMOST DR8 abundances}",
      journal = {\aap},
         year = 2024,
        month = dec,
       volume = {692},
          eid = {A243},
        pages = {A243},
          doi = {10.1051/0004-6361/202451677},
archivePrefix = {arXiv},
       eprint = {2410.15781},
 primaryClass = {astro-ph.GA},
       adsurl = {https://ui.adsabs.harvard.edu/abs/2024A&A...692A.243C}
}

@ARTICLE{Marshall_2007,
       author = {{Marshall}, J.~L.},
        title = "{Mapping the Local Galactic Halo. I. Optical Photometry of Cool Subdwarf Candidates}",
      journal = {\aj},
         year = 2007,
        month = aug,
       volume = {134},
       number = {2},
        pages = {778-786},
          doi = {10.1086/519491},
       adsurl = {https://ui.adsabs.harvard.edu/abs/2007AJ....134..778M}
}

@ARTICLE{Ziegler_2015,
       author = {{Ziegler}, Carl and {Law}, Nicholas M. and {Baranec}, Christoph and {Riddle}, Reed L. and {Fuchs}, Joshua T.},
        title = "{Multiplicity of the Galactic Senior Citizens: A High-resolution Search for Cool Subdwarf Companions}",
      journal = {\apj},
         year = 2015,
        month = may,
       volume = {804},
       number = {1},
          eid = {30},
        pages = {30},
          doi = {10.1088/0004-637X/804/1/30},
archivePrefix = {arXiv},
       eprint = {1411.3336},
 primaryClass = {astro-ph.SR},
       adsurl = {https://ui.adsabs.harvard.edu/abs/2015ApJ...804...30Z}
}

@ARTICLE{Mulders_2025,
       author = {{Parashivamurthy}, Harshitha M. and {Mulders}, Gijs D.},
        title = "{Radius valley scaling among low-mass stars with TESS}",
      journal = {\aap},
         year = 2025,
        month = oct,
       volume = {703},
          eid = {A8},
        pages = {A8},
          doi = {10.1051/0004-6361/202554006},
archivePrefix = {arXiv},
       eprint = {2507.07181},
 primaryClass = {astro-ph.EP},
       adsurl = {https://ui.adsabs.harvard.edu/abs/2025A&A...703A...8P}
}

@ARTICLE{Yee_2023,
       author = {{Yee}, Samuel W. and {Winn}, Joshua N.},
        title = "{The Period Distribution of Hot Jupiters Is Not Dependent on Host Star Metallicity}",
      journal = {\apjl},
         year = 2023,
        month = jun,
       volume = {949},
       number = {2},
          eid = {L21},
        pages = {L21},
          doi = {10.3847/2041-8213/acd552},
archivePrefix = {arXiv},
       eprint = {2305.09488},
 primaryClass = {astro-ph.EP},
       adsurl = {https://ui.adsabs.harvard.edu/abs/2023ApJ...949L..21Y}
}

@ARTICLE{Beleznay_Kunimoto_2022,
       author = {{Beleznay}, Maya and {Kunimoto}, M.},
        title = "{Exploring the dependence of hot Jupiter occurrence rates on stellar mass with TESS}",
      journal = {\mnras},
         year = 2022,
        month = oct,
       volume = {516},
       number = {1},
        pages = {75-83},
          doi = {10.1093/mnras/stac2179},
archivePrefix = {arXiv},
       eprint = {2207.12522},
 primaryClass = {astro-ph.EP},
       adsurl = {https://ui.adsabs.harvard.edu/abs/2022MNRAS.516...75B}
}

@ARTICLE{PAST_2023,
       author = {{Yang}, Jia-Yi and {Chen}, Di-Chang and {Xie}, Ji-Wei and {Zhou}, Ji-Lin and {Dong}, Subo and {Zhu}, Zi and {Zheng}, Zheng and {Liu}, Chao and {Zong}, Weikai and {Luo}, Ali},
        title = "{Planets Across Space and Time (PAST). IV. The Occurrence and Architecture of Kepler Planetary Systems as a Function of Kinematic Age Revealed by the LAMOST-Gaia-Kepler Sample}",
      journal = {\aj},
         year = 2023,
        month = dec,
       volume = {166},
       number = {6},
          eid = {243},
        pages = {243},
          doi = {10.3847/1538-3881/ad0368},
archivePrefix = {arXiv},
       eprint = {2310.20113},
 primaryClass = {astro-ph.EP},
       adsurl = {https://ui.adsabs.harvard.edu/abs/2023AJ....166..243Y}
}

@ARTICLE{Sayeed_2025,
       author = {{Sayeed}, Maryum and {Angus}, Ruth and {Berger}, Travis A. and {Lu}, Yuxi(Lucy) and {Christiansen}, Jessie L. and {Foreman-Mackey}, Daniel and {Ness}, Melissa K.},
        title = "{Exoplanet Occurrence Rate with Age for FGK Stars in Kepler}",
      journal = {\aj},
         year = 2025,
        month = feb,
       volume = {169},
       number = {2},
          eid = {112},
        pages = {112},
          doi = {10.3847/1538-3881/ada8a1},
archivePrefix = {arXiv},
       eprint = {2501.13809},
 primaryClass = {astro-ph.EP},
       adsurl = {https://ui.adsabs.harvard.edu/abs/2025AJ....169..112S}
}

@ARTICLE{Magrini_2022,
       author = {{Magrini}, L. and {Danielski}, C. and {Bossini}, D. and {Rainer}, M. and {Turrini}, D. and {Benatti}, S. and {Brucalassi}, A. and {Tsantaki}, M. and {Delgado Mena}, E. and {Sanna}, N. and {Biazzo}, K. and {Campante}, T.~L. and {Van der Swaelmen}, M. and {Sousa}, S.~G. and {He{\l}miniak}, K.~G. and {Neitzel}, A.~W. and {Adibekyan}, V. and {Bruno}, G. and {Casali}, G.},
        title = "{Ariel stellar characterisation. I. Homogeneous stellar parameters of 187 FGK planet host stars: Description and validation of the method}",
      journal = {\aap},
         year = 2022,
        month = jul,
       volume = {663},
          eid = {A161},
        pages = {A161},
          doi = {10.1051/0004-6361/202243405},
archivePrefix = {arXiv},
       eprint = {2204.08825},
 primaryClass = {astro-ph.SR},
       adsurl = {https://ui.adsabs.harvard.edu/abs/2022A&A...663A.161M}
}

@ARTICLE{Bonomo_2023,
       author = {{Bonomo}, A.~S. and {Dumusque}, X. and {Massa}, A. and {Mortier}, A. and {Bongiolatti}, R. and {Malavolta}, L. and {Sozzetti}, A. and {Buchhave}, L.~A. and {Damasso}, M. and {Haywood}, R.~D. and {Morbidelli}, A. and {Latham}, D.~W. and {Molinari}, E. and {Pepe}, F. and {Poretti}, E. and {Udry}, S. and {Affer}, L. and {Boschin}, W. and {Charbonneau}, D. and {Cosentino}, R. and {Cretignier}, M. and {Ghedina}, A. and {Lega}, E. and {L{\'o}pez-Morales}, M. and {Margini}, M. and {Mart{\'\i}nez Fiorenzano}, A.~F. and {Mayor}, M. and {Micela}, G. and {Pedani}, M. and {Pinamonti}, M. and {Rice}, K. and {Sasselov}, D. and {Tronsgaard}, R. and {Vanderburg}, A.},
        title = "{Cold Jupiters and improved masses in 38 Kepler and K2 small planet systems from 3661 HARPS-N radial velocities. No excess of cold Jupiters in small planet systems}",
      journal = {\aap},
         year = 2023,
        month = sep,
       volume = {677},
          eid = {A33},
        pages = {A33},
          doi = {10.1051/0004-6361/202346211},
archivePrefix = {arXiv},
       eprint = {2304.05773},
 primaryClass = {astro-ph.EP},
       adsurl = {https://ui.adsabs.harvard.edu/abs/2023A&A...677A..33B}
}

@ARTICLE{Penev_2018,
       author = {{Penev}, Kaloyan and {Bouma}, L.~G. and {Winn}, Joshua N. and {Hartman}, Joel D.},
        title = "{Empirical Tidal Dissipation in Exoplanet Hosts From Tidal Spin-up}",
      journal = {\aj},
         year = 2018,
        month = apr,
       volume = {155},
       number = {4},
          eid = {165},
        pages = {165},
          doi = {10.3847/1538-3881/aaaf71},
archivePrefix = {arXiv},
       eprint = {1802.05269},
 primaryClass = {astro-ph.SR},
       adsurl = {https://ui.adsabs.harvard.edu/abs/2018AJ....155..165P}
}

@ARTICLE{Owen_2013,
       author = {{Owen}, James E. and {Wu}, Yanqin},
        title = "{Kepler Planets: A Tale of Evaporation}",
      journal = {\apj},
         year = 2013,
        month = oct,
       volume = {775},
       number = {2},
          eid = {105},
        pages = {105},
          doi = {10.1088/0004-637X/775/2/105},
archivePrefix = {arXiv},
       eprint = {1303.3899},
 primaryClass = {astro-ph.EP},
       adsurl = {https://ui.adsabs.harvard.edu/abs/2013ApJ...775..105O}
}

@ARTICLE{Owen_2017,
       author = {{Owen}, James E. and {Wu}, Yanqin},
        title = "{The Evaporation Valley in the Kepler Planets}",
      journal = {\apj},
         year = 2017,
        month = sep,
       volume = {847},
       number = {1},
          eid = {29},
        pages = {29},
          doi = {10.3847/1538-4357/aa890a},
archivePrefix = {arXiv},
       eprint = {1705.10810},
 primaryClass = {astro-ph.EP},
       adsurl = {https://ui.adsabs.harvard.edu/abs/2017ApJ...847...29O}
}

@ARTICLE{Hamer_Schlaufman_2019,
       author = {{Hamer}, Jacob H. and {Schlaufman}, Kevin C.},
        title = "{Hot Jupiters Are Destroyed by Tides While Their Host Stars Are on the Main Sequence}",
      journal = {\aj},
         year = 2019,
        month = nov,
       volume = {158},
       number = {5},
          eid = {190},
        pages = {190},
          doi = {10.3847/1538-3881/ab3c56},
archivePrefix = {arXiv},
       eprint = {1908.06998},
 primaryClass = {astro-ph.EP},
       adsurl = {https://ui.adsabs.harvard.edu/abs/2019AJ....158..190H}
}

@ARTICLE{Skumanich_1972,
       author = {{Skumanich}, A.},
        title = "{Time Scales for Ca II Emission Decay, Rotational Braking, and Lithium Depletion}",
      journal = {\apj},
         year = 1972,
        month = feb,
       volume = {171},
        pages = {565},
          doi = {10.1086/151310},
       adsurl = {https://ui.adsabs.harvard.edu/abs/1972ApJ...171..565S}
}

@ARTICLE{Barnes_2003,
       author = {{Barnes}, Sydney A.},
        title = "{On the Rotational Evolution of Solar- and Late-Type Stars, Its Magnetic Origins, and the Possibility of Stellar Gyrochronology}",
      journal = {\apj},
         year = 2003,
        month = mar,
       volume = {586},
       number = {1},
        pages = {464-479},
          doi = {10.1086/367639},
archivePrefix = {arXiv},
       eprint = {astro-ph/0303631},
 primaryClass = {astro-ph},
       adsurl = {https://ui.adsabs.harvard.edu/abs/2003ApJ...586..464B}
}

@ARTICLE{Barnes_2007,
       author = {{Barnes}, Sydney A.},
        title = "{Ages for Illustrative Field Stars Using Gyrochronology: Viability, Limitations, and Errors}",
      journal = {\apj},
         year = 2007,
        month = nov,
       volume = {669},
       number = {2},
        pages = {1167-1189},
          doi = {10.1086/519295},
archivePrefix = {arXiv},
       eprint = {0704.3068},
 primaryClass = {astro-ph},
       adsurl = {https://ui.adsabs.harvard.edu/abs/2007ApJ...669.1167B}
}

@ARTICLE{Mamajek_2008,
       author = {{Mamajek}, Eric E. and {Hillenbrand}, Lynne A.},
        title = "{Improved Age Estimation for Solar-Type Dwarfs Using Activity-Rotation Diagnostics}",
      journal = {\apj},
         year = 2008,
        month = nov,
       volume = {687},
       number = {2},
        pages = {1264-1293},
          doi = {10.1086/591785},
archivePrefix = {arXiv},
       eprint = {0807.1686},
 primaryClass = {astro-ph},
       adsurl = {https://ui.adsabs.harvard.edu/abs/2008ApJ...687.1264M}
}

@ARTICLE{Spitzer_1951,
       author = {{Spitzer}, Jr., Lyman and {Schwarzschild}, Martin},
        title = "{The Possible Influence of Interstellar Clouds on Stellar Velocities.}",
      journal = {\apj},
         year = 1951,
        month = nov,
       volume = {114},
        pages = {385},
          doi = {10.1086/145478},
       adsurl = {https://ui.adsabs.harvard.edu/abs/1951ApJ...114..385S}
}

@ARTICLE{Sellwood_Carlberg_1984,
       author = {{Sellwood}, J.~A. and {Carlberg}, R.~G.},
        title = "{Spiral instabilities provoked by accretion and star formation}",
      journal = {\apj},
         year = 1984,
        month = jul,
       volume = {282},
        pages = {61-74},
          doi = {10.1086/162176},
       adsurl = {https://ui.adsabs.harvard.edu/abs/1984ApJ...282...61S}
}

@ARTICLE{Deason_Belukorov_2024,
       author = {{Deason}, Alis J. and {Belokurov}, Vasily},
        title = "{Galactic Archaeology with Gaia}",
      journal = {\nar},
         year = 2024,
        month = dec,
       volume = {99},
          eid = {101706},
        pages = {101706},
          doi = {10.1016/j.newar.2024.101706},
archivePrefix = {arXiv},
       eprint = {2402.12443},
 primaryClass = {astro-ph.GA},
       adsurl = {https://ui.adsabs.harvard.edu/abs/2024NewAR..9901706D}
}

@ARTICLE{Frebel2010,
       author = {{Frebel}, A.},
        title = "{Stellar archaeology: Exploring the Universe with metal-poor stars}",
      journal = {Astronomische Nachrichten},
         year = 2010,
        month = may,
       volume = {331},
       number = {5},
        pages = {474-488},
          doi = {10.1002/asna.201011362},
archivePrefix = {arXiv},
       eprint = {1006.2419},
 primaryClass = {astro-ph.GA},
       adsurl = {https://ui.adsabs.harvard.edu/abs/2010AN....331..474F}
}

@INPROCEEDINGS{Kawate_2026,
       author = {{Kawata}, Daisuke and {Grand}, Robert J.~J. and {Hunt}, Jason A.~S. and {Ciuc{\u{a}}}, Ioana},
        title = "{Milky Way disk}",
    booktitle = {Encyclopedia of Astrophysics, Volume 4},
         year = 2026,
       volume = {4},
        month = jan,
        pages = {38-60},
          doi = {10.1016/B978-0-443-21439-4.00064-X},
archivePrefix = {arXiv},
       eprint = {2412.12252},
 primaryClass = {astro-ph.GA},
       adsurl = {https://ui.adsabs.harvard.edu/abs/2026enap....4...38K}
}

@INPROCEEDINGS{Lebreton_2009,
       author = {{Lebreton}, Yveline and {Montalb{\'a}n}, Josefina},
        title = "{Stellar ages from asteroseismology}",
    booktitle = {The Ages of Stars},
         year = 2009,
       editor = {{Mamajek}, Eric E. and {Soderblom}, David R. and {Wyse}, Rosemary F.~G.},
       series = {IAU Symposium},
       volume = {258},
        month = jun,
        pages = {419-430},
          doi = {10.1017/S1743921309032074},
archivePrefix = {arXiv},
       eprint = {0811.2908},
 primaryClass = {astro-ph},
       adsurl = {https://ui.adsabs.harvard.edu/abs/2009IAUS..258..419L}
}

@ARTICLE{Soderblom_2010,
       author = {{Soderblom}, David R.},
        title = "{The Ages of Stars}",
      journal = {\araa},
         year = 2010,
        month = sep,
       volume = {48},
        pages = {581-629},
          doi = {10.1146/annurev-astro-081309-130806},
archivePrefix = {arXiv},
       eprint = {1003.6074},
 primaryClass = {astro-ph.SR},
       adsurl = {https://ui.adsabs.harvard.edu/abs/2010ARA&A..48..581S}
}

@ARTICLE{Li_2020,
       author = {{Li}, Tanda and {Bedding}, Timothy R. and {Christensen-Dalsgaard}, J{\o}rgen and {Stello}, Dennis and {Li}, Yaguang and {Keen}, Matthew A.},
        title = "{Asteroseismology of 36 Kepler subgiants - II. Determining ages from detailed modelling}",
      journal = {\mnras},
         year = 2020,
        month = jul,
       volume = {495},
       number = {3},
        pages = {3431-3462},
          doi = {10.1093/mnras/staa1350},
archivePrefix = {arXiv},
       eprint = {2006.00901},
 primaryClass = {astro-ph.SR},
       adsurl = {https://ui.adsabs.harvard.edu/abs/2020MNRAS.495.3431L}
}

@ARTICLE{Angus_2019,
       author = {{Angus}, Ruth and {Morton}, Timothy D. and {Foreman-Mackey}, Daniel and {van Saders}, Jennifer and {Curtis}, Jason and {Kane}, Stephen R. and {Bedell}, Megan and {Kiman}, Rocio and {Hogg}, David W. and {Brewer}, John},
        title = "{Toward Precise Stellar Ages: Combining Isochrone Fitting with Empirical Gyrochronology}",
      journal = {\aj},
         year = 2019,
        month = nov,
       volume = {158},
       number = {5},
          eid = {173},
        pages = {173},
          doi = {10.3847/1538-3881/ab3c53},
archivePrefix = {arXiv},
       eprint = {1908.07528},
 primaryClass = {astro-ph.SR},
       adsurl = {https://ui.adsabs.harvard.edu/abs/2019AJ....158..173A}
}

@ARTICLE{Pont_2004,
       author = {{Pont}, Fr{\'e}d{\'e}ric and {Eyer}, Laurent},
        title = "{Isochrone ages for field dwarfs: method and application to the age-metallicity relation}",
      journal = {\mnras},
         year = 2004,
        month = jun,
       volume = {351},
       number = {2},
        pages = {487-504},
          doi = {10.1111/j.1365-2966.2004.07780.x},
archivePrefix = {arXiv},
       eprint = {astro-ph/0401418},
 primaryClass = {astro-ph},
       adsurl = {https://ui.adsabs.harvard.edu/abs/2004MNRAS.351..487P}
}

@ARTICLE{Bryant_2025,
       author = {{Bryant}, Edward M. and {Van Eylen}, Vincent},
        title = "{Determining the impact of post-main-sequence stellar evolution on the transiting giant planet population}",
      journal = {\mnras},
         year = 2025,
        month = nov,
       volume = {544},
       number = {1},
        pages = {1186-1214},
          doi = {10.1093/mnras/staf1771},
archivePrefix = {arXiv},
       eprint = {2511.02896},
 primaryClass = {astro-ph.EP},
       adsurl = {https://ui.adsabs.harvard.edu/abs/2025MNRAS.544.1186B}
}

@ARTICLE{Osborne_2024,
       author = {{Osborne}, H.~L.~M. and {Van Eylen}, V. and {Goffo}, E. and {Gandolfi}, D. and {Nowak}, G. and {Persson}, C.~M. and {Livingston}, J. and {Weeks}, A. and {Pall{\'e}}, E. and {Luque}, R. and {Hellier}, C. and {Carleo}, I. and {Redfield}, S. and {Hirano}, T. and {Garbaccio Gili}, M. and {Alarcon}, J. and {Barrag{\'a}n}, O. and {Casasayas-Barris}, N. and {D{\'\i}az}, M.~R. and {Esposito}, M. and {Knudstrup}, E. and {Jenkins}, J.~S. and {Murgas}, F. and {Orell-Miquel}, J. and {Rodler}, F. and {Serrano}, L. and {Stangret}, M. and {Albrecht}, S.~H. and {Alqasim}, A. and {Cochran}, W.~D. and {Deeg}, H.~J. and {Fridlund}, M. and {Hatzes}, A.~P. and {Korth}, J. and {Lam}, K.~W.~F.},
        title = "{TOI-544 b: a potential water-world inside the radius valley in a two-planet system}",
      journal = {\mnras},
         year = 2024,
        month = feb,
       volume = {527},
       number = {4},
        pages = {11138-11157},
          doi = {10.1093/mnras/stad3837},
archivePrefix = {arXiv},
       eprint = {2310.14908},
 primaryClass = {astro-ph.EP},
       adsurl = {https://ui.adsabs.harvard.edu/abs/2024MNRAS.52711138O}
}

@ARTICLE{Persson_2022,
       author = {{Persson}, Carina M. and {Georgieva}, Iskra Y. and {Gandolfi}, Davide and {Acuna}, Lorena and {Aguichine}, Artem and {Muresan}, Alexandra and {Guenther}, Eike and {Livingston}, John and {Collins}, Karen A. and {Dai}, Fei and {Fridlund}, Malcolm and {Goffo}, Elisa and {Jenkins}, James S. and {Kab{\'a}th}, Petr and {Korth}, Judith and {Levine}, Alan M. and {Serrano}, Luisa M. and {Vines}, Jos{\'e} and {Barragan}, Oscar and {Carleo}, Ilaria and {Colon}, Knicole D. and {Cochran}, William D. and {Christiansen}, Jessie L. and {Deeg}, Hans J. and {Deleuil}, Magali and {Dragomir}, Diana and {Esposito}, Massimiliamo and {Gan}, Tianjun and {Grziwa}, Sascha and {Hatzes}, Artie P. and {Hesse}, Katharine and {Horne}, Keith and {Jenkins}, Jon M. and {Kielkopf}, John F. and {Klagyivik}, P. and {Lam}, Kristine W.~F. and {Latham}, David W. and {Luque}, Rafa and {Orell-Miquel}, Jaume and {Mortier}, Annelies and {Mousis}, Olivier and {Narita}, Noria and {Osborne}, Hannah L.~M. and {Palle}, Enric and {Papini}, Riccardo and {Ricker}, George R. and {Schmerling}, Hendrik and {Seager}, Sara and {Stassun}, Keivan G. and {Van Eylen}, Vincent and {Vanderspek}, Roland and {Wang}, Gavin and {Winn}, Joshua N. and {Wohler}, Bill and {Zambelli}, Roberto and {Ziegler}, Carl},
        title = "{TOI-2196 b: Rare planet in the hot Neptune desert transiting a G-type star}",
      journal = {\aap},
         year = 2022,
        month = oct,
       volume = {666},
          eid = {A184},
        pages = {A184},
          doi = {10.1051/0004-6361/202244118},
archivePrefix = {arXiv},
       eprint = {2208.05797},
 primaryClass = {astro-ph.EP},
       adsurl = {https://ui.adsabs.harvard.edu/abs/2022A&A...666A.184P}
}

@ARTICLE{Gonzalez_1997,
       author = {{Gonzalez}, Guillermo},
        title = "{The stellar metallicity-giant planet connection}",
      journal = {\mnras},
         year = 1997,
        month = feb,
       volume = {285},
       number = {2},
        pages = {403-412},
          doi = {10.1093/mnras/285.2.403},
       adsurl = {https://ui.adsabs.harvard.edu/abs/1997MNRAS.285..403G}
}

@ARTICLE{Fischer_2005,
       author = {{Fischer}, Debra A. and {Valenti}, Jeff},
        title = "{The Planet-Metallicity Correlation}",
      journal = {\apj},
         year = 2005,
        month = apr,
       volume = {622},
       number = {2},
        pages = {1102-1117},
          doi = {10.1086/428383},
       adsurl = {https://ui.adsabs.harvard.edu/abs/2005ApJ...622.1102F}
}

@ARTICLE{Muirhead_2018,
       author = {{Muirhead}, Philip S. and {Dressing}, Courtney D. and {Mann}, Andrew W. and {Rojas-Ayala}, B{\'a}rbara and {L{\'e}pine}, S{\'e}bastien and {Paegert}, Martin and {De Lee}, Nathan and {Oelkers}, Ryan},
        title = "{A Catalog of Cool Dwarf Targets for the Transiting Exoplanet Survey Satellite}",
      journal = {\aj},
         year = 2018,
        month = apr,
       volume = {155},
       number = {4},
          eid = {180},
        pages = {180},
          doi = {10.3847/1538-3881/aab710},
archivePrefix = {arXiv},
       eprint = {1710.00193},
 primaryClass = {astro-ph.SR},
       adsurl = {https://ui.adsabs.harvard.edu/abs/2018AJ....155..180M}
}

@ARTICLE{Cropper_2018,
       author = {{Cropper}, M. and {Katz}, D. and {Sartoretti}, P. and {Prusti}, T. and {de Bruijne}, J.~H.~J. and {Chassat}, F. and {Charvet}, P. and {Boyadjian}, J. and {Perryman}, M. and {Sarri}, G. and {Gare}, P. and {Erdmann}, M. and {Munari}, U. and {Zwitter}, T. and {Wilkinson}, M. and {Arenou}, F. and {Vallenari}, A. and {G{\'o}mez}, A. and {Panuzzo}, P. and {Seabroke}, G. and {Allende Prieto}, C. and {Benson}, K. and {Marchal}, O. and {Huckle}, H. and {Smith}, M. and {Dolding}, C. and {Jan{\ss}en}, K. and {Viala}, Y. and {Blomme}, R. and {Baker}, S. and {Boudreault}, S. and {Crifo}, F. and {Soubiran}, C. and {Fr{\'e}mat}, Y. and {Jasniewicz}, G. and {Guerrier}, A. and {Guy}, L.~P. and {Turon}, C. and {Jean-Antoine-Piccolo}, A. and {Th{\'e}venin}, F. and {David}, M. and {Gosset}, E. and {Damerdji}, Y.},
        title = "{Gaia Data Release 2. Gaia Radial Velocity Spectrometer}",
      journal = {\aap},
         year = 2018,
        month = aug,
       volume = {616},
          eid = {A5},
        pages = {A5},
          doi = {10.1051/0004-6361/201832763},
archivePrefix = {arXiv},
       eprint = {1804.09369},
 primaryClass = {astro-ph.IM},
       adsurl = {https://ui.adsabs.harvard.edu/abs/2018A&A...616A...5C}
}

@ARTICLE{Miglio_2013,
       author = {{Miglio}, A. and {Chiappini}, C. and {Morel}, T. and {Barbieri}, M. and {Chaplin}, W.~J. and {Girardi}, L. and {Montalb{\'a}n}, J. and {Valentini}, M. and {Mosser}, B. and {Baudin}, F. and {Casagrande}, L. and {Fossati}, L. and {Silva Aguirre}, V. and {Baglin}, A.},
        title = "{Galactic archaeology: mapping and dating stellar populations with asteroseismology of red-giant stars}",
      journal = {\mnras},
         year = 2013,
        month = feb,
       volume = {429},
       number = {1},
        pages = {423-428},
          doi = {10.1093/mnras/sts345},
archivePrefix = {arXiv},
       eprint = {1211.0146},
 primaryClass = {astro-ph.GA},
       adsurl = {https://ui.adsabs.harvard.edu/abs/2013MNRAS.429..423M}
}

@ARTICLE{Jofre_2021,
       author = {{Jofr{\'e}}, Paula},
        title = "{Two Sequences in the Age-Metallicity Relation as Seen from [C/N] Abundances in APOGEE}",
      journal = {\apj},
         year = 2021,
        month = oct,
       volume = {920},
       number = {1},
          eid = {23},
        pages = {23},
          doi = {10.3847/1538-4357/ac10c1},
archivePrefix = {arXiv},
       eprint = {2106.16119},
 primaryClass = {astro-ph.GA},
       adsurl = {https://ui.adsabs.harvard.edu/abs/2021ApJ...920...23J}
}

@ARTICLE{Nissen_2020,
       author = {{Nissen}, P.~E. and {Christensen-Dalsgaard}, J. and {Mosumgaard}, J.~R. and {Silva Aguirre}, V. and {Spitoni}, E. and {Verma}, K.},
        title = "{High-precision abundances of elements in solar-type stars. Evidence of two distinct sequences in abundance-age relations}",
      journal = {\aap},
         year = 2020,
        month = aug,
       volume = {640},
          eid = {A81},
        pages = {A81},
          doi = {10.1051/0004-6361/202038300},
archivePrefix = {arXiv},
       eprint = {2006.06013},
 primaryClass = {astro-ph.SR},
       adsurl = {https://ui.adsabs.harvard.edu/abs/2020A&A...640A..81N}
}

@ARTICLE{Anders_2023,
       author = {{Anders}, F. and {Gispert}, P. and {Ratcliffe}, B. and {Chiappini}, C. and {Minchev}, I. and {Nepal}, S. and {Queiroz}, A.~B.~A. and {Amarante}, J.~A.~S. and {Antoja}, T. and {Casali}, G. and {Casamiquela}, L. and {Khalatyan}, A. and {Miglio}, A. and {Perottoni}, H. and {Schultheis}, M.},
        title = "{Spectroscopic age estimates for APOGEE red-giant stars: Precise spatial and kinematic trends with age in the Galactic disc}",
      journal = {\aap},
         year = 2023,
        month = oct,
       volume = {678},
          eid = {A158},
        pages = {A158},
          doi = {10.1051/0004-6361/202346666},
archivePrefix = {arXiv},
       eprint = {2304.08276},
 primaryClass = {astro-ph.GA},
       adsurl = {https://ui.adsabs.harvard.edu/abs/2023A&A...678A.158A}
}

@ARTICLE{Mann_2015,
       author = {{Mann}, Andrew W. and {Feiden}, Gregory A. and {Gaidos}, Eric and {Boyajian}, Tabetha and {von Braun}, Kaspar},
        title = "{How to Constrain Your M Dwarf: Measuring Effective Temperature, Bolometric Luminosity, Mass, and Radius}",
      journal = {\apj},
         year = 2015,
        month = may,
       volume = {804},
       number = {1},
          eid = {64},
        pages = {64},
          doi = {10.1088/0004-637X/804/1/64},
archivePrefix = {arXiv},
       eprint = {1501.01635},
 primaryClass = {astro-ph.SR},
       adsurl = {https://ui.adsabs.harvard.edu/abs/2015ApJ...804...64M}
}

@ARTICLE{Mann_2019,
       author = {{Mann}, Andrew W. and {Dupuy}, Trent and {Kraus}, Adam L. and {Gaidos}, Eric and {Ansdell}, Megan and {Ireland}, Michael and {Rizzuto}, Aaron C. and {Hung}, Chao-Ling and {Dittmann}, Jason and {Factor}, Samuel and {Feiden}, Gregory and {Martinez}, Raquel A. and {Ru{\'\i}z-Rodr{\'\i}guez}, Dary and {Thao}, Pa Chia},
        title = "{How to Constrain Your M Dwarf. II. The Mass-Luminosity-Metallicity Relation from 0.075 to 0.70 Solar Masses}",
      journal = {\apj},
         year = 2019,
        month = jan,
       volume = {871},
       number = {1},
          eid = {63},
        pages = {63},
          doi = {10.3847/1538-4357/aaf3bc},
archivePrefix = {arXiv},
       eprint = {1811.06938},
 primaryClass = {astro-ph.SR},
       adsurl = {https://ui.adsabs.harvard.edu/abs/2019ApJ...871...63M}
}

@ARTICLE{Padova_2000,
       author = {{Girardi}, L. and {Bressan}, A. and {Bertelli}, G. and {Chiosi}, C.},
        title = "{Evolutionary tracks and isochrones for low- and intermediate-mass stars: From 0.15 to 7 M$_{sun}$, and from Z=0.0004 to 0.03}",
      journal = {\aaps},
         year = 2000,
        month = feb,
       volume = {141},
        pages = {371-383},
          doi = {10.1051/aas:2000126},
archivePrefix = {arXiv},
       eprint = {astro-ph/9910164},
 primaryClass = {astro-ph},
       adsurl = {https://ui.adsabs.harvard.edu/abs/2000A&AS..141..371G}
}

@ARTICLE{Fulton_2017,
       author = {{Fulton}, Benjamin J. and {Petigura}, Erik A. and {Howard}, Andrew W. and {Isaacson}, Howard and {Marcy}, Geoffrey W. and {Cargile}, Phillip A. and {Hebb}, Leslie and {Weiss}, Lauren M. and {Johnson}, John Asher and {Morton}, Timothy D. and {Sinukoff}, Evan and {Crossfield}, Ian J.~M. and {Hirsch}, Lea A.},
        title = "{The California-Kepler Survey. III. A Gap in the Radius Distribution of Small Planets}",
      journal = {\aj},
         year = 2017,
        month = sep,
       volume = {154},
       number = {3},
          eid = {109},
        pages = {109},
          doi = {10.3847/1538-3881/aa80eb},
archivePrefix = {arXiv},
       eprint = {1703.10375},
 primaryClass = {astro-ph.EP},
       adsurl = {https://ui.adsabs.harvard.edu/abs/2017AJ....154..109F}
}

@ARTICLE{Vaneylen_2018,
       author = {{Van Eylen}, V. and {Agentoft}, Camilla and {Lundkvist}, M.~S. and {Kjeldsen}, H. and {Owen}, J.~E. and {Fulton}, B.~J. and {Petigura}, E. and {Snellen}, I.},
        title = "{An asteroseismic view of the radius valley: stripped cores, not born rocky}",
      journal = {\mnras},
         year = 2018,
        month = oct,
       volume = {479},
       number = {4},
        pages = {4786-4795},
          doi = {10.1093/mnras/sty1783},
archivePrefix = {arXiv},
       eprint = {1710.05398},
 primaryClass = {astro-ph.EP},
       adsurl = {https://ui.adsabs.harvard.edu/abs/2018MNRAS.479.4786V}
}

@ARTICLE{Mortier_2013,
       author = {{Mortier}, A. and {Santos}, N.~C. and {Sousa}, S. and {Israelian}, G. and {Mayor}, M. and {Udry}, S.},
        title = "{On the functional form of the metallicity-giant planet correlation}",
      journal = {\aap},
         year = 2013,
        month = mar,
       volume = {551},
          eid = {A112},
        pages = {A112},
          doi = {10.1051/0004-6361/201220707},
archivePrefix = {arXiv},
       eprint = {1302.1851},
 primaryClass = {astro-ph.EP},
       adsurl = {https://ui.adsabs.harvard.edu/abs/2013A&A...551A.112M}
}

@ARTICLE{Ciuca_2021,
       author = {{Ciuc{\u{a}}}, Ioana and {Kawata}, Daisuke and {Miglio}, Andrea and {Davies}, Guy R. and {Grand}, Robert J.~J.},
        title = "{Unveiling the distinct formation pathways of the inner and outer discs of the Milky Way with Bayesian Machine Learning}",
      journal = {\mnras},
         year = 2021,
        month = may,
       volume = {503},
       number = {2},
        pages = {2814-2824},
          doi = {10.1093/mnras/stab639},
archivePrefix = {arXiv},
       eprint = {2003.03316},
 primaryClass = {astro-ph.GA},
       adsurl = {https://ui.adsabs.harvard.edu/abs/2021MNRAS.503.2814C}
}

@ARTICLE{casagrande_2011,
       author = {{Casagrande}, L. and {Sch{\"o}nrich}, R. and {Asplund}, M. and {Cassisi}, S. and {Ram{\'\i}rez}, I. and {Mel{\'e}ndez}, J. and {Bensby}, T. and {Feltzing}, S.},
        title = "{New constraints on the chemical evolution of the solar neighbourhood and Galactic disc(s). Improved astrophysical parameters for the Geneva-Copenhagen Survey}",
      journal = {\aap},
         year = 2011,
        month = jun,
       volume = {530},
          eid = {A138},
        pages = {A138},
          doi = {10.1051/0004-6361/201016276},
archivePrefix = {arXiv},
       eprint = {1103.4651},
 primaryClass = {astro-ph.GA},
       adsurl = {https://ui.adsabs.harvard.edu/abs/2011A&A...530A.138C}
}

@ARTICLE{Huber_2012,
       author = {{Huber}, D. and {Ireland}, M.~J. and {Bedding}, T.~R. and {Brand{\~a}o}, I.~M. and {Piau}, L. and {Maestro}, V. and {White}, T.~R. and {Bruntt}, H. and {Casagrande}, L. and {Molenda-{\.Z}akowicz}, J. and {Silva Aguirre}, V. and {Sousa}, S.~G. and {Barclay}, T. and {Burke}, C.~J. and {Chaplin}, W.~J. and {Christensen-Dalsgaard}, J. and {Cunha}, M.~S. and {De Ridder}, J. and {Farrington}, C.~D. and {Frasca}, A. and {Garc{\'\i}a}, R.~A. and {Gilliland}, R.~L. and {Goldfinger}, P.~J. and {Hekker}, S. and {Kawaler}, S.~D. and {Kjeldsen}, H. and {McAlister}, H.~A. and {Metcalfe}, T.~S. and {Miglio}, A. and {Monteiro}, M.~J.~P.~F.~G. and {Pinsonneault}, M.~H. and {Schaefer}, G.~H. and {Stello}, D. and {Stumpe}, M.~C. and {Sturmann}, J. and {Sturmann}, L. and {ten Brummelaar}, T.~A. and {Thompson}, M.~J. and {Turner}, N. and {Uytterhoeven}, K.},
        title = "{Fundamental Properties of Stars Using Asteroseismology from Kepler and CoRoT and Interferometry from the CHARA Array}",
      journal = {\apj},
         year = 2012,
        month = nov,
       volume = {760},
       number = {1},
          eid = {32},
        pages = {32},
          doi = {10.1088/0004-637X/760/1/32},
archivePrefix = {arXiv},
       eprint = {1210.0012},
 primaryClass = {astro-ph.SR},
       adsurl = {https://ui.adsabs.harvard.edu/abs/2012ApJ...760...32H}
}

@ARTICLE{kjeldsen_1995,
       author = {{Kjeldsen}, H. and {Bedding}, T.~R.},
        title = "{Amplitudes of stellar oscillations: the implications for asteroseismology.}",
      journal = {\aap},
         year = 1995,
        month = jan,
       volume = {293},
        pages = {87-106},
          doi = {10.48550/arXiv.astro-ph/9403015},
archivePrefix = {arXiv},
       eprint = {astro-ph/9403015},
 primaryClass = {astro-ph},
       adsurl = {https://ui.adsabs.harvard.edu/abs/1995A&A...293...87K}
}

@ARTICLE{Yu_2018,
       author = {{Yu}, Jie and {Huber}, Daniel and {Bedding}, Timothy R. and {Stello}, Dennis and {Hon}, Marc and {Murphy}, Simon J. and {Khanna}, Shourya},
        title = "{Asteroseismology of 16,000 Kepler Red Giants: Global Oscillation Parameters, Masses, and Radii}",
      journal = {\apjs},
         year = 2018,
        month = jun,
       volume = {236},
       number = {2},
          eid = {42},
        pages = {42},
          doi = {10.3847/1538-4365/aaaf74},
archivePrefix = {arXiv},
       eprint = {1802.04455},
 primaryClass = {astro-ph.SR},
       adsurl = {https://ui.adsabs.harvard.edu/abs/2018ApJS..236...42Y}
}

@ARTICLE{Adibekyan_2021,
       author = {{Adibekyan}, Vardan and {Dorn}, Caroline and {Sousa}, S{\'e}rgio G. and {Santos}, Nuno C. and {Bitsch}, Bertram and {Israelian}, Garik and {Mordasini}, Christoph and {Barros}, Susana C.~C. and {Delgado Mena}, Elisa and {Demangeon}, Olivier D.~S. and {Faria}, Jo{\~a}o P. and {Figueira}, Pedro and {Hakobyan}, Artur A. and {Oshagh}, Mahmoudreza and {Soares}, B{\'a}rbara M.~T.~B. and {Kunitomo}, Masanobu and {Takeda}, Yoichi and {Jofr{\'e}}, Emiliano and {Petrucci}, Romina and {Martioli}, Eder},
        title = "{A compositional link between rocky exoplanets and their host stars}",
      journal = {Science},
         year = 2021,
        month = oct,
       volume = {374},
       number = {6565},
        pages = {330-332},
          doi = {10.1126/science.abg8794},
archivePrefix = {arXiv},
       eprint = {2102.12444},
 primaryClass = {astro-ph.EP},
       adsurl = {https://ui.adsabs.harvard.edu/abs/2021Sci...374..330A}
}

@ARTICLE{Eeles-nolle_2025,
       author = {{Eeles-Nolle}, Fintan and {Armstrong}, David J.},
        title = "{A high stellar multiplicity rate amongst TESS planet candidates in the Neptunian desert using Gaia DR3 astrometry}",
      journal = {\mnras},
         year = 2025,
        month = aug,
       volume = {541},
       number = {2},
        pages = {1419-1433},
          doi = {10.1093/mnras/staf1072},
archivePrefix = {arXiv},
       eprint = {2506.22399},
 primaryClass = {astro-ph.EP},
       adsurl = {https://ui.adsabs.harvard.edu/abs/2025MNRAS.541.1419E}
}





\bsp	
\label{lastpage}
\end{document}